\documentclass[12pt,a4paper]{article}
\usepackage{amsmath,amssymb,bm,ascmac,bbm,mathtools}
\usepackage[dvipdfmx]{graphicx}
\usepackage{color}
\usepackage{authblk}

\newcommand{\pdif}[2]{\frac{\partial #1}{\partial #2}}

\newcommand{\tr}{\text{tr}}

\newcommand{\re}{\text{Re}}
\newcommand{\im}{\text{Im}}

\newcommand{\mbb}{\mathbb}
\newcommand{\mfrak}{\mathfrak}
\newcommand{\la}{\langle}
\newcommand{\ra}{\rangle}
\newcommand{\End}{\text{End}}

\newcommand{\rank}{\text{rank}}
\newcommand{\nn}{\nonumber}
\newcommand{\mn}{{\mu\nu}}
\newcommand{\rs}{{\rho\sigma}}

\newcommand{\lra}{\longrightarrow}

\begin{document}

\title{CFTs on curved spaces}
\author{Ken KIKUCHI}
\affil{Department of Physics and Center for Field Theory and Particle Physics,
Fudan University, 220 Handan Road, 200433 Shanghai, China}
\maketitle

\begin{abstract}
We study conformal field theories (CFTs) on curved spaces including both orientable and unorientable manifolds possibly with boundaries. We first review conformal transformations on curved manifolds. We then compute the identity components of conformal groups acting on various metric spaces using a simple fact; given local coordinate systems be single-valued. Boundary conditions thus obtained which must be satisfied by conformal Killing vectors (CKVs) correctly reproduce known conformal groups. As a byproduct, on $\mbb S^1_l\times\mbb H^2_r$, by setting their radii $l=Nr$ with $N\in\mbb N^\times$, we find (the identity component of) the conformal group enhances, whose persistence in higher dimensions is also argued. We also discuss forms of correlation functions on these spaces using the symmetries. Finally, we study a $d$-torus $\mbb T^d$ in detail, and show the identity component of the conformal group acting on the manifold in general is given by $\text{Conf}_0(\mbb T^d)\simeq U(1)^d$ when $d\ge2$. Using the fact, we suggest some candidates of conformal manifolds of CFTs on $\mbb T^d$ without assuming the presence of supersymmetry (SUSY). In order to clarify which parts of correlation functions are physical, we also discuss renormalization group (RG) and local counterterms on curved spaces.
\end{abstract}

\tableofcontents

\makeatletter
\renewcommand{\theequation}
{\arabic{section}.\arabic{equation}}
\@addtoreset{equation}{section}
\makeatother

\section{Introduction}
Dirac suggested theoretical workers to take indirect ways through generalizations of pure mathematics taking non-Euclidean geometry and non-commutative algebra as examples \cite{Dirac31}. Recalling the history of theoretical physics, human beings actually followed his words; some of the famous (in the sense that we know) examples are supersymmetry (SUSY) and spinors. SUSY was first discussed on flat spaces \cite{SUSY}, while human beings started to discuss the symmetry on curved manifolds \cite{curvedSUSY} motivated by localization \cite{localization}.  Spinors was of course invented by Dirac \cite{spinor}, discussed on flat spaces first, and later theoretical workers started to discuss spinor representations on possibly unorientable manifolds (see, for example, \cite{pin} for excellent reviews) at least partially motivated by symmetry protected topological phases.

Speaking of quantum field theories (QFTs), conformal field theories\footnote{Although it is just a matter of terminology, we call QFTs (defined on a metric space $M$) which enjoy (the identity component of) the conformal symmetry $\text{Conf}_0(M)$ as CFTs for brevity even if the conformal group reduces to (the identity component of) the isometry group $\text{Isom}_0(M)$. This terminology is different from the standard definition of CFTs, i.e., tracelessness of the energy-momentum tensors as operators. We thank Yuji Tachikawa for pointing out this point. However, since we just focus on forms of correlation functions required by (conformal) isometries, there would be no difference.} (CFTs) have always been good playing ground to study QFTs \cite{CFT}. For example, in the Wilsonian renormalization group (RG) \cite{Wilson}, we `integrate out' higher momentum (a.k.a. heavy) modes and deal with just lower momentum (a.k.a. light) modes left in effective theories. Thus, it is intuitively expected that the `number' of degrees of freedom decreases monotonically under RG flows. In fact, this intuition was proved in two and four spacetime dimensions, and they are called $c$-theorem \cite{cthm} and $a$-theorem \cite{athm}, respectively.\footnote{For the case of three dimensions, see \cite{Fthm}.} These quantities, namely $c$ and $a$, are part of the conformal anomalies. Therefore, by studying CFTs and their conformal anomalies, we can judge if a putative RG flow is possible or not. The counting also gives us partial information of phases. In addition, in the context of Wilsonian RG, it is possible to classify QFTs according to their universality class if they flow to some fixed points. Since most (if not all) of the fixed-point theories are described by CFTs, we can extract universal properties of QFTs in the same universality class by exploring CFTs.

Surprisingly, however, most of the study of CFTs were limited to those on conformally flat spacetimes. A typical example of curved conformally flat space is a $d$-sphere $\mbb S^d$. Although CFTs on $\mbb R^d$ suffer from infrared (IR) divergences, a $d$-sphere $\mbb S^d$, which is compact and, thus, free of IR divergence, is a nice place to work on. In particular, sphere partition functions are free of IR divergences. That is why the objects have been studied extensively. See for example \cite{JKLMR,GL,GGK}. Some other exceptions (to the best of our knowledge) of CFTs on curved spaces are discussed on Riemann surfaces motivated by perturbative string theory \cite{pertstring}, on hyperbolic spaces \cite{hyper}\footnote{We thank Lorenzo Di Pietro for bringing these papers to our attention.}, and on real projective spaces \cite{realpro} whose latest papers were motivated by conformal bootstrap \cite{bootstrap}.

In this paper, we would like to follow at least the spirit of Dirac and put CFTs on the same direction of generalization, namely we would like to study CFTs on curved manifolds possibly with boundaries. As a preparation, we review conformal transformations on a given (curved) manifold in section \ref{conf}. Conformal transformations are generated by conformal Killing vectors (CKVs), which are solutions of conformal Killing equations (CKEs). CKVs on some curved manifolds have been studied in the community of gravitational physics and mathematics. For example, CKVs on plane-symmetric spacetime were computed in \cite{KHBK}. For a history of CKVs, see its introduction and references therein. In section \ref{conf}, we also consider comformal maps between two manifolds. We explicitly show that conformal transformations on one manifold are sent to conformal transformations on the other by a conformal map between them. More precisely, the differential (a.k.a. pushforward) of a conformal map sends CKVs to CKVs. In section \ref{ex}, we study some examples. In order to compute conformal groups, one has to impose suitable boundary conditions obeyed by CKVs. To determine the boundary conditions, we use a simple fact; given local coordinate systems be single-valued. When results are available, we find the boundary conditions derived from the fact correctly reproduce known conformal groups. As a byproduct, on $\mbb S^1_l\times\mbb H^2_r$, we find unexpected enhancement of (the identity component of) the conformal group when their radii are set to $l=Nr$ with $N\in\mbb N^\times$. Via conformal maps, the space is related to a flat space with codimension two defect. This fact makes it clear why the space has the conformal group $SO(2,1)\times U(1)$. We give an argument that this phenomenon also takes place in higher-dimensional $\mbb S^1_l\times\mbb H^{d-1}_r$. We also study forms of correlation functions on these spaces using the conformal symmetries, and discuss which part of them are physical by taking local counterterms on curved spaces into account.

Through the study, we found periodic bounary conditions simplify our task of solving CKEs by reducing partial differential equations to algebraic equations. Hence in section \ref{dtorus}, we study a $d$-torus $\mbb T^d$. In particular, we show the identity component of the conformal group acting on the manifold is given by $\text{Conf}_0(\mbb T^d)\simeq U(1)^d\big(\simeq\text{Isom}_0(\mbb T^d)\big)$ when $d\ge2$. We also study CFTs defined on the manifold, and their conformal manifolds \cite{confmfd}. We suggest some candidates of conformal manifolds without assuming SUSY utilizing the compactness of the conformal group. In section \ref{discussion}, we summarize our results, and comment on future directions. In the appendices, we describe details of the proof that the differential of a conformal map sends CKVs to CKVs, explain relations of groups $SU(1,1)$, $SL(2,\mbb R)$, and $SO(2,1)$, discuss general aspects of RG and local counterterms on curved spaces, and describe more examples.

\section{Conformal transformations on curved manifolds}\label{conf}
Consider two $d$-dimensional real metric spaces\footnote{Throughout the paper, we work in the Euclidean signature.} $(M,\gamma)$ and $(N,\gamma')$. A map $\varphi:(M,\gamma)\to(N,\gamma')$ is conformal\footnote{We are not sure where they appeared first, but some results in this section can be found in \cite{Wald}, for example.} iff
\begin{equation}
    (0\neq)\exists\Omega\in C^\infty(M)\ s.t.\ \varphi^*\gamma'=\Omega^2\gamma.\label{confmap}
\end{equation}
Conformal transformations of a given CFT on a metric space $(M,\gamma)$ are special classes of the above, namely the case where a map is defined between the same space\footnote{One may think this condition is too restrictive because the definition rules out a naive transformation caused by changing the scale of the manifold. However, we think the definition is more appropriate for QFTs (without dynamical metric) because we usually do not consider transformations that modify the geometry. So we restrict ourselves to this definition.}
\[ \varphi_M:(M,\gamma)\to(M,\gamma) \]
with
\begin{equation}
    (0\neq)\exists\Omega_M\in C^\infty(M)\ s.t.\ \varphi_M^*\gamma=\Omega_M^2\gamma.\label{conftrans}
\end{equation}
From now on we write
\[ \Omega_M^2=e^{2\sigma_M} \]
for a reason that will be clear in short.

To write down the condition (\ref{conftrans}) more explicitly, let us consider an infinitesimal\footnote{In this analysis, we are restricting ourselves to conformal transformations which are connected to the identity of the group Conf$(M)$, which we will call Conf$_0(M)$, by essentially employing the infinitesimal property of $\xi_M$. It would be interesting to study `large' conformal transformations, which are not connected to the identity.} coordinate transformation\footnote{Since conformal transformations are invertible by definition, and since we are only considering the case of smooth conformal transformations, we have $\text{Conf}(M)\subset\text{Diff}(M)$. On the other hand, since the Weyl transformation  $\gamma_\mn(x)\mapsto e^{2\sigma(x)}\gamma_\mn(x)$ is $not$ a coordinate transformation, this is $not$ an element of the conformal group, although these Weyl transformations are also called conformal transformations in the community of gravitational physics.\label{weyl}}
\begin{equation}
    \varphi_M:x^\mu\mapsto x'^\mu(x):=x^\mu+\xi_M^\mu(x).\label{xi}
\end{equation}
Note that since $x$ is real, we must require $\xi_M$ be also real valued. For the transformation (\ref{xi}) to be conformal (\ref{conftrans}), we must have
\begin{align*}
    \gamma_\rs(x'(x))dx'^\rho(x) dx'^\sigma(x)&=\gamma_\rs(x+\xi_M(x))\pdif{x'^\rho}{x^\mu}(x)\pdif{x'^\sigma}{x^\nu}(x)dx^\mu dx^\nu\\
    &\stackrel!=e^{2\sigma_M(x)}\gamma_\mn(x)dx^\mu dx^\nu,
\end{align*}
or using the linear independence of differential forms, we obtain
\[ \nabla^{(\gamma)}_\mu\xi_{M\nu}(x)+\nabla^{(\gamma)}_\nu\xi_{M\mu}(x)=2\sigma_M(x)\gamma_\mn(x) \]
where $\nabla^{(\gamma)}$ is a covariant derivative constructed with the Levi-Civita (LC) connection calculated with the metric $\gamma$ and we used the fact that $\sigma_M$ is infinitesimal because $\varphi_M$ is so. This is why we employed the notation $\Omega_M^2=e^{2\sigma_M}$. We can determine the unknown function $\sigma_M$ by taking a trace of the equation:
\begin{equation}
    2\sigma_M(x)=\frac2d\nabla^{(\gamma)}\cdot\xi_M(x).\label{2sigma}
\end{equation}
Therefore, we arrive
\begin{equation}
    \nabla^{(\gamma)}_\mu\xi_{M\nu}(x)+\nabla^{(\gamma)}_\nu\xi_{M\mu}(x)=\frac2d\left(\nabla^{(\gamma)}\cdot\xi_M(x)\right)\gamma_\mn(x),\label{CKE}
\end{equation}
which is called the conformal Killing equation (CKE). Solutions $\xi_M$ are called conformal Killing vectors (CKVs).\footnote{Some terminologies: if the conformal factor $2\sigma_M$ is zero, $\xi_M$ is nothing but a Killing vector (KV), and if the factor is nonzero, $\xi_M$ is called a conformal Killing vector. More precisely, if the factor is a constant, $\xi_M$ is called a homothetic vector, and if it is not a constant, $\xi_M$ is called a proper conformal Killing vector. Furthermore, if $\nabla^{(\gamma)}_\mu\nabla^{(\gamma)}_\nu\sigma_M=0$, $\xi_M
$ is called special, and if nonzero, $\xi_M$ is called non-special.} In short, all we have to do to obtain conformal transformations is to solve the CKEs with suitable boundary conditions and the reality $\xi_M^*=\xi_M$ imposed.

Sometimes, it would be convenient to have an alternative form of (\ref{CKE}). Taking a covariant derivative of the equation, we get
\[ \nabla^{(\gamma)}_\rho\nabla^{(\gamma)}_\mu\xi_{M\nu}(x)+\nabla^{(\gamma)}_\rho\nabla^{(\gamma)}_\nu\xi_{M\mu}(x)=\frac2d\gamma_\mn(x)\nabla^{(\gamma)}_\rho\Big(\nabla^{(\gamma)}\cdot\xi_M(x)\Big). \]
Taking linear combinations with cyclically exchanged spacetime indices, and using the definitions of Riemann tensors $[\nabla^{(\gamma)}_\mu,\nabla^{(\gamma)}_\nu]v^\rho\equiv{R^\rho}_{\sigma\mn}v^\sigma$ and $R_\mn:={R^\rho}_{\mu\rho\nu}$, one arrives
\begin{equation}
    \nabla^{(\gamma)2}\xi_{M\mu}(x)+R_\mn\xi_M^\nu(x)=\frac{2-d}d\nabla^{(\gamma)}_\mu\left(\nabla^{(\gamma)}\cdot\xi_M(x)\right),\label{CKE2}
\end{equation}
where $\nabla^{(\gamma)2}:=\gamma^\mn\nabla^{(\gamma)}_\mu\nabla^{(\gamma)}_\nu$. In the case of flat metric $\gamma=\eta$, these CKEs reduce to the familiar forms
\begin{align*}
    \partial_\mu\xi_{M\nu}(x)+\partial_\nu\xi_{M\mu}(x)&=\frac2d\left(\partial\cdot\xi_M(x)\right)\eta_\mn,\\
    \partial^2\xi_{M\mu}(x)&=\frac{2-d}d\partial_\mu\left(\partial\cdot\xi_M(x)\right).
\end{align*}

Finally, we would like to make a comment on the number of constraints imposed by the CKEs. Since the CKE (\ref{CKE}) is symmetric in its spacetime indices, these PDEs naively impose $\frac{d(d+1)}2$ conditions, however, one notices that the trace part is trivially satisfied because we have fixed $\sigma_M(x)$ so. This becomes evident if we rewrite (\ref{CKE}) as
\[ 0=\nabla^{(\gamma)}_\mu\xi_{M\nu}(x)+\nabla^{(\gamma)}_\nu\xi_{M\mu}(x)-\frac2d\left(\nabla^{(\gamma)}\cdot\xi_M(x)\right)\gamma_\mn(x). \]
In fact, taking the trace of both hand sides, we obtain $0=0$, a tautology. Thus, the trace part does not produce new constraints. In other words, one condition of the `diagonal' part can be reproduced from the others. Thus, the CKEs impose
\begin{equation}
    N(d):=\frac{d(d+1)}2-1\label{Nd}
\end{equation}
conditions.\footnote{Logically, some conditions imposed by CKEs can be linearly dependent. For example, when our manifold is given by a product manifolds, and some of the components are identical, sets of constraints on the identical components would be the same. In those cases, the number of constraints imposed by CKEs would become smaller than (\ref{Nd}). If one considers metric spaces with the most general parameters, this degeneracy would not happen. For example, as we will see later, this special cases can be avoided if we choose radii $l_i$ of $\mbb S^1$'s in a $d$-torus $\mbb T^d$ with $d\ge2$ general enough, meaning that ratio squared $(l_i/l_j)^2$ with $i\neq j$ are all irrational. It would be interesting to relax this assumption and study cases with these degeneracies. On the other hand, however, if one considers some special parameters, some conditions (although the conditions are not those on CKV themselves but on their modes) do degenerate, and one can indeed see a phenomenon of conformal symmetry enhancement as we will show through an example in section \ref{secS1xH2}.} If we plug in $d=1$, we get $N(1)=0$, i.e., CKEs are trivially satisfied, and all transformations (\ref{xi}) with suitable boundary conditions and reality imposed generate conformal transformations.

Now, suppose we have found conformal transformations on $(M,\gamma)$, which we call $\varphi_M$, i.e., $\varphi_M:(M,\gamma)\to(M,\gamma)$ with $(0\neq)\exists\Omega_M\in C^\infty(M)\ s.t.\ \varphi_M^*\gamma=\Omega_M^2\gamma$. Given a conformal map $\varphi:(M,\gamma)\to(N,\gamma')$, i.e., $(0\neq)\exists\Omega\in C^\infty(M)\ s.t.\ \varphi^*\gamma'=\Omega^2\gamma$, we can construct another map between $(N,\gamma')$:
\begin{equation}
    \varphi_N:=\varphi\circ\varphi_M\circ\varphi^{-1}.\label{vphiN}
\end{equation}
We would like to ask whether this is a conformal transformation on $(N,\gamma')$, i.e., whether there exists $(0\neq)\Omega_N\in C^\infty(N)$ such that $\varphi_N^*\gamma'=\Omega_N^2\gamma'$.

Equivalently, given CKVs $\xi_M$ on $(M,\gamma)$, which are vector fields $\xi_M\in\mfrak X(M)$, using the pushforward $\varphi_*$, we can get new vector fields $\varphi_*\xi_M\in\mfrak X(N)$. We would like to ask whether the new vector fields are CKVs on $(N,\gamma')$, i.e.,
\[ \nabla^{(\gamma')}_\alpha(\varphi_*\xi_M)_\beta+\nabla^{(\gamma')}_\beta(\varphi_*\xi_M)_\alpha\stackrel?=\frac2d\Big(\nabla^{(\gamma')}\cdot(\varphi_*\xi_M)\Big)\gamma'_{\alpha\beta}. \]

Using the identities
\[ \varphi^{-1}\circ\varphi=id_M,\quad\varphi\circ\varphi^{-1}=id_N, \]
we obtain
\begin{equation}
    (\varphi^{-1})^*\gamma=\Omega^{-2}\circ\varphi^{-1}\cdot\gamma'.\label{pullbackinverse}
\end{equation}
In other words, the inverse of the conformal map is also a conformal map with the inverse conformal factor. Using (\ref{pullbackinverse}) and the property of the pullback $(f\circ g)^*=g^*\circ f^*$, we can easily answer the first question positively:
\begin{align*}
    \varphi_N^*\gamma'&\equiv(\varphi\circ\varphi_M\circ\varphi^{-1})^*\gamma'\\
    &=(\varphi^{-1})^*\circ\varphi_M^*\circ\varphi^*\gamma'\\
    &=\Omega^2\circ\varphi_M\circ\varphi^{-1}\cdot\Omega_M^2\circ\varphi^{-1}\cdot\Omega^{-2}\circ\varphi^{-1}\cdot\gamma',
\end{align*}
namely we have shown that the map (\ref{vphiN}) is a conformal transformation on $(N,\gamma')$ with a conformal factor
\begin{equation}
    \Omega_N^2:=\Omega^2\circ\varphi_M\circ\varphi^{-1}\cdot\Omega_M^2\circ\varphi^{-1}\cdot\Omega^{-2}\circ\varphi^{-1}.\label{OmegaN}
\end{equation}

We can also show that pushforwards of the CKVs on $(M,\gamma)$ are CKVs on $(N,\gamma')$ by studying $\varphi_N^*\gamma'=\Omega_N^2\gamma'$ in components; we write $\varphi^{-1\mu}(y)=x^\mu$, and
\[ \varphi_N:y^\alpha\mapsto y'^\alpha(y):=y^\alpha+\xi_N^\alpha(y), \]
which is infinitesimal because $\varphi_M$ is so. In fact, using the notation, we have
\begin{align*}
    y^\alpha+\xi_N^\alpha(y)&=\varphi^\alpha(\varphi^{-1}(y)+\xi_M(\varphi^{-1}(y)))\\
    &\simeq\varphi^\alpha(\varphi^{-1}(y))+\xi_M^\mu(\varphi^{-1}(y))\pdif{}{x^\mu}\varphi^\alpha(\varphi^{-1}(y))\\
    &=y^\alpha+\xi_M^\mu(\varphi^{-1}(y))\pdif{y^\alpha}{x^\mu}(\varphi^{-1}(y)).
\end{align*}
This confirms the fact that
\[ \xi_N^\alpha(y)=\xi_M^\mu(\varphi^{-1}(y))\pdif{y^\alpha}{x^\mu}(\varphi^{-1}(y)) \]
is infinitesimal.\footnote{Since we are considering only smooth conformal transformations, the Jacobian $\partial y/\partial x$ cannot be singular.} Note that this is nothing but the pushforward of $\xi_M$:
\begin{equation}
    (\varphi_*\xi_M)^\alpha=\xi_M^\mu\pdif{y^\alpha}{x^\mu}=\xi_N^\alpha.\label{pushforward}
\end{equation}
Then, we have
\begin{align*}
    (\text{LHS})&=\gamma'_{\gamma\delta}(y'(y))dy'^\gamma(y)dy'^\delta(y)\\
    &=\gamma'_{\gamma\delta}(y+\xi_N(y))\pdif{y'^\gamma}{y^\alpha}(y)\pdif{y'^\delta}{y^\beta}(y)dy^\alpha dy^\beta\\
    &\simeq\Big[\gamma'_{\alpha\beta}(y)+\nabla^{(\gamma')}_\alpha\xi_{N\beta}(y)+\nabla^{(\gamma')}_\beta\xi_{N\alpha}(y)\Big]dy^\alpha dy^\beta\\
    &=\Omega^2(\varphi_M(\varphi^{-1}(y)))\Omega_M^2(\varphi^{-1}(y))\Omega^{-2}(\varphi^{-1}(y))\gamma'_{\alpha\beta}(y)dy^\alpha dy^\beta\\
    &=\Big[\Omega^2(\varphi^{-1}(y)+\xi_M(\varphi^{-1}(y)))\Omega_M^2(\varphi^{-1}(y))\Omega^{-2}(\varphi^{-1}(y))\Big]\gamma'_{\alpha\beta}(y)dy^\alpha dy^\beta\\
    &\simeq\Big[1+\xi_M^\mu(\varphi^{-1}(y))\pdif{\ln\Omega^2}{x^\mu}(\varphi^{-1}(y))+2\sigma_M(\varphi^{-1}(y))\Big]\gamma'_{\alpha\beta}(y)dy^\alpha dy^\beta,
\end{align*}
thus
\[ \nabla^{(\gamma')}_\alpha\xi_{N\beta}(y)+\nabla^{(\gamma')}_\beta\xi_{N\alpha}(y)=\Big[\xi_M^\mu(\varphi^{-1}(y))\pdif{\ln\Omega^2}{x^\mu}(\varphi^{-1}(y))+\frac2d\nabla^{(\gamma)}\cdot\xi_M(\varphi^{-1}(y))\Big]\gamma'_{\alpha\beta}(y). \]
In the last line, we used (\ref{2sigma}). Since we have already established that $\varphi_N$ is a conformal transformation, what we have computed implies
\[ 2\sigma_N(y)=\frac2d\nabla^{(\gamma')}\cdot\xi_N(y)=\xi_M^\mu(\varphi^{-1}(y))\pdif{\ln\Omega^2}{x^\mu}(\varphi^{-1}(y))+\frac2d\nabla^{(\gamma)}\cdot\xi_M(\varphi^{-1}(y)). \]
Note that we have just $defined$ the overall factor of $\gamma'$ as $2\sigma_N$. To answer our second question that whether pushforwards of CKVs are also CKVs, we have to show the equality
\begin{equation}
    \frac2d\nabla^{(\gamma')}\cdot\xi_N(y)=\xi_M^\mu(\varphi^{-1}(y))\pdif{\ln\Omega^2}{x^\mu}(\varphi^{-1}(y))+\frac2d\nabla^{(\gamma)}\cdot\xi_M(\varphi^{-1}(y)).\label{conffactors}
\end{equation}
In fact we can show this relation. See the appendix \ref{pushexp} for details. The relation establishes our expectation
\begin{equation}
    \nabla^{(\gamma')}_\alpha(\varphi_*\xi_M)_\beta+\nabla^{(\gamma')}_\beta(\varphi_*\xi_M)_\alpha=\frac2d\Big(\nabla^{(\gamma')}\cdot(\varphi_*\xi_M)\Big)\gamma'_{\alpha\beta}.\label{pushCKV}
\end{equation}
In other words, given CKVs $\xi_M$ on $(M,\gamma)$, their pushforwards $\varphi_*\xi_M$ are also CKVs on $(N,\gamma')$. Since the pushforwards are linear and bijective, they preserve vector space structures. Furthermore, since $\xi_M$ and $\xi_N$ are $\varphi$-related, pushforwads also preserve commutator products. Therefore, pushforwards give isomorphisms between conformal algebras:
\begin{equation}
    \mfrak{conf}(M)\cong\mfrak{conf}(N).\label{confMN}
\end{equation}
As an immediate corollary, we get
\[ \dim\text{Conf}(M)=\dim\text{Conf}(N). \]
However, note that the isomorphism between Lie algebras does not mean full conformal groups are isomorphic. We will see an example in which two metric spaces related by a conformal map have different global structures. We would also like to mention that (\ref{confMN}) is a statement about (closed) manifolds $M$ and $N$ which are both completely covered by the conformal map. If the manifolds have boundaries or a conformal map does not cover the whole region, boundary conditions can break the relation. Therefore, it is important to keep track of which parts are covered by conformal maps. We will also see an example of this breaking.

CFTs on conformally flat spaces (possibly some points removed) such as $\mbb S^d_r$ or $\mbb S^1_r\times\mbb H^{d-1}_r$ can be obtained via conformal maps between $\mbb R^d$ (plus or minus some points). The conformal map between $\mbb S^d_r$ and $\mbb R^d\cup\{\infty\}$ is well known under the name of stereographic projection:
\[ \begin{pmatrix}X^1\\X^2\\\vdots\\X^d\\X^{d+1}\end{pmatrix}=\frac r{1+\frac{|x|^2}{4r^2}}\begin{pmatrix}x^1/r\\x^2/r\\\vdots\\x^d/r\\\frac{|x|^2}{4r^2}-1\end{pmatrix}\in\mbb R^{d+1}, \]
where $|x|^2:=(x^1)^2+\cdots+(x^d)^2$. $X$ labels a point on $\mbb S^d_r$:
\[ (X^1)^2+\cdots+(X^{d+1})^2=r^2. \]
In fact, the coordinate system manifests conformal flatness:
\[ ds_{\mbb S^d_r}^2=ds_{\mbb R^{d+1}}^2\Big|_{r;\text{fixed}}=\sum_{j=1}^{d+1}(dX^j)^2\Big|_{r;\text{fixed}}=\left(\frac1{1+\frac{|x|^2}{4r^2}}\right)^2\sum_{j=1}^d(dx^j)^2. \]
To see the conformal flatness of $\mbb S^1_r\times\mbb H^{d-1}_r$, it is useful to employ a coordinate system
\[
\begin{split}
    ds_{\mbb R^d}^2&=dt^2+d\rho^2+\rho^2ds_{\mbb S^{d-2}_1}^2,\\
    t\in(-&\infty,+\infty),\quad\rho\in[0,+\infty).
\end{split}
\]
Then, the conformal map between $\mbb R^d$ and $\mbb S^1_r\times\mbb H^{d-1}_r$ is given by
\[ t=r\frac{\sin\tau}{\cosh\eta+\cos\tau},\quad\rho=r\frac{\sinh\eta}{\cosh\eta+\cos\tau}. \]
The inverse map is given by
\[ \coth\eta=\frac{r^2+t^2+\rho^2}{2r\rho},\quad\cot\tau=\frac{r^2-t^2-\rho^2}{2rt}. \]
Using the transformation, we obtain
\[
\begin{split}
    dt^2+d\rho^2+\rho^2ds_{\mbb S^{d-2}_1}^2&=\left(\frac1{\cosh\eta+\cos\tau}\right)^2\Big\{r^2d\tau^2+r^2\left[d\eta^2+\sinh^2\eta ds_{\mbb S^{d-2}_1}^2\right]\Big\},\\
    &\tau\in[0,2\pi),\quad\eta\in[0,+\infty).
\end{split}
\]
This conformal map was used, for example, in \cite{CHM}. Also, the conformal map between $\mbb S^d_r$ (minus the great sphere $\mbb S^{d-2}_{\phi=\pi}$) and $\mbb S^1_r\times\mbb H^{d-1}_r$ is known (see e.g. \cite{HZ}); in the coordinate system\footnote{To make the domain of $\eta$ nonnegative, $[0,+\infty)$, we changed the domain of $\phi$ from $[0,\pi/2]$ to $[\pi/2,\pi]$. Then, $\eta\in[0,+\infty)$ corresponds to $\phi\in[\pi/2,\pi)$.}
\[
\begin{split}
    ds_{\mbb S^d_r}^2/r^2&=d\phi^2+\sin^2\phi d\tau^2+\cos^2\phi ds_{\mbb S^{d-2}_1}^2,\\
    &\phi\in[\pi/2,\pi],\quad\tau\in[0,2\pi),
\end{split}
\]
the map is simply given by
\[ \sinh\eta=-\cot\phi. \]
In fact, we obtain
\[ ds_{\mbb S^d_r}^2/r^2=\frac1{\cosh^2\eta}ds_{\mbb S^1_r\times\mbb H^{d-1}_r}^2/r^2. \]
Therefore, using these conformal maps, one can easily translate results of CFTs on $\mbb R^d$ (or its one-point compactification $\mbb R^d\cup\{\infty\}$) to those on $\mbb S^1_r\times\mbb H^{d-1}_r$ or $\mbb S^d_r$ once suitable boundary conditions are imposed (and nontrivial operator mixings caused by local counterterms including background metric \cite{GGIKKP} are taken into account).

\section{Examples}\label{ex}
It would be instructive to study a few metric spaces to understand the formalism of the previous section. So we explore CFTs on some manifolds including both orientable and unorientable ones, possibly with boundaries, in this section. For more examples, see appendix \ref{moreex}.

We study the examples in the following steps; (1) fix a metric space, (2) specify boundary conditions, (3) use some boundary conditions to write down an ansatz, (4) solve the CKE imposing the boundary conditions and the reality, (5) compute the Lie algebra formed by generators, and (6) consider forms of $n$-point functions, and whether they are physical by taking local counterterms into account. In (2), we use a simple fact that a given local coordinate system be single-valued. Therefore, if two points $p,p'\in M$ are identified, they must have the same value in any given local coordinate systems $x(p)\stackrel!=x(p')$. For $p$ and $p'$ are identified in $M$, the conditions do not contradict the local nature of our analysis even though we call them boundary conditions for brevity. Since conformal transformations are subset of coordinate transformations $x\mapsto x'$, we consider two local coordinate systems $x$ and $x'$. So we have two different ways to express $x'(p')$, and these two expressions give us boundary conditions CKVs must obey. In (5), if possible, we comment on the global structures of the conformal groups.

\subsection{$\mbb S^1_l\times\mbb H^1_r$}
As a warm-up, let us study $\mbb S^1_l\times\mbb H^1_r$.\footnote{One may notice that this example is trivial. Indeed it is, however, we will study anyway because we would like to use the result in the next example.} The metric is given by
\begin{equation}
\begin{split}
    ds^2=l^2d\tau^2&+r^2d\eta^2,\\
    \tau\in[0,2\pi),\quad&\eta\in[0,+\infty).
\end{split}\label{S1xH1}
\end{equation}
Since LC connections are trivial, CKE (\ref{CKE}) reduces to\footnote{From now on, we denote CKVs $\xi$ instead of $\xi_M$ because on which metric space we are working on would be clear.}
\begin{equation}
    \partial_\mu\xi_\nu(x)+\partial_\nu\xi_\mu(x)=\gamma_\mn(x)\left(\frac1{l^2}\partial_\tau\xi_\tau(x)+\frac1{r^2}\partial_\eta\xi_\eta(x)\right).\label{CKES1xH1}
\end{equation}
Case analysis yields
\begin{equation}
\begin{split}
    0&=\partial_\tau\xi_\tau(x)-\left(\frac lr\right)^2\partial_\eta\xi_\eta(x),\\
    0&=\partial_\tau\xi_\eta(x)+\partial_\eta\xi_\tau(x).
\end{split}\label{CKES1xH1'}
\end{equation}
We should solve these equations with a boundary condition
\begin{equation}
    \forall\eta\in[0,+\infty),\quad\xi_\mu(\tau+2\pi,\eta)=\xi_\mu(\tau,\eta).\label{S1xH1bc}
\end{equation}
The most general form of $\xi$ consistent with the boundary condition would be given by
\begin{equation}
    \xi_\mu(\tau,\eta)=\sum_{m\in\mbb Z}e^{im\tau}\xi_\mu^{(m)}(\eta).\label{S1xH1ansatz}
\end{equation}
To make this real, we must impose
\begin{equation}
    \forall m\in\mbb Z,\forall\eta\in[0,+\infty),\quad\xi_\mu^{(m)}(\eta)=\xi_\mu^{(-m)*}(\eta).\label{xi_muS1xH1}
\end{equation}
Substituting the ansatz in (\ref{CKES1xH1'}), we obtain
\[ \partial_\eta\begin{pmatrix}\xi_\tau^{(m)}(\eta)\\\xi_\eta^{(m)}(\eta)\end{pmatrix}=im\begin{pmatrix}0&-1\\\left(\frac rl\right)^2&0\end{pmatrix}\begin{pmatrix}\xi_\tau^{(m)}(\eta)\\\xi_\eta^{(m)}(\eta)\end{pmatrix}. \]
We can diagonalize the matrix on the RHS as
\[ \begin{pmatrix}0&-1\\a&0\end{pmatrix}=SJS^{-1} \]
where
\[ S:=\begin{pmatrix}-\frac i{\sqrt a}&\frac i{\sqrt a}\\1&1\end{pmatrix},\quad J:=\begin{pmatrix}-i\sqrt a&0\\0&i\sqrt a\end{pmatrix}. \]
Thus, the equation can be solved with ease:
\[ S^{-1}\begin{pmatrix}\xi_\tau^{(m)}(\eta)\\\xi_\eta^{(m)}(\eta)\end{pmatrix}=\begin{pmatrix}e^{mr\eta/l}&0\\0&e^{-mr\eta/l}\end{pmatrix}\begin{pmatrix}c_\tau^{(m)}\\c_\eta^{(m)}\end{pmatrix}, \]
where $c_\mu^{(m)}$ are complex constants. The reality condition (\ref{xi_muS1xH1}) is now given by
\begin{equation}
    \forall m\in\mbb Z,\quad c_\eta^{(m)}=c_\tau^{(-m)*}.\label{c_muS1xH1}
\end{equation}
Using this condition, we arrive
\begin{equation}
    \begin{pmatrix}\xi_\tau(\tau,\eta)\\\xi_\eta(\tau,\eta)\end{pmatrix}=\sum_{m\in\mbb Z}e^{im\tau}\begin{pmatrix}\frac{il}r\left(-e^{mr\eta/l}c_\tau^{(m)}+e^{-mr\eta/l}c_\tau^{(-m)*}\right)\\e^{mr\eta/l}c_\tau^{(m)}+e^{-mr\eta/l}c_\tau^{(-m)*}\end{pmatrix}.\label{xiS1xH1'}
\end{equation}

We have not yet imposed any boundary condition in the $\eta$-direction. To preserve the `boundary' $\eta=0$, it would be natural to require the $\eta$-component of the CKV vanish at $\eta=0$:\footnote{Depending on the problems we consider, we also have to impose suitable boundary conditions at $\eta\to+\infty$, but we do not consider the problem in this paper.}
\begin{equation}
    \forall\tau\in[0,2\pi),\quad\xi_\eta(\tau,\eta=0)\stackrel!=0.\label{bceta=0}
\end{equation}
The boundary condition implies
\[ \sum_{m\in\mbb Z}e^{im\tau}\Big\{c_\tau^{(m)}+c_\tau^{(-m)*}\Big\}=0, \]
or using the orthogonality of the basis $e^{im\tau}$, we obtain
\begin{equation}
    \forall m\in\mbb Z,\quad c_\tau^{(m)}+c_\tau^{(-m)*}=0.\label{ctaubc}
\end{equation}

Then, (\ref{xiS1xH1'}) further reduces to
\begin{equation}
    \begin{pmatrix}\xi_\tau(\tau,\eta)\\\xi_\eta(\tau,\eta)\end{pmatrix}=\sum_{m\in\mbb Z}e^{im\tau}\begin{pmatrix}-\frac{2il}rc_\tau^{(m)}\cosh\frac{mr\eta}l\\
    2c_\tau^{(m)}\sinh\frac{mr\eta}l\end{pmatrix}.\label{xiS1xH1}
\end{equation}

In short, we have obtained an infinite number of conformal transformations on $M=\mbb S^1_l\times\mbb H^1_r$. This result is expected because it is well known that $\mbb R^2$ (and not $\mbb S^2$, in the sense that the regularity of CKVs at infinity is not imposed) enjoys infinite dimensional conformal algebra, and our coordinate system (\ref{S1xH1}) is just a rescaled version of the familiar coordinate system to describe the complex plane $\mbb C^1\approx\mbb R^2$ employed in the radial quantization. This becomes clear if we introduce a coordinate
\[ z:=r\eta+il\tau,\quad\bar z=r\eta-il\tau. \]
Then, the metric (\ref{S1xH1}) can be rewritten as
\[ ds^2=\frac12(dzd\bar z+d\bar zdz). \]
In this coordinate system, linear combinations of the CKVs are `holomorphic':
\[ \xi^z:=r\xi^\eta+il\xi^\tau=\frac2r\sum_{m\in\mbb Z}c_\tau^{(m)}e^{mz/l},\quad\xi^{\bar z}:=r\xi^\eta-il\xi^\tau=-\frac2r\sum_{m\in\mbb Z}c_\tau^{(m)}e^{-m\bar z/l}. \]
Now that it is clear that $\xi^z=\xi^z(z)$ is holomorphic, and it generates holomorphic transformations, which has infinite number of degrees of freedom. $\xi^z$ has the correct periodic property $\xi^z(z+2\pi il)=\xi^z(z)$. $\xi^{\bar z}=\xi^{\bar z}(\bar z)$ has exactly the same properties once $z$ is replaced with $\bar z$. It generates an infinite number of antiholomorphic transformations.

We define
\[ \partial:=\frac12\left(\frac1r\pdif{}\eta-\frac il\pdif{}\tau\right),\quad\bar\partial:=\frac12\left(\frac1r\pdif{}\eta+\frac il\pdif{}\tau\right), \]
so that
\[ \partial z=1,\quad\partial\bar z=0,\quad\bar\partial z=0,\quad\bar\partial\bar z=1. \]
Then, conformal transformations on our manifold $\mbb S^1_l\times\mbb H^1_r$ are generated by
\[ \xi(\tau,\eta)=\xi^\mu(\tau,\eta)\partial_\mu=\xi^z(z)\partial+\xi^{\bar z}(\bar z)\bar\partial\equiv\xi(z)+\bar\xi(\bar z), \]
where
\[ \xi(z):=\xi^z(z)\partial,\quad\bar\xi(\bar z):=\xi^{\bar z}(\bar z)\bar\partial. \]
So generators are identified as $P_m=z^m\partial,\bar P_{\bar m}=\bar z^{\bar m}\bar\partial$ with $m,\bar m\in\mbb N$. Since any holomorphic and antiholomorphic generators commute
\[ [P_n,\bar P_{\bar m}]=0, \]
the conformal group Conf$_0(\mbb S^1_l\times\mbb H^1_r)$ is given by a product of groups consist of holomorphic transformations, which is generated by $\xi(z)$, and of antiholomorphic transformations, which is generated by $\bar\xi(\bar z)$.

Forms of $n$-point functions in CFTs on the manifold are thus well-known.

\subsection{$\mbb H^2_r$}\label{secH2}
Next, let us study a hyperbolic plane $\mbb H^2_r$. The line element is given by
\begin{equation}
\begin{split}
    ds^2&=r^2(d\eta^2+\sinh^2\eta d\tau^2)\\
    \eta&\in[0,+\infty),\quad\tau\in[0,2\pi).
\end{split}\label{H2}
\end{equation}
The space has LC connections
\begin{equation}
    \Gamma_{\eta\tau}^\tau=\coth\eta=\Gamma_{\tau\eta}^\tau,\quad\Gamma_{\tau\tau}^\eta=-\sinh\eta\cosh\eta,\quad\text{the others}=0.\label{LCH2}
\end{equation}
Hence the CKE reduces to
\begin{equation}
\begin{split}
    0&=-\sinh^2\eta\partial_\eta\xi_\eta+\partial_\tau\xi_\tau+\sinh\eta\cosh\eta\xi_\eta,\\
    0&=\partial_\eta\xi_\tau+\partial_\tau\xi_\eta-2\coth\eta\xi_\tau.
\end{split}\label{CKEH2}
\end{equation}
We first consider the `interior' $\eta\in(0,+\infty)$. Because of the periodic identification $\tau+2\pi\sim\tau$, we have to solve the PDEs with a boundary condition
\begin{equation}
    \forall\eta\in[0,+\infty),\quad\xi_\mu(\eta,\tau+2\pi)=\xi_\mu(\eta,\tau).\label{H2bc}
\end{equation}
The most general form of $\xi_\mu$ is thus given by
\begin{equation}
    \xi_\mu(\eta,\tau)=\sum_{m\in\mbb Z}e^{im\tau}\xi_\mu^{(m)}(\eta),\label{H2ansatz}
\end{equation}
on which the reality condition is given by
\[ \forall m\in\mbb Z,\forall\eta\in[0,+\infty),\quad\xi_\mu^{(-m)*}(\eta)=\xi_\mu^{(m)}(\eta). \]
Substituting the ansatz in the CKE (\ref{CKEH2}), one obtains
\begin{equation}
    \partial_\eta\begin{pmatrix}\xi_\eta^{(m)}\\\xi_\tau^{(m)}\end{pmatrix}=\begin{pmatrix}\coth\eta&\frac{im}{\sinh^2\eta}\\-im&2\coth\eta\end{pmatrix}\begin{pmatrix}\xi_\eta^{(m)}\\\xi_\tau^{(m)}\end{pmatrix}.\label{CKEH2'}
\end{equation}

It may be possible to solve this PDE directly, but an easy way\footnote{This computation of CKVs may not be completely correct. Possible problems are in the treatment of boundaries. A natural boundary condition $\forall\tau\in[0,2\pi),\xi_\eta(\eta=0,\tau)\stackrel!=0$ is overkill. On the other hand, although the boundary condition (\ref{bceta=0}) seems overkill because $\eta'=0$ corresponds to $\eta=+\infty$, the boundary condition correctly kills putative CKVs such as $\sinh\eta\partial_\eta$. At present, we do not understand why the boundary conditions work, but decided to present this computation for educational reasons; this computation gives an example of solving CKE via a conformal map and that of conformal algebras ceasing to be isomorphic due to boundary conditions, and lets us keep track of CKVs explicitly. If one is only interested in conformal algebras and correlation functions, one can jump to the last part of this subsection where we generalize our results to $\mbb H^d_r$.} is to use a conformal map $\mbb S^1_r\times\mbb H^1_r\to\mbb H^2_r$ given by
\begin{equation}
    \sinh\eta'=\frac1{\sinh\eta}.\label{S1xH1toH2}
\end{equation}
In fact, one can show
\[ \begin{split}
    ds^2_{\mbb S^1_r\times\mbb H^1_r}&=r^2d\tau^2+r^2d\eta'^2,\\
    \tau\in[0,2\pi)&,\quad\eta'\in[0,+\infty)
\end{split} \]
is conformal to (\ref{H2}):
\begin{align*}
    ds^2_{\mbb S^1_r\times\mbb H^1_r}&=r^2d\tau^2+r^2d\eta'^2\\
    &=\frac{r^2}{\sinh^2\eta}(d\eta^2+\sinh^2\eta d\tau^2).
\end{align*}
Note that the conformal map only covers $\eta'\in(0,+\infty)$ or $\eta\in(0,+\infty)$. Thus, this is a conformal map between an infinite cylinder (or equivalently a two-sphere with two poles removed) and $\mbb H^2_r\backslash\{\eta=0\}$. We will comment on the `boundary' point later, and consider the `interior,' i.e., $\eta\in(0,+\infty)$, first.

The Jacobian is given by
\[ \pdif{y^\alpha}{x^\mu}=\begin{pmatrix}\pdif\eta\tau&\pdif\eta{\eta'}\\\pdif\tau\tau&\pdif\tau{\eta'}\end{pmatrix}=\begin{pmatrix}0&-\sinh\eta\\1&0\end{pmatrix}. \]
Then, the pushforward of the solution (\ref{xiS1xH1}) yields
\begin{equation}
    \begin{pmatrix}\xi_\eta(\eta,\tau)\\\xi_\tau(\eta,\tau)\end{pmatrix}=\sum_{m\in\mbb Z}e^{im\tau}\begin{pmatrix}-2c_\tau^{(m)}\sinh\eta\sinh m\eta'(\eta)\\-2ic_\tau^{(m)}\sinh^2\eta\cosh m\eta'(\eta)\end{pmatrix}.\label{xiH2'}
\end{equation}
One can easily show that this vector field solves (\ref{CKEH2'}). This gives an example of (\ref{confMN}) because $\mbb S^1_r\times\mbb H^1_r\backslash\{\eta'=0\}$ (or equivalently a two-sphere without two poles, or an infinite cylinder) is conformal to $\mbb H^2_r\backslash\{\eta=0\}$ via the conformal map (\ref{S1xH1toH2}).

Let us consider the `boundary.' Since the conformal map (\ref{S1xH1toH2}) is not well-defined at `boundaries,' the solution (\ref{xiH2'}) is valid only in the interior $\eta\in(0,+\infty)$. However, if we shift to the local coordinate system $(\eta,\tau)$, the solution (\ref{xiH2'}) stands on its own feet, and we can extend the domain to the whole $\mbb H^2_r$. Since we are interested in globally well-defined CKVs, the solution has to be extented to the whole $\eta$ including the point $\eta=0$. This can be done by naively extending the domain of the solution to $\eta\in[0,+\infty)$, and imposing a suitable boundary condition at $\eta=0$. Since the $\mbb S^1$ labeled by $\tau$ shrinks to a point at $\eta=0$, we cannot have a vector field with nonzero $\tau$-component there. So it is natural to impose a boundary condition
\begin{equation}
    \forall\tau\in[0,2\pi),\quad\xi_\tau(\eta=0,\tau)\stackrel!=0.\label{H2bceta}
\end{equation}
This boundary condition kills degrees of freedom generated by $c_\tau^{(m)}$ with $|m|\ge2$ because $\cosh m\eta'(\eta)$ are too singular for those $m$. Thus, globally well-defined CKVs are parametrized with three real constants
\[ b_\tau^{(0)},\quad a_\tau^{(1)}(=-a_\tau^{(-1)}),\quad b_\tau^{(1)}(=b_\tau^{(-1)}), \]
and the final solution is given by
\begin{equation}
    \begin{pmatrix}\xi_\eta(\eta,\tau)\\\xi_\tau(\eta,\tau)\end{pmatrix}=b_\tau^{(0)}\begin{pmatrix}0\\2\sinh^2\eta\end{pmatrix}+a_\tau^{(1)}\begin{pmatrix}-4\cos\tau\\4\sinh\eta\cosh\eta\sin\tau\end{pmatrix}+b_\tau^{(1)}\begin{pmatrix}4\sin\tau\\4\sinh\eta\cosh\eta\cos\tau\end{pmatrix}.\label{xiH2}
\end{equation}

Let us work out the group as in the previous example. The CKV can be written as
\begin{align*}
	\xi(\eta,\tau)&=\xi^\mu(\eta,\tau)\partial_\mu\\
	&=\frac{2b_\tau^{(0)}}{r^2}\partial_\tau+\frac{4a_\tau^{(1)}}{r^2}\left(-\cos\tau\partial_\eta+\coth\eta\sin\tau\partial_\tau\right)+\frac{4b_\tau^{(1)}}{r^2}\left(\sin\tau\partial_\eta+\coth\eta\cos\tau\partial_\tau\right).
\end{align*}
So if we define\footnote{In the usual embedding of $\mbb H^2_r$ in $\mbb R^{2,1}$
\[ X^0=r\cosh\eta,\quad X^1=r\sinh\eta\cos\tau,\quad X^2=r\sinh\eta\sin\tau, \]
these vector fields are given by
\[ U=X^1\pdif{}{X^2}-X^2\pdif{}{X^1},\quad V=-X^0\pdif{}{X^1}-X^1\pdif{}{X^0},\quad W=X^0\pdif{}{X^2}+X^2\pdif{}{X^0}, \]
which are well-known KVs of $\mbb H^2_r$.}
\[ U:=\partial_\tau,\quad V:=-\cos\tau\partial_\eta+\coth\eta\sin\tau\partial_\tau,\quad W:=\sin\tau\partial_\eta+\coth\eta\cos\tau\partial_\tau, \]
these form the Lie algebra $\mfrak{so}(2,1)\cong\mfrak{su}(1,1)$
\[ [U,V]=W,\quad[V,W]=-U,\quad[W,U]=V, \]
locally giving
\begin{equation}
    \text{Conf}_0(\mbb H^2)\simeq SO(2,1)\simeq\text{Isom}_0(\mbb H^2_r),\label{confH2}
\end{equation}
in which $U(1)$, constant shifts in the $\tau$-direction, generated by $U$ is contained as a subgroup as expected. This gives an example of the comment below (\ref{confMN}); although subspaces $\mbb S^1_r\times\mbb H^1_r\backslash\{\eta'=0\}$ and $\mbb H^2_r\backslash\{\eta=0\}$ are conformal (so $\mfrak{conf}(\mbb S^1_r\times\mbb H^1_r\backslash\{\eta'=0\})\cong\mfrak{conf}(\mbb H^2_r\backslash\{\eta=0\})$), adding a point $\{\eta=0\}$ breaks the relation. Therefore, it is important to keep track of which parts are covered by conformal maps.

To judge whether the group is $SO(2,1)$ or $SU(1,1)$, let us study the nontrivial element $-id:(\eta,\tau)\mapsto(-\eta,-\tau)$ of the center $\mbb Z_2\subset SU(1,1)$. One finds this is not an element of the (full) conformal group because it does not preserve the domain. So the group would be $SO(2,1)$ rather than $SU(1,1)$:
\[ SO(2,1)\subset\text{Conf}(\mbb H^2). \]
This result was expected because the space is conformal to semi-infinite cylinder or further to the upper half-plane (plus $\infty$), the conformal group must be locally identical to $SL(2,\mbb R)$, which is half of the conformal group on $\mbb R^2\cup\{\infty\}$ preserved by the conformal boundary condition \cite{cbc} $T(z)=\bar T(\bar z)$ at $z=\bar z$ in the complex coordinate system. (More precisely, the group is isomorphic to $PSL(2,\mbb R)\simeq SO(2,1)$.)

The conformal WT relations are given by
\begin{equation}
    0\stackrel!=T\la O_1(x_1)\cdots O_n(x_n)\ra,\label{confWTH2}
\end{equation}
where $T=U,V,W$. Thus, the general form of $n$-point functions is
\begin{equation}
    \la O_1(\eta_1,\tau_1)\cdots O_n(\eta_n,\tau_n)\ra=f(\eta_1,\eta_2,\cdots,\eta_n;\tau_{12},\tau_{23},\cdots,\tau_{n-1,n}).\label{nptfuncH2}
\end{equation}
For example, one-point functions are given by
\begin{equation}
    \la O(\eta,\tau)\ra=\text{const.},\label{1ptfuncH2}
\end{equation}
but it is difficult to determine forms of higher-point functions.

To compute higher-point functions on $\mbb H^2_r$, we take an indirect way; use a conformal map from the upper half-plane (we include $\infty$) to the hyperbolic plane.\footnote{This procedure to compute correlation functions on hyperbolic spaces was used in the latest paper of \cite{hyper}, and our results are consistent with theirs.} On the upper half-plane\footnote{This notation may not be standard, but anyway we use this notation because it turns out that $\mbb H^2_r$ (or more generally $\mbb H^d_r$) is conformal to $\mbb R^2_+$ (or $\mbb R^d_+$). We also write $\mbb R^d_\ge:=\{(x^1,\dots,x^d)\in\mbb R^d\cup\{\infty\}|x^d\ge0\}$.} $\mbb R^2_+:=\{(x^1,x^2)\in\mbb R^2\cup\{\infty\}|x^2>0\}$ plus $x^2=0$, i.e., $\mbb R^2_\ge$, we can compute CKVs as on $\mbb{RP}^d$ in appendix \ref{moreex}, i.e., employing the expression
\[ \xi^\mu(x)=a^\mu+m^\mn x_\nu+\lambda x^\mu+(2b^\nu x_\nu x^\mu-|x|^2b^\mu), \]
one imposes boundary condition
\begin{equation}
    \forall x^1\in\mbb R\cup\{\infty\},\quad\xi^2(x^1,x^2=0)\stackrel!=0\label{R2+bc}
\end{equation}
in order not to move the boundary $x^2=0$. Then, the CKV is given by
\begin{equation}
    \xi(x)=a^1\partial_1+\lambda x^\mu\partial_\mu+b^1(2x_1x^\mu\partial_\mu-|x|^2\partial_1).\label{R2+gen}
\end{equation}
The conformal group thus has translation in the $x^1$ direction, dilation, and special conformal transformation in the $x^1$ direction. Defining
\[ P:=\partial_1,\quad D:=-x^\mu\partial_\mu,\quad K:=2x_1x^\mu\partial_\mu-|x|^2\partial_1, \]
one can compute
\[ [P,D]=-P,\quad[P,K]=-2D,\quad[D,K]=-K, \]
which is isomorphic\footnote{This can be seen by defining
\[ U_\pm:=\frac1{\sqrt2}(P\pm K). \]
In fact, $\{U_+,U_-,D\}$ obey
\[ [U_+,U_-]=D,\quad[U_-,D]=-U_+,\quad[D,U_+]=U_-. \]} to $\mfrak{so}(2,1)\cong\mfrak{su}(1,1)$. Since $-id$ does not preserve the domain, it is not an element of the (full) conformal group, and the group must be $SO(2,1)$:\footnote{As long as we impose the same boundary condition (\ref{R2+bc}), conformal groups are the same with or without $x^2=0$.}
\begin{equation}
    SO(2,1)\subset\text{Conf}(\mbb R^2_+).\label{confR2+}
\end{equation}
Using the symmetries, one can obtain one- and two-point functions (see for example the first three papers of \cite{defect} and also \cite{MRZ} for a recent progress):
\[ \la O_\Delta(x^1,x^2)\ra=\frac{f_\Delta}{(2x^2)^\Delta},\quad\la O_{\Delta_1}(x_1^1,x_1^2)O_{\Delta_2}(x_2^1,x_2^2)\ra=\frac{g_{12}(\zeta)}{(2x_1^2)^{\Delta_1}(2x_2^2)^{\Delta_2}}, \]
where $x_j^\mu$ is the $\mu$-th component of the $j$-th insertion point, and
\[ \zeta:=\frac{(x_{12}^1)^2+(x_{12}
^2)^2}{4x_1^2x_2^2} \]
is the cross ratio, where $x_{jk}^\mu:=x_j^\mu-x_k^\mu$.

Since there are no local counterterms which can shift the one-point functions and nonlocal parts of the two-point functions, these are physical.

We would like to translate these results to correlation functions on $\mbb H^2_r$ using conformal maps. Under a conformal map
\[
\begin{array}{ccccc}
    \varphi&:&(M,\gamma)&\to&(N,\gamma')\\
    &&\rotatebox{90}{$\in$}&&\rotatebox{90}{$\in$}\\
    &&x&\mapsto&y(x)
\end{array}
\]
with $\varphi^*\gamma'=\Omega^2\gamma$, scalar operators with scaling dimension $\Delta$ transform as \cite{OP}
\begin{equation}
    O'_\Delta(y(x))=\Omega^{-\Delta}(x)O_\Delta(x).\label{ODeltaconfmap}
\end{equation}
This relation enables us to compute correlation functions on $\mbb H^2$. To accomplish the purpose, we need a concrete form of the map, and have to work out the conformal factor $\Omega$. It is convenient to introduce a complex coordinate system
\[ z:=x^1+ix^2, \]
and consider a conformal map from the upper half-plane to the semi-infinite cylinder (or $\mbb S^1\times\mbb H^1$) of radius $r$ by\footnote{To match mass dimensions, we assume there is a dimensionful parameter in CFTs on $\mbb R^2_+$, and dimensionful parameters such as $z$ are normalized with it. If one does not like the assumption, one can use a cleaner conformal map explained later, which is free of this issue.}
\begin{equation}
    w(z):=-ir\ln\left(\frac{1+iz}{1-iz}\right),\label{sliver}
\end{equation}
called the Sliver frame in string field theory. Then, real and imaginary parts of $w$ can be naturally identified with our familiar coordinates:
\begin{align*}
    w=&r(\tau+i\eta'),\\
    \tau\in[-\pi,\pi),&\quad\eta'\in[0,+\infty).
\end{align*}
After that we can go to the hyperbolic space easily via the conformal map (\ref{S1xH1toH2}). The first conformal map has a conformal factor\footnote{These conformal factors can also be computed using (\ref{Omega2}).}
\begin{equation}
    \Omega^2(x)=\frac{4r^2}{(1+z^2)(1+\bar z^2)}=\frac{4r^2}{[1+(x^1)^2-(x^2)^2]^2+4(x^1)^2(x^2)^2},\label{OmegaR2+toS1xH1}
\end{equation}
and the second has
\begin{equation}
    \Omega^2(\eta'(x))=\frac1{\sinh^2\eta'(z)}=\frac{[(1+x^2)^2+(x^1)^2][(1-x^2)^2+(x^1)^2]}{4(x^2)^2}.\label{OmegaS1xH1toH2}
\end{equation}
As we saw in section \ref{conf}, conformal factors of successive conformal maps simply multiply. Thus, the total conformal factor we wanted to compute is given by
\begin{equation}
    \Omega^2(x)=\frac{r^2}{(x^2)^2}.\label{OmegaR2+toH2}
\end{equation}
Therefore, one- and two-point functions on $\mbb H^2_r$ are given by
\begin{equation}
    \la O'_\Delta(y(x))\ra=\frac{f_\Delta}{(2r)^\Delta},\quad\la O'_{\Delta_1}(y_1(x_1))O'_{\Delta_2}(y_2(x_2))\ra=\frac{g_{12}(\zeta)}{(2r)^{\Delta_1+\Delta_2}}.\label{12ptfuncH2}
\end{equation}
Note that the one-point functions are constants, and consistent with (\ref{1ptfuncH2}). However, since $\mbb H^2$ has nontrivial Riemann tensors, we can write a local counterterm
\[ S\ni\int d^2x\sqrt\gamma f(\lambda)R(x), \]
making the one-point functions of operators with Weyl weight $(-2)$ unphysical unless the local counterterm is forbidden by symmetries one would like to preserve. In other words, operators with Weyl weight $(-2)$ mix with the identity operator. Nonlocal parts of the two- and higher-point functions are physical.

For later use, we would also like to get an expression in terms of $\mbb H^2$ coordinate systems. Inverting (\ref{sliver}), one obtains
\begin{equation}
    z=\tan\left(\frac{\tau+i\eta'(\eta)}2\right),\label{ztaueta}
\end{equation}
where $\eta'(\eta)$ is given by
\begin{equation}
    \eta'(\eta)=\ln\left(\frac{1+\cosh\eta}{\sinh\eta}\right).\label{eta'eta}
\end{equation}
Thus, the local coordinate system in the upper half-plane is written as
\begin{equation}
\begin{split}
    x^1&=\re\left[\tan\left(\frac{\tau+i\eta'(\eta)}2\right)\right]=\frac{\sinh\eta\sin\tau}{\cosh\eta+\sinh\eta\cos\tau},\\
    x^2&=\im\left[\tan\left(\frac{\tau+i\eta'(\eta)}2\right)\right]=\frac1{\cosh\eta+\sinh\eta\cos\tau}.
\end{split}\label{x1x2}
\end{equation}
Note that $x^2=0$ corresponds to $\eta=+\infty$. Thus, $\mbb H^2_r$ is conformal to $\mbb R^2_+,$ not $\mbb R^2_\ge$. The relation (\ref{x1x2}) reduces the conformal factor (\ref{OmegaR2+toH2}) to
\begin{equation}
    \Omega^2(\eta,\tau)=r^2(\cosh\eta+\sinh\eta\cos\tau)^2.\label{OmegaR2+toH2'}
\end{equation}
Hence, one- and two-point functions on $\mbb H^2_r$ are given by
\begin{equation}
    \la O_\Delta(\eta,\tau)\ra=\frac{f_\Delta}{(2r)^\Delta},\quad\la O_{\Delta_1}(\eta_1,\tau_1)O_{\Delta_2}(\eta_2,\tau_2)\ra=\frac{g_{12}(\zeta(\eta_1,\eta_2,\tau_1,\tau_2))}{(2r)^{\Delta_1+\Delta_2}},\label{12ptfuncH2'}
\end{equation}
where
\begin{align*}
    \zeta(\eta_1,\eta_2,\tau_1,\tau_2)&=\frac{(\cosh\eta_1+\sinh\eta_1\cos\tau_1)(\cosh\eta_2+\sinh\eta_2\cos\tau_2)}4\\
    &~~\times\Bigg[\left(\frac{\sinh\eta_1\sin\tau_1}{\cosh\eta_1+\sinh\eta_1\cos\tau_1}-\frac{\sinh\eta_2\sin\tau_2}{\cosh\eta_2+\sinh\eta_2\cos\tau_2}\right)^2\\
    &~~~~+\left(\frac1{\cosh\eta_1+\sinh\eta_1\cos\tau_1}-\frac1{\cosh\eta_2+\sinh\eta_2\cos\tau_2}\right)^2\Bigg].
\end{align*}
Although the cross ratio looks complicated, it can be simplified dramatically. Recalling the usual embedding of $\mbb H^2_r$ in $\mbb R^{2,1}$, we can write
\[ X^0=r\cosh\eta,\quad X^1=r\sinh\eta\cos\tau,\quad X^2=r\sinh\eta\sin\tau. \]
Then, the cross ratio reduces to
\begin{equation}
\begin{split}
    \zeta&=-\frac1{2r^2}\left(X_1^1X_2^1+X_1^2X_2^2-X_1^0X_2^0+r^2\right)\\
    &=-\frac12\left(\sinh\eta_1\sinh\eta_2\cos\tau_{12}-\cosh\eta_1\cosh\eta_2+1\right).
\end{split}\label{zetaH2}
\end{equation}
The constant shift is needed to make $\zeta$ vanish when $X_1=X_2$. Now that it is clear that the cross ratio, so that the two-point functions are invariant under $SO(2,1)$. In fact, one can show the correlators satisfy the conformal WT relations (\ref{confWTH2}).

Repeating the procedure, one can compute higher-point functions on $\mbb H^2_r$ including local operators in nontrivial representations of $SO(2,1)$ from those on the upper half-plane (plus $\infty$).

The results can be generalized to higher dimensions. To study $\mbb H^d_r$, it is convenient to employ a coordinate system \cite{AGMOO}
\[ \begin{pmatrix}X^0\\X^1\\\vdots\\X^{d-1}\\X^d\end{pmatrix}=\begin{pmatrix}\frac{x^d}2+\frac1{2x^d}(r^2+\Vec x^2)\\rx^1/x^d\\\vdots\\rx^{d-1}/x^d\\\frac{x^d}2-\frac1{2x^d}(r^2-\Vec x^2)\end{pmatrix}=r\begin{pmatrix}\cosh\eta\\\sinh\eta\cos\theta^1\\\vdots\\\sinh\eta\sin\theta^1\cdots\sin\theta^{d-2}\cos\theta^{d-1}\\\sinh\eta\sin\theta^1\cdots\sin\theta^{d-1}\end{pmatrix}, \]
where $\Vec x\in\mbb R^{d-1}\cup\{\infty\},x^d>0$. Using the first coordinate system, one can easily show
\[ (X^0)^2-\Vec X^2=r^2, \]
where $\Vec X^2=(X^1)^2+\cdots+(X^d)^2$, implying $X\in\mbb H^d_r$, and
\[ ds^2_{\mbb H^d_r}=-(dX^0)^2+d\Vec X^2=\left(\frac r{x^d}\right)^2[d\Vec x^2+(dx^d)^2], \]
showing the upper half-space $\mbb R^d_+$ is conformal to $\mbb H^d_r$ with the conformal factor
\[ \Omega^2=\left(\frac r{x^d}\right)^2. \]
An existence of the conformal map implies\footnote{In the embedding coordinate $X$ above, the (C)KVs are given by
\[ X^j\pdif{}{X^k}-X^k\pdif{}{X^j}\quad(1\le j<k\le d),\quad X^0\pdif{}{X^j}+X^j\pdif{}{X^0}\quad(1\le j\le d), \]
as usual.}
\begin{equation}
    \mfrak{conf}(\mbb H^d_r)\cong\mfrak{so}(d,1)\cong\mfrak{isom}(\mbb H^d_r).\label{confHd}
\end{equation}
Employing the conformal map $\varphi:(\mbb R^d_+,\delta)\to(\mbb H^d_r,\gamma)$, one can translate one- and two-point functions on $\mbb R^d_+$
\[ \la O_\Delta(x)\ra=\frac{f_\Delta}{(2x^d)^\Delta},\quad\la O_{\Delta_1}(x_1)O_{\Delta_2}(x_2)\ra=\frac{g_{12}(\zeta)}{(2x_1^d)^{\Delta_1}(2x_2^d)^{\Delta_2}} \]
to those on $\mbb H^d_r$:
\begin{equation}
    \la O_\Delta(x)\ra=\frac{f_\Delta}{(2r)^\Delta},\quad\la O_{\Delta_1}(x_1)O_{\Delta_2}(x_2)\ra=\frac{g_{12}(\zeta)}{(2r)^{\Delta_1+\Delta_2}},\label{12ptfuncHd}
\end{equation}
where
\begin{equation}
\begin{split}
    \zeta&=\frac{(\Vec x_1-\Vec x_2)^2+(x_{12}^d)^2}{4x_1^dx_2^d}\\
    &=-\frac1{2r^2}(\Vec X_1\cdot\Vec X_2-X_1^0X_2^0+r^2)\\
    &=-\frac12\Big(\sinh\eta_1\sinh\eta_2[\cos\theta_1^1\cos\theta_2^1+\cdots+\sin\theta_1^1\sin\theta_2^1\cdots\sin\theta_1^{d-3}\sin\theta_2^{d-3}\cos\theta_1^{d-2}\cos\theta_2^{d-2}\\
    &\hspace{100pt}+\sin\theta_1^1\sin\theta_2^1\cdots\sin\theta_1^{d-2}\sin\theta_2^{d-2}\cos\theta_{12}^{d-1}]-\cosh\eta_1\cosh\eta_2+1\Big)
\end{split}\label{zetaHd}
\end{equation}
is the cross ratio. One can easily compute this realizing $\Omega_jx_j^\mu=X_j^\mu$ with $\mu=1,\dots,d-1$ and $X_j^0-X_j^d=r\Omega_j$. This form is consistent with our result in two dimensions, and was expected because the `Minkowski metric' $\eta_{ab}=\text{diag}(+1,\cdots,+1,-1)$ with $a=1,\dots,d+1$ is the only invariant two-tensor of $SO(d,1)$, the conformal group of $\mbb R^d_+$. Taking local counterterms into account, one learns the one-point functions are physical if $d$ is odd. If $d$ is even, all one-point functions of local operators but those with Weyl weights $(-d)$ are physical. Nonlocal parts of higher-point functions are physical.

Finally, we would like to briefly comment on conformal manifolds of the CFTs on $\mbb H^d_r$. As we have seen, CFTs on $\mbb H^d_r$ can be obtained from defect CFTs on $\mbb R^d_+$. Since the latter space enjoys dilation, conformal manifolds are spanned by exactly marginal scalar operators $O_{\Delta=d}$. If the reflection positivity \cite{OS} in the Euclidean space applies also to its subspaces such as $\mbb R^d_+$, then two-point functions of the operators define the Zamolodchikov metrics $g(\lambda)$. Since nontrivial operator mixings take place on curved spaces, the Zamolodchikov metrics on $\mbb H^d_r$ and $\mbb R^d_+$ would not necessarily match. It would be interesting to study properties of $g(\lambda)$ on $\mbb H^d_r$.

\subsection{$\mbb S^1_l\times\mbb H^2_r$}\label{secS1xH2}
Let us study $\mbb S^1_l\times\mbb H^2_r$, which turns out to show conformal symmetry enhancement if we tune the radii. This manifold has a line element
\begin{equation}
\begin{split}
    ds^2&=l^2d\tau^2+r^2(d\eta^2+\sinh^2\eta d\theta^2)\\
    \tau\in[0,&2\pi),\quad\eta\in[0,+\infty),\quad\theta\in[0,2\pi).\label{S1xH2}
\end{split}
\end{equation}
We emphasize that $\tau,\theta$ are always taken to be $2\pi$ periodic whatever the radii $l,r$ are. Its LC connections are computed as
\begin{equation}
    \Gamma_{\eta\theta}^\theta=\coth\eta=\Gamma_{\theta\eta}^\theta,\quad\Gamma_{\theta\theta}^\eta=-\sinh\eta\cosh\eta,\quad\text{the others}=0.\label{LCS1xH2}
\end{equation}
We first consider the `interior' $\eta\in(0,+\infty)$. Because of the periodic identifications
\[ \tau+2\pi\sim\tau,\quad\theta+2\pi\sim\theta, \]
the most general form of the CKVs are given by
\begin{equation}
    \xi_\mu(\tau,\eta,\theta)=\sum_{m,n\in\mbb Z}e^{i(m\tau+n\theta)}\xi_\mu^{(m,n)}(\eta),\label{S1xH2ansatz}
\end{equation}
on which the reality condition is given by
\begin{equation}
    \forall(m,n)\in\mbb Z^2,\forall\eta\in[0,+\infty),\quad\xi_\mu^{(-m,-n)*}(\eta)=\xi_\mu^{(m,n)}(\eta).\label{xi_muS1xH2}
\end{equation}
Then, the CKE can be reorganized as
\begin{equation}
    \partial_\eta\begin{pmatrix}\xi_\tau^{(m,n)}(\eta)\\\xi_\eta^{(m,n)}(\eta)\\\xi_\theta^{(m,n)}(\eta)\end{pmatrix}=\begin{pmatrix}0&-im&0\\im\left(\frac rl\right)^2&0&0\\0&-in&2\coth\eta\end{pmatrix}\begin{pmatrix}\xi_\tau^{(m,n)}(\eta)\\\xi_\eta^{(m,n)}(\eta)\\\xi_\theta^{(m,n)}(\eta)\end{pmatrix}\label{CKES1xH21}
\end{equation}
with two constraints
\begin{equation}
\begin{split}
    0&=m\xi_\theta^{(m,n)}(\eta)+n\xi_\tau^{(m,n)}(\eta),\\
    0&=-im\left(\frac rl\right)^2\xi_\tau^{(m,n)}(\eta)+\frac{in}{\sinh\eta}\xi_\theta^{(m,n)}(\eta)+\coth\eta\xi_\eta^{(m,n)}(\eta).
\end{split}\label{CKES1xH22}
\end{equation}
We solve the CKE using case analysis.

\underline{$n=0$} In this case, (\ref{CKES1xH21}) can be solved easily by diagonalizing the upper left two by two submatrix as before to find
\begin{equation}
    \begin{pmatrix}\xi_\tau^{(m,0)}(\eta)\\\xi_\eta^{(m,0)}(\eta)\\\xi_\theta^{(m,0)}(\eta)\end{pmatrix}=\begin{pmatrix}-\frac{il}r\left(e^{rm\eta/l}c_\tau^{(m,0)}-e^{-rm\eta/l}c_\tau^{(-m,0)*}\right)\\e^{rm\eta/l}c_\tau^{(m,0)}+e^{-rm\eta/l}c_\tau^{(-m,0)*}\\c_\theta^{(m,0)}\sinh^2\eta\end{pmatrix}\label{xiS1xH2''}
\end{equation}
with $c_\theta^{(-m,0)*}=c_\theta^{(m,0)}$. We still have to impose (\ref{CKES1xH22}). Since we are considering $n=0$, the first condition yields
\[ c_\theta^{(m\neq0,0)}=0. \]
Taking the reality condition into account, only the real part of the $c_\theta^{(0,0)}$, denoted as $a_\theta^{(0,0)}$, can be nonzero:
\begin{equation}
    c_\theta^{(0,0)}=a_\theta^{(0,0)}.\label{ctheta00S1xH2}
\end{equation}
The second condition turns out to play an interesting role; by inserting the solution (\ref{xiS1xH2''}), we find
\begin{equation}
    0=-\frac{rm}l\left(e^{rm\eta/l}c_\tau^{(m,0)}-e^{-rm\eta/l}c_\tau^{(-m,0)*}\right)\sinh\eta+\left(e^{rm\eta/l}c_\tau^{(m,0)}+e^{-rm\eta/l}c_\tau^{(-m,0)*}\right)\cosh\eta.\label{CKES1xH23}
\end{equation}
This must hold for all $\eta$ (at least away from the `boundary' $\eta=0$). Since the complex constants $c_\theta^{(m,0)}$ cannot change the power of $\eta$, coefficients of each power of $\eta$ should vanish by themselves. Let us first consider the case $m\neq0$. 

In this case, if $r/l$ is irrational, or $r/l>1$, all four factors $\displaystyle{\exp\Big[\left(\pm\frac{rm}l\pm1\right)\eta\Big]}$ are independent. So four coefficients of them should vanish separately:
\begin{align*}
    c_\tau^{(m\neq0,0)}\left(-\frac{rm}l+1\right)\stackrel!=&0,\\
    c_\tau^{(-m\neq0,0)*}\left(\frac{rm}l+1\right)\stackrel!=&0,\\
    c_\tau^{(m\neq0,0)}\left(\frac{rm}l+1\right)\stackrel!=&0,\\
    c_\tau^{(-m\neq0,0)*}\left(-\frac{rm}l+1\right)\stackrel!=&0,
\end{align*}
leading to
\[ c_\tau^{(m\neq0,0)}=0. \]

If $l=Nr$ with $N\in\mbb N^\times$, some factors of (\ref{CKES1xH23}) coincide, leading to weaker conditions. In fact, at $m=\pm N$, two factors become $e^{0\cdot\eta}=1$, and the condition reduces to
\[ c_\tau^{(N,0)}+c_\tau^{(-N,0)*}\stackrel!=0, \]
leading to
\begin{equation}
    a_\tau^{(-N,0)}=-a_\tau^{(N,0)},\quad b_\tau^{(-N,0)}=b_\tau^{(N,0)},\label{ctauN0S1xH2}
\end{equation}
where $c_\tau^{(m,n)}\equiv a_\tau^{(m,n)}+ib_\tau^{(m,n)}$ with real $a_\tau^{(m,n)},b_\tau^{(m,n)}$. We still have to consider boundary conditions, however, we will see these solutions survive and thus $\mbb S^1_{Nr}\times\mbb H^2_r$ enjoys larger conformal group. The result was expected for $N=1$ because the manifold is conformally flat,\footnote{More precisely, $\mbb S^1_r\times\mbb H^2_r$ is conformal to $\mbb R^3\cup\{\infty\}\backslash(\mbb R^1\cup\{\infty\})$ as we will see later.} but the enhancement for $N>1$ have not appeared in literature to the best of our knowledge.

When $m=0$, (\ref{CKES1xH23}) simplifies and give
\[ c_\tau^{(0,0)*}=-c_\tau^{(0,0)}, \]
so it is pure imaginary, and the solution is parametrized by
\begin{equation}
    c_\tau^{(0,0)}=ib_\tau^{(0,0)}.\label{ctau00S1xH2}
\end{equation}
Since no radius dependence appears in $m=0$ case, this result is true for any $l$.

Combining all the results, for the case $n=0$, we arrive
\begin{equation}
    \begin{pmatrix}\xi_\tau(\tau,\eta,\theta)\\\xi_\eta(\tau,\eta,\theta)\\\xi_\theta(\tau,\eta,\theta)\end{pmatrix}\ni\begin{pmatrix}\frac{2l}rb_\tau^{(0,0)}\\0\\a_\theta^{(0,0)}\sinh^2\eta\end{pmatrix}\label{xiS1xH2n=0gen}
\end{equation}
for irrational $r/l$ or $r/l>1$, and
\begin{equation}
    \begin{pmatrix}\xi_\tau(\tau,\eta,\theta)\\\xi_\eta(\tau,\eta,\theta)\\\xi_\theta(\tau,\eta,\theta)\end{pmatrix}\ni\begin{pmatrix}2Nb_\tau^{(0,0)}\\0\\a_\theta^{(0,0)}\sinh^2\eta\end{pmatrix}+\begin{pmatrix}4N\cosh\eta\left(a_\tau^{(N,0)}\sin N\tau+b_\tau^{(N,0)}\cos N\tau\right)\\4\sinh\eta\left(a_\tau^{(N,0)}\cos N\tau-b_\tau^{(N,0)}\sin N\tau\right)\\0\end{pmatrix}\label{xiS1xH2n=0l=Nr}
\end{equation}
for $l=Nr$ with $N\in\mbb N^\times$.

\underline{$n\neq0$} Next, let us study the case $n\neq0$. In this case, the first condition in (\ref{CKES1xH22}) reduces to
\begin{equation}
    \xi_\tau^{(m,n\neq0)}(\eta)=-\frac mn\xi_\theta^{(m,n\neq0)}(\eta).\label{xi_tauS1xH2}
\end{equation}
Using this, we reduce the second condition to
\begin{equation}
    \xi_\eta^{(m,n\neq0)}(\eta)=-i\Big[\frac1n\left(\frac{rm}l\right)^2\tanh\eta+\frac n{\sinh\eta\cosh\eta}\Big]\xi_\theta^{(m,n\neq0)}(\eta).\label{xi_etaS1xH2}
\end{equation}

Let us first consider the case $m\neq0$. By inserting (\ref{xi_tauS1xH2}) and (\ref{xi_etaS1xH2}) in (\ref{CKES1xH21}), one finds
\begin{equation}
\begin{split}
    \partial_\eta\xi_\theta^{(m\neq0,n\neq0)}(\eta)&=\coth\eta\xi_\theta^{(m\neq0,n\neq0)}(\eta),\\
    \Big[\left(\frac{rm}l\right)^2\tanh\eta+\frac{n^2}{\sinh\eta\cosh\eta}\Big]\xi_\theta^{(m\neq0,n\neq0)}(\eta)&=\coth\eta\xi_\theta^{(m\neq0,n\neq0)}(\eta),\\
    \partial_\eta\Big\{\Big[\left(\frac{rm}l\right)^2\tanh\eta+\frac{n^2}{\sinh\eta\cosh\eta}\Big]\xi_\theta^{(m\neq0,n\neq0)}(\eta)\Big\}&=\left(\frac{rm}l\right)^2\xi_\theta^{(m\neq0,n\neq0)}(\eta).
\end{split}\label{CKES1xH24}
\end{equation}
The first line can be solved with ease:
\begin{equation}
    \xi_\theta^{(m\neq0,n\neq0)}(\eta)=c_\theta^{(m\neq0,n\neq0)}\sinh\eta,\label{xi_thetaS1xH2}
\end{equation}
with complex constant $c_\theta^{(m\neq0,n\neq0)}$. Substituting the solution in the second constraint of (\ref{CKES1xH24}), we get
\[ 0=c_\theta^{(m\neq0,n\neq0)}\Big\{\left[\left(\frac{rm}l\right)^2-1\right]\cosh^2\eta+\left[n^2-\left(\frac{rm}l\right)^2\right]\Big\}. \]
Since this must hold for arbitrary $\eta$, to have nontrivial conformal transformations, each square bracket must vanish independently:
\begin{equation}
    1=\left(\frac{rm}l\right)^2=n^2.\label{CKES1xH25}
\end{equation}
Only when these conditions are satisfied, $c_\theta^{(m\neq0,n\neq0)}$ can be nonzero. One notices at once that this condition does not have a solution if $r/l>1$, but only when $l=Nr$ with $N\in\mbb N^\times$. Therefore, only $c_\theta^{(\pm N,\pm1)}$ can be nonzero.

We still have to consider the last equation of (\ref{CKES1xH24}). With the help of the second equation of (\ref{CKES1xH24}) and the solution (\ref{xi_thetaS1xH2}), the equation reduces to
\[ 0=c_\theta^{(m\neq0,n\neq0)}\sinh\eta\left[\left(\frac{rm}l\right)^2-1\right], \]
which is automatically satisfied if (\ref{CKES1xH25}) holds.

Thus, for the case $m\neq0$, we found
\[ \begin{pmatrix}\xi_\tau^{(m\neq0,n\neq0)}(\eta)\\\xi_\eta^{(m\neq0,n\neq0)}(\eta)\\\xi_\theta^{(m\neq0,n\neq0)}(\eta)\end{pmatrix}=c_\theta^{(m\neq0,n\neq0)}\begin{pmatrix}-\frac mn\sinh\eta\\-\frac in\cosh\eta\\\sinh\eta\end{pmatrix}, \]
where $c_\theta^{(m\neq0,n\neq0)}$ can be nonzero only when $l=Nr$ with $N\in\mbb N^\times$ and $(m,n)=(\pm N,\pm1)$. We still have to impose the reality condition (\ref{xi_muS1xH2}) on this, and one finds
\[ a_\theta^{(\pm N,\pm1)}=a_\theta^{(\mp N,\mp1)},\quad b_\theta^{(\pm N,\pm1)}=-b_\theta^{(\mp N,\mp1)}. \]
These give
\begin{equation}
\begin{split}
    \begin{pmatrix}\xi_\tau(\tau,\eta,\theta)\\\xi_\eta(\tau,\eta,\theta)\\\xi_\theta(\tau,\eta,\theta)\end{pmatrix}&\ni2a_\theta^{(N,1)}\begin{pmatrix}-N\cos(N\tau+\theta)\sinh\eta\\\sin(N\tau+\theta)\cosh\eta\\\cos(N\tau+\theta)\sinh\eta\end{pmatrix}+2b_\theta^{(N,1)}\begin{pmatrix}N\sin(N\tau+\theta)\sinh\eta\\\cos(N\tau+\theta)\cosh\eta\\-\sin(N\tau+\theta)\sinh\eta\end{pmatrix}\\
    &\hspace{15pt}+2a_\theta^{(N,-1)}\begin{pmatrix}N\cos(N\tau-\theta)\sinh\eta\\-\sin(N\tau-\theta)\cosh\eta\\\cos(N\tau-\theta)\sinh\eta\end{pmatrix}+2b_\theta^{(N,-1)}\begin{pmatrix}-N\sin(N\tau-\theta)\sinh\eta\\-\cos(N\tau-\theta)\cosh\eta\\-\sin(N\tau-\theta)\sinh\eta\end{pmatrix}.
\end{split}\label{xiS1xH2mnneq0}
\end{equation}

Finally, let us study the case $m=0$. In this case, the first two PDEs of (\ref{CKES1xH21}) implies
\begin{equation}
\begin{split}
    \xi_\tau^{(0,n\neq0)}(\eta)&=c_\tau^{(0,n\neq0)}=\text{const.},\\
    \xi_\eta^{(0,n\neq0)}(\eta)&=c_\eta^{(0,n\neq0)}=\text{const.}.
\end{split}\label{xiS1xH2m=0}
\end{equation}
The first condition of (\ref{CKES1xH22}) impose
\[ c_\tau^{(0,n\neq0)}=0. \]
Substituting (\ref{xiS1xH2m=0}) in the second condition of (\ref{CKES1xH21}), it reduces to
\begin{equation}
    \xi_\theta^{(0,n\neq0)}(\eta)=\frac in\cosh\eta\sinh\eta c_\eta^{(0,n\neq0)}.\label{xi_thetaS1xH2m=0}
\end{equation}
With these, the last PDE of (\ref{CKES1xH21}) reduces to
\[ 0=\frac inc_\eta^{(0,n\neq0)}(n^2-1). \]
Thus, only $c_\eta^{(0,\pm1)}$ can be nonzero. Collecting the results, we get
\[ \begin{pmatrix}\xi_\tau^{(0,n\neq0)}(\eta)\\\xi_\eta^{(0,n\neq0)}(\eta)\\\xi_\theta^{(0,n\neq0)}(\eta)\end{pmatrix}=\begin{pmatrix}0\\c_\eta^{(0,n\neq0)}\\\frac inc_\eta^{(0,n\neq0)}\cosh\eta\sinh\eta\end{pmatrix}, \]
where only $c_\eta^{(0,\pm1)}$ can be nonzero. Imposing the reality condition (\ref{xi_muS1xH2}) on this, one finds
\[ a_\eta^{(0,\pm1)}=a_\eta^{(0,\mp1)},\quad b_\eta^{(0,\pm1)}=-b_\eta^{(0,\mp1)}. \]
Thus, from the $n\neq0$ case, regardless of $l$ we get
\begin{equation}
\begin{split}
    \begin{pmatrix}\xi_\tau(\tau,\eta,\theta)\\\xi_\eta(\tau,\eta,\theta)\\\xi_\theta(\tau,\eta,\theta)\end{pmatrix}&\ni2a_\eta^{(0,1)}\begin{pmatrix}0\\\cos\theta\\-\sin\theta\cosh\eta\sinh\eta\end{pmatrix}+2b_\eta^{(0,1)}\begin{pmatrix}0\\-\sin\theta\\-\cos\theta\cosh\eta\sinh\eta\end{pmatrix}.
\end{split}\label{xiS1xH2nneq0}
\end{equation}

We realize that for irrational $r/l$ or $r/l>1$, the sum of (\ref{xiS1xH2n=0gen}) and (\ref{xiS1xH2nneq0}) is our full CKV, and for $l=Nr$ with $N\in\mbb N^\times$, the sum of (\ref{xiS1xH2n=0l=Nr}), (\ref{xiS1xH2mnneq0}), and (\ref{xiS1xH2nneq0}) is. We still have to consider the `boundary' $\eta=0$. As before, by restricting ourselves to globally well-defined CKVs, we extend the domain of the solution to the whole $\mbb S^1_l\times\mbb H^2_r$ by imposing a suitable boundary condition at $\eta=0$. As in the previous example,
\[ \forall\tau\in[0,2\pi),\forall\theta\in[0,2\pi),\quad\xi_\theta(\tau,\eta=0,\theta)\stackrel!=0 \]
is a natural boundary condition, however, this is automatically satisfied by the solutions. Furthermore, in order not to move the `boundary' $\eta=0$, we have to impose
\[ \forall\tau\in[0,2\pi),\forall\theta\in[0,2\pi),\quad\xi_\eta(\tau,\eta=0,\theta)\stackrel!=0, \]
and this kills all solutions coming from $n\neq0$. To summarize, our final CKV is given by
\begin{align}
    &\begin{pmatrix}\xi_\tau(\tau,\eta,\theta)\\
    \xi_\eta(\tau,\eta,\theta)\\
    \xi_\theta(\tau,\eta,\theta)\end{pmatrix}\\\nn
    &=\begin{cases}\begin{pmatrix}\frac{2l}rb_\tau^{(0,0)}\\0\\a_\theta^{(0,0)}\sinh^2\eta\end{pmatrix}&\left(\displaystyle{\frac rl;\text{irrational or }\frac rl>1}\right)\\
    \begin{pmatrix}2Nb_\tau^{(0,0)}\\0\\a_\theta^{(0,0)}\sinh^2\eta\end{pmatrix}+\begin{pmatrix}4N\cosh\eta\left(a_\tau^{(N,0)}\sin N\tau+b_\tau^{(N,0)}\cos N\tau\right)\\4\sinh\eta\left(a_\tau^{(N,0)}\cos N\tau-b_\tau^{(N,0)}\sin N\tau\right)\\0\end{pmatrix}&(l=Nr\text{ with }N\in\mbb N^\times)\end{cases}.\label{xiS1xH2}
\end{align}
Note that $\xi_\mu$ is periodic in $\tau$ as we imposed, however, the period is given by $2\pi/N$ when $l=Nr$. Although dimensions of $\text{Conf}_0(\mbb S^1_{Nr}\times\mbb H^2_r)$ are the same for different $N$, the group gets `smaller' in this sense as $N$ gets larger. When $l=Nr$, the CKV has nontrivial conformal factor
\begin{equation}
    2\sigma_M=\frac{8\cosh\eta}{r^2}\left(a_\tau^{(N,0)}\cos N\tau-b_\tau^{(N,0)}\sin N\tau\right),\label{2sigmaS1xH2}
\end{equation}
and those generated by $a_\tau^{(N,0)},b_\tau^{(N,0)}$ are proper.

We would like to make a small digression to comment on these points. If we replace all $\tau$ in CKE by $\tau':=N\tau$, the partial differential equations are exactly the same as the case $N=1$. This formal replacement also works even when $l/r\notin\mbb N$. (In that case, one has to replace $\tau$ with $l\tau/r$.) This replacement can also be viewed as rescalings of `momenta' $m\to m':=Nm$. This interpretation is also possible even when $l/r\notin\mbb N$, however, in that case, putative solutions with nonzero `momenta' $m\in\mbb Z$ in $N=1$ cease to solve the CKE because $lm/r$ are no longer integers; as we emphasized in the beginning, CKVs must always be periodic in $\tau$ (and $\theta$) with period $2\pi$, and `momenta' $m$ always have to be integral. In case $l=Nr$, the rescaled `momenta' $m'$ are also integers, and continues to be solutions of CKE by replacing $m$ with $m'=Nm$.\footnote{This argument also opens possibilities that even if one replaces $\mbb S^1$'s included in spaces we have studied by their $N$-fold covers, the conformal groups may be the same.} This is actually what we found. This enhancement is difficult to see directly in flat spaces, and this would be the reason why this phenomenon of conformal symmetry enhancement has not appeared in literature (to the best of our knowledge\footnote{The closest we could find are \cite{CHM,HZ}. In these papers, conformal maps between $\mbb S^1_l\times\mbb H^{d-1}_r$ were found. The existence of the conformal maps signals an enhancement of the conformal group. We thank Zohar Komargodski for explaining this implication to us. However, explicit computation of CKVs or conformal groups does not seem to appear in literature.}). Since this argument is independent of dimensions, the conformal symmetry enhancement would also take place in higher dimensions.\footnote{In higher dimensions, it is logically possible that solutions with nonzero `momenta' $m$ are more entangled with additional directions, and this simple replacement may not preserve all solutions of $N=1$. However, it is hard to believe this happens because both CKE and periodic boundary conditions are the same with the case $N=1$ once $\tau$ is replaced with $\tau'$. In order to understand this point better, we are now studying $d=4$.}

There are more physical interpretations. As we mentioned in section \ref{conf}, there is a conformal map between $\mbb S^d$ (minus the great sphere at $\phi=\pi$, $\mbb S^{d-2}_{\phi=\pi}$) and $\mbb S^1\times\mbb H^{d-1}$:
\[ \sinh\eta=-\cot\phi. \]
Using the map, we obtain
\[
\begin{split}
	ds_{\mbb S^1_l\times\mbb H^2_r}^2&=\frac1{\sin^2\phi}\left(r^2d\phi^2+l^2\sin^2\phi d\tau^2+r^2\cos^2\phi d\theta^2\right),\\
	\phi&\in[\pi/2,\pi),\quad\tau\in[0,2\pi),\quad\theta\in[0,2\pi).
\end{split}
\]
(Note that $\eta=+\infty$ corresponds to the great circle $\mbb S^1_\theta$ parametrized by $\theta$, and which includes the poles $(\phi,\theta)=(\pi,0),(\pi,\pi)$.) Especially, if $l=Nr$ with $N\in\mbb N^\times$, one can easily recognize the space as an $N$-fold covering of the sphere minus the great circle $\mbb S^1_\theta$. This becomes clear if one rewrites in terms of $\tau'$
\[
\begin{split}
	ds_{\mbb S^1_{Nr}\times\mbb H^2_r}^2&=\frac{r^2}{\sin^2\phi}\left(d\phi^2+\sin^2\phi d\tau^{'2}+\cos^2\phi d\theta^2\right),\\
	\phi&\in[\pi/2,\pi),\quad\tau'\in[0,2N\pi),\quad\theta\in[0,2\pi).
\end{split}
\]
Another conformal map sends the space to the flat space minus codimension two defect. In the coordinate system above, the codimension two defect is placed at $\phi=\pi$. It is the great circle $\mbb S^1_\theta$. It turns out that $l/r$ can be understood as the number of `sheets' we glue along the defect to construct our manifold. So when $l/r<1$, there is a deficit angle, (possibly) leading to smaller conformal groups. And if $l/r\in\mbb N^\times$, if one goes around the codimension two defect once, one goes to the next sheet, and from the $N$-th sheet, one comes back to the first sheet, resulting in the $2N\pi$ period of $\tau'$. We will give an easy way to visualize this picture in any dimensions later.

Since these spaces have been extensively studied to compute R\'enyi entropies \cite{Renyi}, many further interpretations are known. See an excellent paper \cite{NY}, for example. One of them is a coupling to a backgrond $\mbb Z_N$ gauge field with holonomy $e^{2\pi ik/N}$ with $k=0,1,\dots,N-1$ around the defect. Although situations are slightly different, turning on some background fields is a standard procedure to cancel, in some sense, some deformations $-$ for example constructing conformal Killing spinors on curved spaces \cite{curvedSUSY} $-$ it may be possible to interpret the presence of codimension two defect as coupling a background $\mbb Z_N$ gauge field to CKVs, not to an $N$-fold copy of the theory. It is desirable to persue this possibility.

Going back to the story of CFTs, let us work out the conformal groups. We start from the general case, meaning that $r/l$ is irrational or $r/l>1$. In this case, the CKV can be written
\[ \xi(\tau,\eta,\theta)=\xi^\mu(\tau,\eta,\theta)\partial_\mu=\frac{2b_\tau^{(0,0)}}{lr}\partial_\tau+\frac{a_\theta^{(0,0)}}{r^2}\partial_\theta. \]
The generators
\[ P_\tau:=\partial_\tau,\quad P_\theta:=\partial_\theta \]
clearly form $U(1)$'s consist of constant shifts in $\tau$- and $\theta$-directions:
\begin{equation}
    \text{Conf}_0(\mbb S^1_l\times\mbb H^2_r)\simeq U(1)\times U(1)\simeq\text{Isom}_0(\mbb S^1_l\times\mbb H^2_r)\quad\left(\frac rl;\text{irrational or }\frac rl>1\right).\label{confS1xH2gen}
\end{equation}
One cannot refine the result even if one considers reflections. Thus, local operators on the manifold are labeled by two integer $U(1)$ `charges' $(q_\tau,q_\theta)$, and $n$-point functions must have net `charges' zero:
\[ \sum_{j=1}^nq_\mu^{(j)}=0, \]
where $\mu=\tau,\theta$, and $q_\mu^{(j)}$ is the `charge' $q_\mu$ of the $j$th operator. More concisely, these are expressed by conformal WT relations
\begin{equation}
	0\stackrel!=P_\mu\la O_1(x_1)\cdots O_n(x_n)\ra.\label{confWTS1xH2gen}
\end{equation}
Therefore, $n$-point functions are functions of $\eta_j$ and differences of $\tau$'s and $\theta$'s:
\begin{equation}
	\la O_1(\tau_1,\eta_1,\theta_1)\cdots O_n(\tau_n,\eta_n,\theta_n)\ra=f(\tau_{12},\tau_{23},\cdots,\tau_{n-1,n};\eta_1,\cdots,\eta_n;\theta_{12},\theta_{23},\cdots,\theta_{n-1,n}).\label{nptfuncS1xH2}
\end{equation}
In particular, some lower-point functions are given by
\begin{equation}
\begin{split}
	\la O(\tau,\eta,\theta)\ra&=f(\eta)\delta_{q_\tau,0}\delta_{q_\theta,0},\\
	\la O_1(\tau_1,\eta_1,\theta_1)O_2(\tau_2,\eta_2,\theta_2)\ra&=g(\tau_{12};\eta_1,\eta_2;\theta_{12})\delta_{q_\tau^{(1)}+q_\tau^{(2)},0}\delta_{q_\theta^{(1)}+q_\theta^{(2)},0},\\
	\la O_1(\tau_1,\eta_1,\theta_1)O_2(\tau_2,\eta_2,\theta_2)O_3(\tau_3,\eta_3,\theta_3)\ra&=h(\tau_{12},\tau_{23};\eta_1,\eta_2,\eta_3;\theta_{12},\theta_{23})\\
&~~~~~~~~\times\delta_{q_\tau^{(1)}+q_\tau^{(2)}+q_\tau^{(3)},0}\delta_{q_\theta^{(1)}+q_\theta^{(2)}+q_\theta^{(3)},0}.
\end{split}\label{S1xH2123pt}
\end{equation}
Let us examine which parts of these correlators are physical. We have nontrivial Riemann tensors:
\[ R_{\eta\theta\eta\theta}=-r^2\sinh^2\eta,\quad R_{\eta\eta}=-1,\quad R_{\theta\theta}=-\sinh^2\eta,\quad R=-\frac2{r^2}. \]
We can use these to construct local counterterms, however, since our space has odd dimension, we must use the LC tensor $\epsilon^{\mn\rho}$. When we do not have background gauge field, all we can write is then
\[ S\ni\int d^3x\sqrt\gamma\epsilon^{\mn\rho}c_{IJK}(\lambda)\partial_\mu\lambda^I(x)\partial_\nu\lambda^J(x)\partial_\rho\lambda^K(x), \]
where $c_{IJK}$ is antisymmetric under exchange of indices. If we have background gauge fields, we can also write
\[ S\ni\int d^3x\sqrt\gamma\epsilon^{\mn\rho}c^a_IF_\mn^a(x)D_\rho\lambda^I(x), \]
where $a$ is a gauge index, $F_\mn$ the field strength, and $D_\mu$ the covariant derivative containing both the LC connection and the gauge connection. These terms appeared in the study of three-dimensional conformal anomaly \cite{3dtrace}. Of course, one can also have Chern-Simons (CS) terms as local counterterms. These CS counterterms were studied in detail in the last paper of \cite{Fthm} and \cite{CDFKS}. From this consideration we learn the one-point functions are physical because there is no local counterterm which can shift the correlators. Similarly, nonlocal parts of higher-point functions are physical. Some parts of contact terms of operators $O_I$ coupled to $\lambda^I$, current operators, and energy-momentum tensors fail to be physical. See the papers cited above for more details.

Next, let us study the case $l=Nr$ with $N\in\mbb N^\times$. Since the conformal group is larger, we can say more about correlation functions. In this case, the CKV reduces to
\begin{align*}
    \xi(\tau,\eta,\theta)&=\xi^\mu(\tau,\eta,\theta)\partial_\mu\\
    &=\frac{2b_\tau^{(0,0)}}{Nr^2}\partial_\tau+\frac{a_\theta^{(0,0)}}{r^2}\partial_\theta+\frac{4a_\tau^{(N,0)}}{r^2}\left(\frac1N\cosh\eta\sin N\tau\partial_\tau+\sinh\eta\cos N\tau\partial_\eta\right)\\
    &~~~~~~~~~~~~~~~~~~~~~~~~~~~~~+\frac{4b_\tau^{(N,0)}}{r^2}\left(\frac1N\cosh\eta\cos N\tau\partial_\tau-\sinh\eta\sin N\tau\partial_\eta\right).
\end{align*}
Defining
\begin{align*}
    U:=\frac1N\partial_\tau,&\quad P_\theta:=\partial_\theta,\\
    V:=\frac1N\cosh\eta\sin N\tau\partial_\tau+\sinh\eta\cos N\tau\partial_\eta,&\quad W:=\frac1N\cosh\eta\cos N\tau\partial_\tau-\sinh\eta\sin N\tau\partial_\eta,
\end{align*}
one finds commutation relations
\[ [P_\theta,U]=[P_\theta,V]=[P_\theta,W]=0,\quad[U,V]=W,\quad[V,W]=-U,\quad[W,U]=V. \]
Just like as we saw in the previous example, $\{U,V,W\}$ forms the Lie algebra $\mfrak{so}(2,1)\cong\mfrak{su}(1,1)$. Thus, we obtain
\[ (\{U,V,W,P_\theta\},+,\stackrel{\mbb R}\cdot,[\cdot,\cdot])\cong\mfrak{so}(2,1)\oplus\mfrak u(1)\cong\mfrak{su}(1,1)\oplus\mfrak u(1), \]
locally forming
\begin{equation}
    \text{Conf}_0(\mbb S^1_{Nr}\times\mbb H^2_r)\simeq SO(2,1)\times U(1),\label{confS1xH2l=Nr}
\end{equation}
in which $U(1)$ generated by $U$, constant shifts in the $\tau$-direction, is contained as a subgroup of $SO(2,1)$ as expected. It is interesting to study $-id$. Since $U,V,W$ only touch $\tau$ and $\eta$, it acts like
\[ -id:(\tau,\eta)\mapsto(-\tau,-\eta). \]
This preserves the line element (\ref{S1xH2}) but not the domain. So the group would be $SO(2,1)$ rather than $SU(1,1)$:
\[ SO(2,1)\times U(1)\subset\text{Conf}(\mbb S^1_{Nr}\times\mbb H^2_r). \]
Note that although the group is the same, the $SO(2,1)$ subgroup of the conformal group is $not$ the isometry group of $\mbb H^2$ because the subgroup acts on $(\tau,\eta)$ and not on $(\eta,\theta)$. It is amusing to find such a `twist' appears. In addition, it is also interesting to find not only the conformal group enhances but also it develops to be noncompact when the ratio $l/r$ is set to a (nonzero) natural number. Therefore, local operators in CFTs on $\mbb S^1_{Nr}\times\mbb H^2_r$ are labeled by representations $\rho$ of $SO(2,1)$, which are in general continuous, and integer $U(1)$ `charges' $q$. To respect the conformal symmetry, $n$-point functions must form the trivial representation of $SO(2,1)$ and have net $U(1)$ `charge' zero:
\[ (\rho^{(1)},\cdots,\rho^{(n)})=1,\quad\sum_{j=1}^nq^{(j)}=0, \]
where $\rho^{(j)}$ and $q^{(j)}$ are representations of $SO(2,1)$ and $U(1)$ of the $j$th operator, respectively, and the bracket denotes an appropriate contraction of the $SO(2,1)$ representations. More concisely, the conformal WT relations are given by
\begin{equation}
    0\stackrel!=T\la O_1(x_1)\cdots O_n(x_n)\ra,\label{confWTS1xH2l=Nr}
\end{equation}
where $T=U,V,W,$ and $P_\theta$. We can work out the constraints explicitly for lower-point functions. For example, only operators which belong to the trivial representation of the $SO(2,1)$ and have $q=0$ can have nonzero one-point functions:
\begin{equation}
	\la O(\tau,\eta,\theta)\ra=\begin{cases}\text{const.}&(\rho,q)=(1,0),\\0&\text{otherwise}.\end{cases}\label{S1xH21pt}
\end{equation}
As in the case $l\neq Nr$, the one-point functions are physical.

Since $SO(2,1)$ has an invariant two-tensor $\eta_{ab}=\text{diag}(+1,+1,-1)$ and a completely antisymmetric three-tensor $\epsilon_{abc}$, there are some nontrivial two- and three-point functions. Writing singlets $O_j$, and vectors (or fundamentals) $-$ which are three-dimensional $-$ $O_j^a$, we can have nontrivial two-point functions
\[ \la O_1O_2\ra,\quad\eta_{ab}\la O_1^aO_2^b\ra \]
with
\[ q^{(1)}+q^{(2)}=0 \]
where $q^{(j)}$ is the $U(1)$ `charge' of the $j$th operator, and three-point functions
\[ \la O_1O_2O_3\ra,\quad\eta_{ab}\la O_1O_2^aO_3^b\ra,\quad\epsilon_{abc}\la O_1^aO_2^bO_3^c\ra \]
with
\[ q^{(1)}+q^{(2)}+q^{(3)}=0. \]
Nonlocal parts of these higher-point functions are physical, and some parts of contact terms are unphysical.

To compute higher-point functions, we use the same method as in subsection \ref{secH2}, namely we use conformal maps. For simplicity, we limit ourselves to scalar operators. As we mentioned above, our space $\mbb S^1_{Nr}\times\mbb H^2_r$ is conformal to an $N$-fold cover of the sphere minus codimension two great circle $\mbb S^1_\theta$ on which poles are located. Since the round sphere $\mbb S^3$ is conformal to $\mbb R^3\cup\{\infty\}$, and poles correspond to the origin $0$ and the infinity $\infty$, the codimension two defect, $\mbb S^1_\theta$, is conformally mapped to a straight line through $0$ and $\infty$. We take the defect as the $x^3$ axis $\mbb R^1\cup\{\infty\}$. Therefore, our space is conformal to an $N$-fold cover of $\mbb R^3\cup\{\infty\}\backslash(\mbb R^1\cup\{\infty\})$.\footnote{Since the domain of the defect is removed from our space, there is no degrees of freedom supported on the defect, such as defect local operators. In this sense, our theory is simpler than the usual defect CFTs.} We will denote the space $\mbb R^3_N\cup\{\infty\}\backslash(\mbb R^1\cup\{\infty\})$. Instead of working directly with the $N$-fold cover, we will take a quicker way; we will work out the case $N=1$, and then replace $\tau$ with $\tau'$ at last based on our arguement above. One can easily compute the CKV; so as to leave the $x^3$ axis invariant, by imposing boundary condition
\begin{equation}
    \forall x^3\in\mbb R^1\cup\{\infty\},\quad\xi^1(x^1=0,x^2=0,x^3)\stackrel!=0\stackrel!=\xi^2(x^1=0,x^2=0,x^3)\label{R3-Rbc}
\end{equation}
on our familiar solution on $\mbb R^3\cup\{\infty\}$
\[ \xi^\mu(x)=a^\mu+m^\mn x_\nu+\lambda x^\mu+(2b^\nu x_\nu x^\mu-|x|^2b^\mu), \]
one finds
\begin{equation}
    \xi(x)=a^3\partial_3+m^{12}(x^2\partial_1-x^1\partial_2)+\lambda x^\mu\partial_\mu+b^3(2x^3x^\mu\partial_\mu-|x|^2\partial_3).\label{R3-Rgen}
\end{equation}
Defining the generators
\[ P:=\partial_3,\quad M:=x^2\partial_1-x^1\partial_2,\quad D:=-x^\mu\partial_\mu,\quad K:=2x^3x^\mu\partial_\mu-|x|^2\partial_3, \]
one can easily show
\[ [M,P]=[M,D]=[M,K]=0,\quad[P,D]=-P,\quad[D,K]=-K,\quad[K,P]=2D, \]
which is isomorphic to $\mfrak{so}(2,1)\oplus\mfrak u(1)\cong\mfrak{su}(1,1)\oplus\mfrak u(1)$, as expected. We realize that this space is nothing but $\mbb R^3$ with codimension two conformal defect, which preserves $SO(2,1)\times SO(2)$ subgroup of the ambient conformal group $SO(4,1)$. We also comment that since $-id$ preserves both the metric and the domain, this is an element of the full conformal group. Thus, the group must be
\[ SU(1,1)\times SO(2)\subset\text{Conf}(\mbb R^3\cup\{\infty\}\backslash(\mbb R^1\cup\{\infty\}))\simeq\text{Conf}(\mbb R^3_N\cup\{\infty\}\backslash(\mbb R^1\cup\{\infty\})). \]
Although conformal algebras must be preserved under conformal maps, global structures of conformal groups can be different, and this provides an example.

Before we compute correlation functions, let us construct a conformal map from $\mbb R^3\cup\{\infty\}\backslash(\mbb R^1\cup\{\infty\})$ to $\mbb S^1_r\times\mbb H^2_r$. We find it convenient to do this in two steps; we first map $\mbb R^3\cup\{\infty\}\backslash(\mbb R^1\cup\{\infty\})$ to $\mbb S^3_r\backslash\mbb S^1_{\phi=\pi}$, and then to $\mbb S^1_r\times\mbb H^2_r$. The first conformal map is given by\footnote{In this coordinate system, two $U(1)$'s for generic $l$ correspond to rotation in the $x^1-x^2$ plane, which is parameterized by $\tau$, and (compact) translation in the $x^3$-direction, which is parameterized by $\theta$. The latter is compact because $\infty$ is included.}
\[ \begin{pmatrix}X^1\\X^2\\X^3\\X^4\end{pmatrix}=\frac r{1+\frac{|x|^2}{r^2}}\begin{pmatrix}x^1/r\\x^2/r\\x^3/r\\\frac{|x|^2}{r^2}-1\end{pmatrix}=r\begin{pmatrix}\sin\phi\cos\tau\\\sin\phi\sin\tau\\\cos\phi\cos\theta\\\cos\phi\sin\theta\end{pmatrix}, \]
and the second by $\sinh\eta=-\cot\phi$. The first conformal map has the conformal factor
\[ \Omega^2=\left(\frac1{1+\frac{|x|^2}{r^2}}\right)^2=\left(\frac12(1-\cos\phi\sin\theta)\right)^2, \]
and the second has
\[ \Omega^2=\cosh^2\eta=\frac1{\sin^2\phi} \]
as we reviewed in section \ref{conf}. Note that the defect $\mbb R^1\cup\{\infty\}=\{x^1=0=x^2\}$ is mapped to the great circle $\mbb S^1_{\phi=\pi}$, on which poles are located, parametrized by $\theta$, as desired. Then, the second conformal map sends $\mbb S^1_{\phi=\pi}$ to the $\mbb S^1$ at $\eta=+\infty$. Therefore, $\mbb R^3_N\cup\{\infty\}\backslash(\mbb R^1\cup\{\infty\})$ is conformal to $\mbb S^1_{Nr}\times\mbb H^2_r$.

Using these conformal maps, we send correlation functions on $\mbb R^3\cup\{\infty\}\backslash(\mbb R^1\cup\{\infty\})$ to our space. Fortunately, a convenient way to compute correlation functions in defect CFTs is known \cite{embed}. Using the method, one can easily compute one- and two-point functions of ambient (or bulk) local operators
\begin{equation}
    \la O_\Delta(x)\ra=\frac{f_\Delta}{\left[(x^1)^2+(x^2)^2\right]^{\frac{\Delta}2}},\quad\la O_{\Delta_1}(x_1)O_{\Delta_2}(x_2)\ra=\frac{g_{12}(\xi_1,\xi_2)}{[(x_1^1)^2+(x_1^2)^2]^{\frac{\Delta_1}2}[(x_2^1)^2+(x_2^2)^2]^{\frac{\Delta_2}2}},\label{12ptR3-R}
\end{equation}
where\footnote{These cross ratios were first discovered by Billo et al in \cite{embed}. We thank Marco Meireni for a correspondence. In order to prevent confusion, however, we basically follow the notation of Kobayashi and Nishioka in \cite{embed}, and denote two cross ratios $\xi_1$ and $\xi_2$.}
\[ \xi_1:=\frac{x_1^1x_2^1+x_1^2x_2^2}{[(x_1^1)^2+(x_1^2)^2]^{1/2}[(x_2^1)^2+(x_2^2)^2]^{1/2}},\quad\xi_2:=\frac{(x_{12}^3)^2+(x_1^1)^2+(x_1^2)^2+(x_2^1)^2+(x_2^2)^2}{[(x_1^1)^2+(x_1^2)^2]^{1/2}[(x_2^1)^2+(x_2^2)^2]^{1/2}} \]
are two cross ratios. The correlators are nonvanishing only when their total $SO(2)\simeq U(1)$ `charges' are zero. These are mapped to one- and two-point functions on $\mbb S^1_{Nr}\times\mbb H^2_r$:
\begin{equation}
    \la O_\Delta(x)\ra=\frac{f_\Delta}{r^\Delta},\quad\la O_{\Delta_1}(x_1)O_{\Delta_2}(x_2)\ra=\frac{g_{12}(\xi_1,\xi_2)}{r^{\Delta_1+\Delta_2}},\label{12ptS1xH2}
\end{equation}
where
\[ \xi_1=\cos N\tau_{12},\quad\xi_2=2\cosh\eta_1\cosh\eta_2(1-\tanh\eta_1\tanh\eta_2\cos\theta_{12}). \]
We have replaced $\tau$ by $\tau'=N\tau$. The one-point functions obviously satisfy the conformal WT relations (\ref{confWTS1xH2l=Nr}), however, one notices that these cross ratios are not invariant under our conformal generators. Isometries $P_\theta$ and $U$ do annihilate the cross ratios but $V$ and $W$ do not.\footnote{Of course, there are some freedoms to construct new cross ratios out of the two, but it does not seem to work. This would imply pullbacks of cross ratios by conformal maps are not necessarily cross ratios. We tried to find this claim in literature but could not find.} So the conformal invariance of the two-point functions are not automatic. Since correlation functions must satisfy the conformal WT relations (\ref{confWTS1xH2l=Nr}), it may be possible to constrain $g_{12}$ further as a function of $\xi_1$ and $\xi_2$. The one-point functions and nonlocal parts of the two-point functions are physical.

As all elements $-$ conformal groups, conformal maps, and correlation functions $-$ can be generalized to higher dimensions, we can easily extend the results to CFTs on $\mbb S^1_r\times\mbb H^{d-1}_r$.\footnote{To be conservative, we just consider the case $N=1$ below, but we believe our results also hold for larger $N$ if one replaces $\tau$ by $\tau'=N\tau$.} The conformal algebra is given by
\begin{equation}
    \mfrak{conf}(\mbb S^1_r\times\mbb H^{d-1}_r)\cong\mfrak{so}(d-1,1)\oplus\mfrak{so}(2)\label{confS1xHd-1}
\end{equation}
because the space is conformal to $\mbb R^d\cup\{\infty\}\backslash(\mbb R^{d-2}\cup\{\infty\})$. In $d$-dimensions, the first conformal map has the conformal factor
\[ \Omega^2=\left(\frac1{1+\frac{|x|^2}{r^2}}\right)^2=\left(\frac12(1-\cos\phi\sin\theta^1\sin\theta^2\cdots\sin\theta^{d-2})\right)^2 \]
in the coordinate system
\[ \begin{pmatrix}X^1\\X^2\\\vdots\\X^d\\X^{d+1}\end{pmatrix}=\frac r{1+\frac{|x|^2}{r^2}}\begin{pmatrix}x^1/r\\x^2/r\\\vdots\\x^d/r\\\frac{|x|^2}{r^2}-1\end{pmatrix}=r\begin{pmatrix}\sin\phi\cos\tau\\\sin\phi\sin\tau\\\vdots\\\cos\phi\sin\theta^1\cdots\sin\theta^{d-3}\cos\theta^{d-2}\\\cos\phi\sin\theta^1\cdots\sin\theta^{d-3}\sin\theta^{d-2}\end{pmatrix}. \]
The correlation functions in the defect CFTs are still given by (\ref{12ptR3-R}) if we place the codimension two defect at $x^1=0=x^2$,\footnote{Using this configuration, one can easily visualize moves around the codimension two defect; the codimension two defect is placed at the origin of the $x^1-x^2$ plane. Thus, going around the origin in this plane corresponds to circling the defect. The $N$-fold cover of $\mbb R^d\cup\{\infty\}\backslash(\mbb R^{d-2}\cup\{\infty\})$ is constructed by gluing $N$ copies of $\mbb R^d\cup\{\infty\}\backslash(\mbb R^{d-2}\cup\{\infty\})$ along the codimension two defect $\mbb R^{d-2}\cup\{\infty\}$.} but the second cross ratio becomes
\[ \xi_2:=\frac{(x_{12}^3)^2+\cdots+(x_{12}^d)^2+(x_1^1)^2+(x_1^2)^2+(x_2^1)^2+(x_2^2)^2}{[(x_1^1)^2+(x_1^2)^2]^{1/2}[(x_2^1)^2+(x_2^2)^2]^{1/2}} \]
Therefore, one- and two-point functions on $\mbb S^1_r\times\mbb H^{d-1}_r$ are given by
\begin{equation}
    \la O_\Delta(x)\ra=\frac{f_\Delta}{r^\Delta},\quad\la O_{\Delta_1}(x_1)O_{\Delta_2}(x_2)\ra=\frac{g_{12}(\xi_1,\xi_2)}{r^{\Delta_1+\Delta_2}},\label{12ptS1xHd-1}
\end{equation}
where
\begin{align*}
    \xi_1&=\cos\tau_{12},\\
    \xi_2&=2\cosh\eta_1\cosh\eta_2\Big(1-\tanh\eta_1\tanh\eta_2[\cos\theta_1^1\cos\theta_2^1+\sin\theta_1^1\sin\theta_2^1\cos\theta_1^2\cos\theta_2^2\\
    &\hspace{115pt}+\cdots+\sin\theta_1^1\sin\theta_2^1\cdots\sin\theta_1^{d-4}\sin\theta_2^{d-4}\cos\theta_1^{d-3}\cos\theta_2^{d-3}\\
    &\hspace{130pt}+\sin\theta_1^1\sin\theta_2^1\cdots\sin\theta_1^{d-3}\sin\theta_2^{d-3}\cos\theta_{12}^{d-2}]\Big).
\end{align*}
If $d$ is odd, the one-point functions are physical. If $d$ is even, local operators with Weyl weight $(-d)$ can mix with the identity operator, and their one-point functions are ambiguous. The other one-point functions are physical. Nonlocal parts of higher-point functions are physical in any dimensions.

Finally, the same comment as in section \ref{secH2} can be repeated here; if the reflection positivity also applies to our subspace $\mbb R^d\cup\{\infty\}\backslash(\mbb R^{d-2}\cup\{\infty\})$ (or its $N$-fold cover), two-point functions would be nonnegative. Thus, two-point functions of exactly marginal operators would define the Zamolodchikov metric on $\mbb S^1_r\times\mbb H^{d-1}_r$ a priori different from that on $\mbb R^d\cup\{\infty\}\backslash(\mbb R^{d-2}\cup\{\infty\})$, giving a distance to conformal manifolds.

\subsection{$\mbb{RP}^2\times\mbb S^1_r$}
In the same way, we can also study unorientable manifolds. As an example, let us study $\mbb{RP}^2\times\mbb S^1_r$. For more exapmles of unorientable spaces, see appendix \ref{moreex}. Even if the manifold is unorientable, we can locally define a metric. The line element on this manifold is locally given by
\begin{equation}
\begin{split}
	ds^2=(dy^1)^2+(dy^2)^2+r^2d\theta^2,\\
	(y^1,y^2)\in\mbb R^2\cup\{\infty\},\quad\theta\in[0,2\pi).
\end{split}\label{RP2xS1}
\end{equation}
Since LC connections are trivial, the CKE reduces to
\begin{equation}
	\partial_\mu\xi_\nu+\partial_\nu\xi_\mu=\frac23\left(\partial_1\xi_1+\partial_2\xi_2+\frac1{r^2}\partial_\theta\xi_\theta\right)\gamma_\mn.\label{CKERP2xS1}
\end{equation}
The periodic identification $\theta+2\pi\sim\theta$ gives us an anzatz
\begin{equation}
	\xi_\mu(y^1,y^2,\theta)=\sum_{m\in\mbb Z}e^{im\theta}\xi_\mu^{(m)}(y^1,y^2),\label{RP2xS1ansatz}
\end{equation}
on which the reality condition is given by
\[ \forall m\in\mbb Z,\forall(y^1,y^2)\in\mbb R^2\cup\{\infty\},\quad\xi_\mu^{(m)}(y^1,y^2)=\xi_\mu^{(-m)*}(y^1,y^2). \]
Substituting the ansatz in (\ref{RP2xS1ansatz}) and using the orthogonality of basis $e^{im\theta}$, one obtains
\begin{equation}
\begin{split}
	0&=-im\xi_\theta^{(m)}+r^2\partial_1\xi_1^{(m)},\\
	\partial_1\xi_1^{(m)}&=\partial_2\xi_2^{(m)},\\
	0&=im\xi_1^{(m)}+\partial_1\xi_\theta^{(m)},\\
	0&=im\xi_2^{(m)}+\partial_2\xi_\theta^{(m)},\\
	0&=\partial_1\xi_2^{(m)}+\partial_2\xi_1^{(m)}.
\end{split}\label{CKERP2xS1'}
\end{equation}
We solve these PDEs employing case analysis.

\underline{$m=0$} In this case, the first four of the equations imply
\begin{align*}
	\xi_1^{(0)}=\xi_1^{(0)}(y^2),\quad\xi_2^{(0)}=\xi_2^{(0)}(y^1),\\
	\xi_\theta^{(0)}=a_\theta^{(0)}=\text{const.}\in\mbb R.
\end{align*}
The last equation then reduces to
\[ \partial_1\xi_2^{(0)}(y^1)=-\partial_2\xi_1^{(0)}(y^2). \]
LHS is a function of $y^1$, while RHS is a function of $y^2$, so for this equality to hold identically, it must be a constant:
\[ \partial_1\xi_2^{(0)}(y^1)=C=-\partial_2\xi_1^{(0)}(y^2). \]
This PDEs can be solved at once:
\[ \begin{pmatrix}\xi_1^{(0)}(y^2)\\\xi_2^{(0)}(y^1)\end{pmatrix}=\begin{pmatrix}-Cy^2+a_1^{(0)}\\Cy^1+a_2^{(0)}\end{pmatrix}, \]
where $C,a_j^{(0)}$ are real (constants) by the reality condition. If we write $C$ in terms of antisymmetric two by two matrix $m^{jk}$ with $m^{12}=-C$, we obtained
\begin{equation}
	\begin{pmatrix}\xi_1^{(0)}(y^1,y^2)\\\xi_2^{(0)}(y^1,y^2)\\\xi_\theta^{(0)}(y^1,y^2)\end{pmatrix}=\begin{pmatrix}m^{12}y^2+a_1^{(0)}\\m^{21}y^1+a_2^{(0)}\\a_\theta^{(0)}\end{pmatrix}.\label{xiRP2xS1m=0}
\end{equation}

\underline{$m\neq0$} Next, let us study the case $m\neq0$. The first two equations of (\ref{CKERP2xS1'}) yield
\[ \xi_\theta^{(m\neq0)}(y^1,y^2)=\frac{r^2}{im}\partial_1\xi_1^{(m\neq0)}(y^1,y^2)=\frac{r^2}{im}\partial_2\xi_2^{(m\neq0)}(y^1,y^2). \]
Plugging this into the third equation, one obtains
\begin{equation}
	\xi_1^{(m\neq0)}(y^1,y^2)=\left(\frac rm\right)^2\partial_1^2\xi_1^{(m\neq0)}(y^1,y^2)=\left(\frac rm\right)^2\partial_1\partial_2\xi_2^{(m\neq0)}(y^1,y^2).\label{xi1RP2xS1}
\end{equation}
The first equality implies
\[ \xi_1^{(m\neq0)}(y^1,y^2)=e^{\epsilon_1my^1/r}X^{(m\neq0)}(y^2), \]
where $\epsilon_1=\pm1$. Similarly, substituting $\xi_\theta^{(m\neq0)}$ in the fourth equation of (\ref{CKERP2xS1'}), we obtain
\begin{equation}
	\xi_2^{(m\neq0)}(y^1,y^2)=\left(\frac rm\right)^2\partial_2^2\xi_2^{(m\neq0)}(y^1,y^2)=\left(\frac rm\right)^2\partial_2\partial_1\xi_1^{(m\neq0)}(y^1,y^2).\label{xi2RP2xS1}
\end{equation}
The first equality yields
\[ \xi_2^{(m\neq0)}(y^1,y^2)=e^{\epsilon_1my^2/r}Y^{(m\neq0)}(y^1), \]
where $\epsilon_2=\pm1$. Substituting thus obtained $\xi_j^{(m\neq0)}$ in the second equations of (\ref{xi1RP2xS1}) and (\ref{xi2RP2xS1}), we arrive
\[ X^{(m\neq0)}(y^2)=e^{\epsilon_2my^2/r}c_1^{(m\neq0)},\quad Y^{(m\neq0)}(y^1)=\epsilon_1\epsilon_2e^{\epsilon_1my^1/r}c_1^{(m\neq0)}, \]
or
\[ \xi_1^{(m\neq0)}(y^1,y^2)=e^{m(\epsilon_1y^1+\epsilon_2y^2)/r}c_1^{(m\neq0)},\quad\xi_2^{(m\neq0)}(y^1,y^2)=\epsilon_1\epsilon_2e^{m(\epsilon_1y^1+\epsilon_2y^2)/r}c_1^{(m\neq0)}, \]
where $c_1^{(m\neq0)}$ are constants. However, substituting these in the last equation of (\ref{CKERP2xS1'}), one notices that only $c_1^{(m\neq0)}=0$ is allowed. Thus, we do not get nontrivial conformal transformations from the case $m\neq0$.

To summarize, our full CKV is given by (\ref{xiRP2xS1m=0}). We still have to impose a boundary condition originating from the identification of $\mbb{RP}^2$. This space is given by
\begin{equation}
    \mbb{RP}^2:=(\mbb R^2\cup\{\infty\})/\sim,\label{RP2}
\end{equation}
where two points $p$ and $p'$ are identified $p\sim p'$ if
\begin{equation}
    y^\mu(p')=-\frac{y^\mu(p)}{\delta_\rs y^\rho(p)y^\sigma(p)}.\label{simRP2}
\end{equation}
Since conformal transformations are subset of diffeomorphisms, we consider a coordinate transformation 
\[ x\mapsto x'(x):=x+\xi(x), \]
and there are two ways to express $x'(p')$. Here we use the single-valuedness of given local coordinates systems. Calculating $x'(p')$ in two ways, we obtain
\begin{align*}
    x'(p')&\equiv(y^{'1}(p'),y^{'2}(p'),\theta'(p'))=\left(-\frac{y^{'1}(p)}{|y'(p)|^2},-\frac{y^{'2}(p)}{|y'(p)|^2},\theta'(p)\right)\\
    &\equiv\left(-\frac{y^1(p)+\xi^1(x(p))}{|y(p)+\xi(x(p))|^2},-\frac{y^2(p)+\xi^2(x(p))}{|y(p)+\xi(x(p))|^2},\theta(p)+\xi^\theta(x(p))\right)\\
    &\equiv(y^1(p')+\xi^1(x(p')),y^2(p')+\xi^2(x(p')),\theta(p')+\xi^\theta(x(p')))\\
    &=\left(-\frac{y^1(p)}{|y(p)|^2}+\xi^1(x(p')),-\frac{y^2(p)}{|y(p)|^2}+\xi^2(x(p')),\theta(p)+\xi^\theta(x(p'))\right),
\end{align*}
or using $x(p')=(-\frac{y^1(p)}{|y(p)|^2}
,-\frac{y^2(p)}{|y(p)|^2},\theta(p))$, we obtain a boundary condition
\begin{equation}
	\xi^j\left(-\frac{y^1}{|y|^2},-\frac{y^2}{|y|^2},\theta\right)=\frac{y^j}{|y|^2}-\frac{y^j+\xi^j(y^1,y^2,\theta)}{|y+\xi(y^1,y^2,\theta)|^2},\quad\xi^\theta\left(-\frac{y^1}{|y|^2},-\frac{y^2}{|y|^2},\theta\right)=\xi^\theta(y^1,y^2,\theta).\label{RP2xS1bc}
\end{equation}
The second boundary condition is trivially satisfied by our solution. The first condition is the same as in $\mbb{RP}^d$, and we can conclude immediately that constant shifts $a_j^{(0)}$ cannot survive because their counterpart, special conformal transformations, are absent, while rotations $m^{jk}$ do survive. For more details, see appendix \ref{moreex}. Therefore, the CKV on $\mbb{RP}^2\times\mbb S^1_r$ is given by
\begin{equation}
	\begin{pmatrix}\xi_1(y^1,y^2,\theta)\\\xi_2(y^1,y^2,\theta)\\\xi_\theta(y^1,y^2,\theta)\end{pmatrix}=\begin{pmatrix}m^{12}y^2\\m^{21}y^1\\a_\theta^{(0)}\end{pmatrix}.\label{xiRP2xS1}
\end{equation}
This can also be written
\[ \xi(y^1,y^2,\theta)=\xi^\mu(y^1,y^2,\theta)\partial_\mu=m^{12}(y^2\partial_1-y^1\partial_2)+\frac{a_\theta^{(0)}}{r^2}\partial_\theta. \]
Thus, one can see
\[ P_y:=y^2\partial_1-y^1\partial_2,\quad P_\theta:=\partial_\theta \]
generate
\begin{equation}
	\text{Conf}_0(\mbb{RP}^2\times\mbb S^1_r)\simeq U(1)\times U(1)\simeq\text{Isom}_0(\mbb{RP}^2\times\mbb S^1_r).\label{confRP2xS1}
\end{equation}
Reflections do not refine the result. Local operators are hence labeled by two integer $U(1)$ `charges' $(q_y,q_\theta)\in\mbb Z^2$, and to respect the conformal symmetry, net `charges' of $n$-point functions have to be zero:
\[ \sum_{j=1}^nq_\mu^{(j)}=0, \]
where $q_\mu^{(j)}$ is the $U(1)$ `charge' of the $j$th operator with $\mu=y,\theta$. More concisely, these are summarized by conformal WT relations
\begin{equation}
	0\stackrel!=P_\mu\la O_1(x_1)\cdots O_n(x_n)\ra.\label{confWTRP2xS1}
\end{equation}
Thus, $n$-point functions are given by
\begin{equation}
	\la O_1(x_1)\cdots O_n(x_n)\ra=f(y_j\cdot y_k;\theta_{12},\theta_{23},\cdots,\theta_{n-1,n}).\label{nptfuncRP2xS1}
\end{equation}
In particular, some lower-point functions are given by
\begin{equation}
\begin{split}
	\la O(y^1,y^2,\theta)\ra&=\frac {f_O}{(1+|y|^2)^\Delta}\delta_{q_y,0}\delta_{q_\theta,0},\\
	\la O_1(y_1^1,y_1^2,\theta_1)O_2(y_2^1,y_2^2,\theta_2)\ra&=\frac{(1+|y_1|^2)^{(-\Delta_1+\Delta_2)/2}(1+|y_2|^2)^{(\Delta_1-\Delta_2)/2}}{|y_1-y_2|^{2\cdot(\Delta_1+\Delta_2)/2}}\\
	&~~~~\times g(\eta;\theta_{12})\delta_{q_y^{(1)}+q_y^{(2)},0}\delta_{q_\theta^{(1)}+q_\theta^{(2)},0},
\end{split}\label{RP2xS1123pt}
\end{equation}
where $O_j$ is a scalar field with weight $(h_1^{(j)},h_2^{(j)},h_\theta^{(j)})$ and `scaling dimension' $\Delta_j=h_1^{(j)
}+h_2^{(j)}$, $f_O$ a constant, and
\[ \eta:=\frac{|y_1-y_2|^2}{(1+|y_1|^2)(1+|y_2|^2)} \]
is the crosscap cross ratio.

Analysis of the local counterterms are the same as before. One concludes the one-point functions and nonlocal parts of higher-point functions are physical. Some parts of contact terms among operators $O_I$ coupled to $\lambda^I$ and current operators are unphysical.

\section{CFTs on a $d$-torus $\mbb T^d$}\label{dtorus}
We observed that some boundary conditions, such as periodic ones, make the analysis remarkably easier by reducing problems of solving partial differential equations to algebraic equations. Hence in this section, we would like to study CFTs on a manifold with periodic boundary conditions in every direction, namely a $d$-torus. Its metric is given by
\begin{equation}
\begin{split}
    ds^2=l_1^2d\theta_1^2&+\cdots+l_d^2d\theta_d^2,\\
    \theta_1\in[0,2\pi),&\cdots,\theta_d\in[0,2\pi).
\end{split}\label{Td}
\end{equation}
Since LC connections are trivial, CKE (\ref{CKE}) reduces to
\[ \partial_\mu\xi_\nu+\partial_\nu\xi_\mu=\frac2d\left(\frac1{l_1^2}\partial_{\theta_1}\xi_{\theta_1}+\cdots+\frac1{l_d^2}\partial_{\theta_d}\xi_{\theta_d}\right)\gamma_\mn. \]
Because of the periodic boundary conditions, $\xi$ must have a form
\begin{equation}
    \xi_\mu(\theta_1,\dots,\theta_d)=\sum_{m_1,\dots,m_d\in\mbb Z}e^{i(m_1\theta_1+\cdots+m_d\theta_d)}c_\mu^{(m_1,\dots,m_d)}\label{Tdansatz}
\end{equation}
where $c_\mu^{(m_1,\dots,m_d)}$ are complex constants with the reality condition
\begin{equation}
    \forall(m_1,\dots,m_d)\in\mbb Z^d,\quad c_\mu^{(m_1,\dots,m_d)}=c_\mu^{(-m_1,\dots,-m_d)*}.\label{Tdreality}
\end{equation}
Naively, `diagonal' part $(\mu,\nu)=(\theta_j,\theta_j)$ yields $d$ conditions, however, as we remarked in section \ref{conf}, the trace part is trivial, and we can forget about one of them, say the $d$th condition. Thus, we get $(d-1)$ nontrivial conditions from the `diagonal' part. `Nondiagonal' part $(\mu,\nu)=(\theta_i,\theta_j)$ with $i\neq j$ yields $\frac{d(d-1)}2$ conditions. Collecting all of them in a matrix form, we get\footnote{From now on we denote $\mathbf m:=(m_1,\dots,m_d)\in\mbb Z^d$.}
\begin{equation}
    0_{N(d)}=\begin{pmatrix}(d-1)m_1&-\left(\frac{l_1}{l_2}\right)^2m_2&\cdots&\cdots&-\left(\frac{l_1}{l_d}\right)^2m_d\\-\left(\frac{l_2}{l_1}\right)^2m_1&(d-1)m_2&-\left(\frac{l_2}{l_3}\right)^2m_3&\dots&-\left(\frac{l_2}{l_d}\right)^2m_d\\\vdots&&\ddots&&\vdots\\-\left(\frac{l_{d-1}}{l_1}\right)^2m_1&\dots&-\left(\frac{l_{d-1}}{l_{d-1}}\right)^2m_{d-2}&(d-1)m_{d-1}&-\left(\frac{l_{d-1}}{l_d}\right)^2m_d\\m_2&m_1&0&\cdots&0\\\vdots&&\ddots&&\vdots\\0&\cdots&0&m_d&m_{d-1}\end{pmatrix}\begin{pmatrix}c_{\theta_1}^{(\mathbf m)}\\\vdots\\c_{\theta_d}^{(\mathbf m)}\end{pmatrix},\label{Tdconf}
\end{equation}
where the matrix on the RHS is an $N(d)\times d$ real matrix. We can consider the matrix $M$\footnote{We hope the reader does not get confused by our abuse of the letter $M$ for a manifold and the matrix.} as a map
\begin{equation}
\begin{array}{ccccc}
    M:&\mbb Z^d&\longrightarrow&M(N(d),d;\mbb R)&\\
    &\rotatebox{90}{$\in$}&&\rotatebox{90}{$\in$}&\\
    &\mathbf m:=(m_1,\cdots,m_d)&\longmapsto&M(\mathbf m)&:\quad\mbb C^d\to\mbb C^{N(d)}.
\end{array}\label{Mm}
\end{equation}
Therefore, CKEs have reduced to an algebraic condition
\begin{equation}
    \text{for a fixed }\mathbf m\in\mbb Z^d,\quad\begin{pmatrix}c_{\theta_1}^{(\mathbf m)}\\\vdots\\c_{\theta_d}^{(\mathbf m)}\end{pmatrix}\in\ker M(\mathbf m).\label{ker}
\end{equation}

The algebraic condition (\ref{ker}) may cause a confusion because $M(\mathbf m)$ is a real matrix while $d$-vectors $\begin{pmatrix}c_{\theta_1}^{(\mathbf m)}\\\vdots\\ c_{\theta_d}^{(\mathbf m)}\end{pmatrix}$ are complex valued. Since $c_\mu^{(\mathbf m)}$ are complex, it is convenient to write
\[ c_\mu^{(\mathbf m)}\equiv a_\mu^{(\mathbf m)}+ib_\mu^{(\mathbf m)} \]
with real $a_\mu^{(\mathbf m)}$ and $b_\mu^{(\mathbf m)}$. In this language, the reality condition (\ref{Tdreality}) is given by
\[ a_\mu^{(\mathbf m)}=a_\mu^{(-\mathbf m)},\quad b_\mu^{(\mathbf m)}=-b_\mu^{(-\mathbf m)}. \]
The decomposition reduces the algebraic condition (\ref{ker}) to
\begin{equation}
    \text{for a fixed }\mathbf m\in\mbb Z^d,\quad\begin{pmatrix}a_{\theta_1}^{(\mathbf m)}\\\vdots\\ a_{\theta_d}^{(\mathbf m)}\end{pmatrix}\in\ker M(\mathbf m)\&\begin{pmatrix}b_{\theta_1}^{(\mathbf m)}\\\vdots\\ b_{\theta_d}^{(\mathbf m)}\end{pmatrix}\in\ker M(\mathbf m).\label{kerab}
\end{equation}
Thus, more precisely, the algebraic condition (\ref{ker}) is a set of $2N(d)$ real conditions on $2d$ real variables $a_\mu^{(\mathbf m)}$ and $b_\mu^{(\mathbf m)}$ (for a fixed $\mathbf m\in\mbb Z^d$). Employing the decomposition, we can safely discuss just real objects. This decomposition should be understood below.

Returning to our journey to find conformal transformations, $0_d\in\mbb C^d$ is a trivial element of the kernel, but it does not give a nontrivial conformal transformation. To obtain a nontrivial conformal transformation, we need nontrivial element of the kernel. In other words, we need $\dim_{\mbb R}\ker M(\mathbf m)\ge1$. Using the rank-nullity theorem, we have
\[ \rank M(\mathbf m)+\dim_{\mbb R}\ker M(\mathbf m)=d, \]
or solving this for the dimension of the kernel,\footnote{Let us call $k:=\dim_{\mbb R}\ker M(\mathbf m)$. Then, the kernel can be recognized as a Grassmannian $Gr(k,d)$ because $\ker M(\mathbf m)$ span a $k$-dimensional subvector space of $\mbb C^d$, or more precisely $a_\mu^{(\mathbf m)}$ and $b_\mu^{(\mathbf m)}$ each spans a $k$-dimensional subvector space of $\mbb R^d$. These are $\mbb R^d$ and not some compact spaces like $\mbb T^d$ because if one expands (\ref{Tdansatz}), $a_\mu^{(\mathbf m)}$ and $b_\mu^{(\mathbf m)}$ are multiplied by trigonometric functions, and the constants can take arbitrarily large numbers without being identified at some finite number.} we get
\begin{equation}
    \dim_{\mbb R}\ker M(\mathbf m)=d-\rank M(\mathbf m).\label{dimkerM}
\end{equation}
This equality ensures an existence of nontrivial conformal transformation iff the rank of $M(\mathbf m)$ is smaller than $d$:
\begin{equation}
    \rank M(\mathbf m)<d\iff\begin{pmatrix}c_{\theta_1}^{(\mathbf m)}\\\vdots\\c_{\theta_d}^{(\mathbf m)}\end{pmatrix}\text{(with reality) generates nontrivial conformal transformation.}\label{rankconf}
\end{equation}
Therefore, our journey to find conformal transformations have reduced to compute the rank of matrices $M(\mathbf m)$. For a general $\mathbf m\in\mbb Z^d$, $\rank M(\mathbf m)=d$, and we cannot get a nontrivial conformal transformation. We get a nontrivial transformation if $\mathbf m\in\mbb Z^d$ is chosen so that $\rank M(\mathbf m)<d$.

The discussion above gives us a criterion to judge when nontrivial conformal transformations appear, however, to construct explicit transformations we have to solve $\ker M(\mathbf m)$. This would be a difficult task in general, however, we can accomplish the goal in lower dimensions. Possibly with a help of Mathematica, we could solve the problem in $d=2,3,4,$ and $5$.\footnote{When $d=6$, Mathematica stops computation.} Interestingly, we always find
$\text{Conf}_0(\mbb T^d)\simeq U(1)^d$. Hence it is natural to guess that $\text{Conf}_0(\mbb T^d)\cong U(1)^d$ for $d\ge2$. In fact, we can prove the conjecture. We first present the proof in subsection \ref{proofTd}. Using the knowledge, we discuss deformation problems of CFTs on $\mbb T^d$. Especially, we discuss exactly ``marginal'' deformations. We suggest some candidates of conformal manifolds without assuming SUSY.

\subsection{Proof of Conf$_0(\mbb T^d)\simeq U(1)^d$}\label{proofTd}
In this subsection, we prove a conjecture
\[ \text{Conf}_0(\mbb T^d)\simeq U(1)^d\quad(d\ge2), \]
led from our observation in $d=2,3,4,$ and $5$. In dimensions $d>2$, we have $N(d)>d$.\footnote{$N(d)\ge d$ is equivalent to $(d+1)(d-2)\ge0$.} Thus, the number of constraints are larger than the number of degrees of freedom, and there are no nontrivial solutions in general.\footnote{The situation is reminiscent of \cite{LS}. If some conditions are linearly dependent, the actual number of constraints reduces, and the conformal group would be enhanced.} Since constraints become more and more severe in higher dimensions in the sense that $N(d)-d$ gets larger and larger, the statement that $\rank M(\mathbf m)=0$ or $d$ seems to hold. Indeed, we can prove the claim when the radii of the $d$-torus are general enough, or more precisely, the ratio squared of radii $(l_i/l_j)^2$ with $i\neq j$ are irrational. So in this paper we assume all the ratio squared are irrational.\footnote{We tried some combinations of radii relaxing the assumption, but we always got the same result. So although we failed to prove the result without the assumption, we speculate that the result would be also true even if we relax the assumption.}

To prove the claim, it is enough to show the linear (in)dependence of $d$ columns of the matrix $M(\mathbf m)$. It is crucial that components of the first upper $(d-1)$ rows of the matrix take irrational values, while the other components in lower rows are integer valued. Let us denote the $j$th column $e_j$. The $d$ columns are linearly independent iff
\[ \sum_{j=1}^dd_je_j=0_{N(d)} \]
requires the coefficients vanish $d_j=0$.

Let us first study the case with all $m_j$'s are nonzero. Suppose we could find a set of nonzero coefficients $\{d_j\}_{j=1,\dots,d}$. To make linear combinations of components in the upper $(d-1)$ rows vanish, (i) the solution $d_j$ must be irrational.\footnote{One may complain ratios of irrational numbers are not always irrational. This is true, but due to the structure of $M(\mathbf m)$, ratios of components in the same row are again given by ratio squared $(l_j/l_k)^2$, and by our assumption, they are all irrational. It would be interesting to relax the assumption. Then, some ratio squared become rational, possibly leading to linear dependence of columns, and to an enhancement of the conformal group acting on the $d$-torus with such a special radii.} On the other hand, to make linear combinations of the lower components which are integer valued vanish, (ii) the solutions must be rational numbers. These two conditions cannot be satisfied simultaneously. Thus, for general radii $l_j$s, in the sense that all the ratio squared are irrational, the $d$ columns are linearly independent when $m_j\neq0$. For such $\mathbf m\in\mbb Z^d$, we have $\rank M(\mathbf m)=d$, or equivalently $\dim_{\mbb R}\ker M(\mathbf m)=0$. We do not get nontrivial conformal transformations for these $\mathbf m\in\mbb Z^d$.

When some $m_j$'s are zero, many components of $e_j$'s vanish, so there is a possibility that some of them are linearly dependent. When more than one $m_j$'s are nonzero, there are necessarily irrational numbers in the upper component while all lower components are integers, and we can repeat the argument in the previous paragraph to conclude $d$ columns are linearly independent for general radii $l_j$'s. When all but one $m_j$'s are zero, we can explicitly read $\text{rank}M(\mathbf m)=d$ from (\ref{Tdconf}) because every $m_j$ appears at least once in every column in different rows, meaning linear independence of $e_j$'s. Finally, when all $m_j$'s are zero, we obviously have $\rank M(\mathbf m=0_d)=0$, or equivalently $\dim_{\mbb R}\ker M(\mathbf m=0_d)=d$.

To summarize, we have seen (when all the ratio squared are irrational)
\begin{equation}
    \text{rank}M(\mathbf m)=\begin{cases} d&(\mathbf m\neq0_d)\\0&(\mathbf m=0_d)\end{cases}.\label{rankM}
\end{equation}
This proves
\begin{equation}
    \dim_{\mbb R}\ker M(\mathbf m)=\begin{cases}0&(\mathbf m\neq0_d)\\d&(\mathbf m=0_d)\end{cases},\label{dimker}
\end{equation}
so that
\begin{equation}
    \xi_\mu(\theta_1,\dots,\theta_d)=c_\mu^{(0,\dots,0)}=a_\mu^{(0,\dots,0)},\label{xiTd}
\end{equation}
and we conclude
\begin{equation}
    \text{Conf}_0(\mbb T^d)\simeq U(1)^d\simeq\text{Isom}_0(\mbb T^d)\quad(d\ge2).\label{confTd}
\end{equation}
We used $d\ge2$ because for the case $d=1$, the matrix $M(\mathbf m)$ in (\ref{Tdconf}) is not defined.\footnote{Formally, it is a $N(1)\times1=0\times1$ matrix.}

As an immediate consequence, we learn that $Gr(k,n)$ is empty unless $k=0$ or $d$, i.e.,
\[ Gr(k,d)=\emptyset\iff k\neq0,d. \]
In other words, from (\ref{dimker}), for each $\mathbf m\neq0_d$, we obtain a short exact sequence of complex fields, or more precisely, recalling the decomposition around (\ref{kerab}), a short exact sequence of real fields:
\begin{equation}
    0\lra\mbb R^d\stackrel{M(\mathbf m)}\lra\mbb R^{N(d)}\lra\mbb R^{N(d)}/\mbb R^d\lra0.\label{SEC}
\end{equation}
It may be interesting to study (co)homologies obtained from the short exact sequence.

As we saw before, operators in CFTs defined on $\mbb T^d$ are thus labeled by $d$ sets of integer $U(1)$ `charges' $(q_1,\dots,q_d)$. The conformal WT relations are given by
\begin{equation}
    0\stackrel!=P_\mu\la O_1(x_1)\dots O_n(x_n)\ra.\label{confWTTd}
\end{equation}
Therefore, we have
\begin{equation}
    \la O_1(x_1)\cdots O_n(x_n)\ra=f(x_{12},x_{13},\dots,x_{n-1,n}).\label{nptfuncTd}
\end{equation}
In particular,
\begin{equation}
\begin{split}
    \la O(\theta_1,\dots,\theta_d)\ra&=\text{const.}\times\prod_{j=1}^d\delta_{q_j,0},\\
    \la O_1(\theta_1,\dots,\theta_d)O_2(\theta'_1,\dots,\theta'_d)\ra&=g(\theta_j-\theta'_j)\prod_{j=1}^d\delta_{q_j^{(1)}+q_j^{(2)},0},\\
    \la O_1(\theta_1,\dots,\theta_d)O_2(\theta'_1,\dots,\theta'_d)O_3(\theta''_1,\dots,\theta''_d)\ra&=h(\theta_j-\theta'_j,\theta'_j-\theta''_j)\prod_{j=1}^d\delta_{q_j^{(1)}+q_j^{(2)}+q_j^{(3)},0}.
\end{split}\label{Td123pt}
\end{equation}
Since there is no local counterterm which can shift the one-point functions, the constants are interesting observables. Nonlocal parts of higher-point functions are also physical.

\subsection{Conformal manifolds of CFTs on $\mbb T^d$}
Deformation problems of CFTs
\[ S_\text{CFT}\mapsto S:=S_\text{CFT}+\lambda\int d^dx\sqrt\gamma O(x) \]
are important in QFTs. Since the operator $O$ has to belong to a representation of the original conformal group, the operator is labeled by $d$ $U(1)$ `charges' $(q_1,\dots,q_d)\in\mbb Z^d$ when we consider CFTs on $\mbb T^d$. If some of $q_j$'s are nonzero, the deformation breaks the conformal symmetry (to discrete subgroups in general\footnote{For example, if the operator belongs to $(2,0,\dots,0)$, the operator is still invariant under $\pi$-shift of $\theta_1$, i.e., $\mbb Z_2$. More generally, if the operator belongs to $(q_1,0,\dots,0)$, the first $U(1)$ is broken to $\mbb Z_{q_1}$.}). For example, if precisely one of the $q_j$'s, say $q_1$, is nonzero, (the identity component of) the conformal group is broken to $\mbb Z_{q_1}\times U(1)^{d-1}$. Therefore, it is necessary for the deformation operators to belong to the trivial representation $(q_1,\dots,q_d)=(0,\dots,0)$ of $U(1)^d$ in order to preserve the conformal group. Let us denote such operators in the trivial representation $\{O_{I'}\}$. Of course, the identity operator is an element of the set $id\in\{O_{I'}\}$.

Do the deformations with $\{O_{I'}\}$ preserve the conformal symmetry $U(1)^d$? Since the representations are labeled by discrete quantum numbers, it may seem that the representations cannot be modified under the continuous deformations, however, there is another trigger which can spoil `marginal'\footnote{As we saw in examples in this paper, metric spaces which admit conformal groups with dilations are rare. Thus, marginality may not be a good property to characterize a family of CFTs in general. From now on we will call these operators which parametrize families of CFTs ``conformal operators'' for brevity.} property of the trivial operators, multiplet recombinations \cite{EMD} (see also \cite{KS}). When there are flavor symmetries, and some of the trivial operators belong to nontrivial representations of the flavor groups, the operators break some flavor symmetries under which they are charged. Thus, away from the conformal point $\lambda^{I'}=0$, the corresponding currents (if existed in the first place \cite{HO}\footnote{It would be interesting to reconsider the arguments of multiplet recombination with currents \cite{recombi} on curve manifolds $M$ with nontrivial $H_{d-2}(M)$, on which Noether currents may not exist.}) cease to be conserved, and the deformation operator and the broken current would combine to form a nontrivial representation, leading to possible conformal symmetry breaking. To avoid the possible multiplet recombinations, we further restrict to operators belonging to the trivial representations of flavor symmetries. Let us denote such operators $\{O_I\}\subset\{O_{I'}\}$. Then, the deformations
\begin{equation}
    S_\text{CFT}\mapsto S:=S_\text{CFT}+\lambda^I\int_{\mbb T^d}d^dx\sqrt\gamma O_I(x)\label{EMD}
\end{equation}
would preserve the conformal group $U(1)^d$.

One caveat in the discussion above is that we have ignored the possibility of `large' conformal transformations. There may exist `large' conformal transformations in $\text{Conf}(\mbb T^d)$, and in that case some operators $O_I$ would transform nontrivially under these elements of the group. If such operators are included in the deformation (\ref{EMD}), they break the full conformal symmetry $\text{Conf}(\mbb T^d)$ to its identity component $U(1)^d$. Furthermore, the integer `charges' may jump under the `large' conformal transformations. Since we do not know how to find `large' conformal transformations systematically, we leave these points as open questions.

Conformal manifolds associated to CFTs on (conformally) flat spaces are naturally equipped with a metric known as the Zamolodchikov metric $g_{IJ}(\lambda)$, which is defined through two-point functions of the conformal operators \cite{cthm}:
\[ \la O_I(x)O_J(x')\ra_{\mbb R^d}\equiv\frac{g_{IJ}(\lambda)}{(x-x')^{2d}}. \]
Since the coefficient function $g(\lambda)$ is symmetric and nonnegative thanks to the reflection positivity \cite{OS}, it can be recognized as a metric on the theory space. However, on general Riemannian spaces, to the best of our knowledge, it is not known whether the reflection positivity holds. Spacetime parts of the two-point functions are not positive in general, neither. For example, on our $\mbb T^d$, there is no guarantee that the two-point functions
\begin{equation}
    \la O_I(\theta_1,\dots,\theta_d)O_J(\theta'_1,\dots,\theta'_d)\ra\equiv g_{IJ}(\lambda)f(\theta_1-\theta'_1,\dots,\theta_d-\theta'_d)\label{gIJ}
\end{equation}
have positive $f$. All we can say is that the coefficient function $g(\lambda)$ is symmetric due to the Bose symmetry. It would be interesting to study properties of $g(\lambda)$.

The reason why we could easily find candidates of conformal operators without assuming SUSY is because (the identity component of) the conformal group is compact. Our example does not answer the original problem of finding conformal manifolds of non-SUSY CFTs defined on conformally flat spacetimes, but this observation makes clear that why it is difficult to find such candidates; on conformally flat manifolds, conformal groups are noncompact, and a priori, no known mechanism prevents continuous quantum corrections except SUSY.

It is easy to raise concrete examples of such trivial operators which would label families of CFTs. On $\mbb T^d$, we have a real scalar field $\phi$ which belongs to the trivial representation $(q_1,\dots,q_d)=(0,\dots,0)$ of $U(1)^d$. Using the field, we can get many trivial operators. Suppose we have $N_f$ of them $\phi_j$ where $j=1,\dots,N_f$ with degenerated mass $m$. Then, the theory
\begin{equation}
    S=\int_{\mbb T^d}d^dx\sqrt\gamma\Big(\frac12\gamma^\mn\delta^{jk}\partial_\mu\phi_j(x)\partial_\nu\phi_k(x)+\frac12m^2\delta^{jk}\phi_j(x)\phi_k(x)\Big)\label{realscalar}
\end{equation}
enjoys an $SO(N_f)$ flavor symmetry. Interestingly, the (diagonal) mass term $\delta^{jk}\phi_j\phi_k$ is also an element of $\{O_I\}$, thus the mass term can be exactly `marginal.' We can construct many operators which belong to $\{O_I\}$. These are some of the examples:
\begin{itemize}
    \item $O=(\delta^{jk}\phi_j\phi_k)^3$ on $\mbb T^3$,
    \item $O=(\delta^{jk}\phi_j\phi_k)^2$ on $\mbb T^4$.
\end{itemize}
These would be candidates of conformal operators. Generalization to various degenerated masses $m,\widetilde m,\dots$ is trivial. In that case, the flavor symmetry would be given by $SO(N_f)\times SO(\widetilde{N_f})\times\dots$ where $N_f$, $\widetilde{N_f},\dots$ are the numbers of real scalars with mass $m$, $\widetilde m$, and so on. It is desirable to check these suggestions explicitly, however, since we do not know about `large' conformal transformations, we leave this point as a future work.

Since all information we used about $\mbb T^d$ is the compactness of the conformal group $U(1)^d$, one can also easily find candidates of conformal operators on spaces with compact conformal groups. Within the examples we studied, they include $\mbb S^1\times[0,1],\text{KB,MS},\mbb S^1_{l(\neq Nr)}\times\mbb H^2_r,$ and $\mbb{RP}^2\times\mbb S^1$. Conformal manifolds of $\mbb S^1_l\times\mbb H^2_r$ are especially worth studying. Since the conformal group becomes noncompact by setting $l=Nr$ with $N\in\mbb N^\times$, it would be interesting to see how conformal manifolds change by rescaling $l$. More concretely, since the compact conformal group is contained as a subgroup of the larger conformal group, conformal manifolds associated to $\mbb S^1_{Nr}\times\mbb H^2_r$ must be a subspace of the one associated to $\mbb S^1_{l(\neq Nr)}\times\mbb H^2_r$.

\section{Discussion}\label{discussion}
In this paper, we tried to generalize the study of CFTs by considering CFTs on both orientable and unorientable curved manifolds possibly with boundaries. As a preparation, we first reviewed conformal transformations on curved manifolds. Using the formalism, we then studied various CFTs defined on lower dimensional manifolds possibly with boundaries. We saw boundary conditions derived from a simple fact $-$ given local coordinate systems be single-valued $-$ correctly reproduces known conformal groups. Using the method, we also produced a few new results. One of them is a phenomenon of conformal symmetry enhancement. In general, the conformal group of $\mbb S^1_l\times\mbb H^2_r$ is given by $U(1)\times U(1)$, however, if we set $l=Nr$ with $N\in\mbb N^\times$, we found that the space enjoys a larger conformal group $SO(2,1)\times U(1)$. This symmetry can be understood as the conformal symmetry of $\mbb R^3\cup\{\infty\}$ with codimension two defect. Equipped with this interpretation, it is obvious that $\mbb S^1_r\times\mbb H^{d-1}_r$ has the conformal group $SO(d-1,1)\times SO(2)$, and using our argument to replace $\tau$ by $\tau'=N\tau$, it is believed that the conformal symmetry enhancement also takes place in higher dimensions; the conformal group associated to $\mbb S^1_{Nr}\times\mbb H^{d-1}_r$ with $N\in\mbb N^\times$ would also be given by $SO(d-1,1)\times SO(2)$, which is larger than those associated to the manifolds with $l/r\notin\mbb N$. We also determined forms of correlation functions in these examples. Taking local counterterms on curved spaces into account, we also discussed which parts of the correlators are physical.

Through the examples, we found suitable boundary conditions make the analysis remarkably simple by reducing problems of solving partial differential equations to those of solving algebraic equations. Thus, in the last section we focused on a manifold with periodic boundary conditions in every direction, namely a $d$-torus $\mbb T^d$. We found Conf$_0(\mbb T^d)\simeq U(1)^d$ for the cases $d=2,3,4,$ and $5$ (with generic radii). From the results, we conjectured $\text{Conf}_0(\mbb T^d)\simeq U(1)^d$ for $d\ge2$, and in fact, we could prove the claim. Employing the fact that representations of $U(1)^d$ are labeled by discrete quantum numbers, i.e., compactness of the conformal group, we also suggested some candidates of conformal manifolds without assuming SUSY.

We have left many interesting (and important) questions:
\begin{itemize}
\item `large' conformal transformations; we have mainly focused on identity components Conf$_0(M)$ of the conformal groups on metric spaces $(M,\gamma)$. It would be interesting to study conformal transformations which are not in the identity component. One way would be to write down all Lie algebras which are isomorphic to the conformal algebras in question, and see whether the nontrivial center elements are elements of the (full) conformal groups, as we did in this paper.
\item direct computation of correlators; in some cases, we employed conformal maps between flat spaces to compute correlation functions of CFTs on curved spaces because direct computation was difficult. It is desirable to develop a direct way. It would remove a bottleneck to find conformal maps, and further in some cases they are necessary because not all metric spaces can be related to flat spaces.
\item (conformal) defects \cite{defect}; we have seen $\mbb R^d\cup\{\infty\}$ with codimension one conformal defect (i.e., boundary) is conformal to $\mbb H^d$, and $\mbb R^d\cup\{\infty\}$ with codimension two conformal defect $\mbb R^{d-2}\cup\{\infty\}$ is conformal to $\mbb S^1\times\mbb H^{d-1}$. So a natural question is what is a (curved) space $M$ which is conformal to $\mbb R^d\cup\{\infty\}$ with codimension three (or higher) conformal defect. At least we know the metric space must have the conformal algebra
\[ \mfrak{conf}(M)\cong\mfrak{so}(d-2,1)\oplus\mfrak{so}(3). \]
Also, conformal defects in curved spaces themselves would be interesting to study. For instance, a $d$-torus $\mbb T^d$ with $q$-torus $\mbb T^q$ as a codimension $(d-q)$ defect may be interesting objects to study. The defects obviously have conformal symmetries $\text{Conf}_0(\mbb T^q)\simeq U(1)^q$.\footnote{Those cases with $q=0,1$ are special. They would need separate treatments.}
\item conformal anomaly; as in the case of CFTs defined on conformally flat even-dimensional spaces, there is a possibility that the conformal symmetries suffer from conformal anomalies \cite{confanom}. Since part of the usual conformal anomaly, i.e., the $a$-anomaly, counts the `number' of effective degrees of freedom in a theory \cite{cthm,athm}, it is interesting to investigate whether a similar statement also holds on manifolds which are not conformal to $\mbb R^d$ or $\mbb R^{d-1,1}$ (possibly with $\infty$). One would be able to tackle this problem using the local renormalization group (LRG) mentioned in the appendix \ref{massLRG}.
\item gravity dual; having AdS/CFT correspondence \cite{Malda97} in mind, it is natural to ask what are gravity duals of CFTs on curved spaces, if such a correspondence exists. If we employ the matching of the symmetries and dimension independent structure of the known correspondence as guiding principles, in case of $\mbb T^d$, gravitational theories defined also on $\mbb T^d$ is one of a natural candidate of the dual because we have $\text{Isom}_0(\mbb T^d)\simeq\text{Conf}_0(\mbb T^d)$ ($d\ge2)$. Studies in this direction would shed light on how crucial the matching of symmetries is in the AdS/CFT correspondence.
\end{itemize}
We would like to address these problems in the future.

\section*{Acknowledgement}
We would like to thank Masazumi Honda, Taishi Katsuragawa, Zohar Komargodski, and Yang Zhou for helpful comments on the manuscript. We are also grateful for ``East Asia Joint Workshop on Fields and Strings 2018'' held at KIAS and ``KEK Theory Workshop 2018'' held at KEK, where part of the results were presented.

\appendix
\setcounter{section}{0}
\renewcommand{\thesection}{\Alph{section}}
\setcounter{equation}{0}
\renewcommand{\theequation}{\Alph{section}.\arabic{equation}}
\section{Derivation of (\ref{conffactors})}\label{pushexp}
Taking the determinant of the definition of the conformal map
\[ \gamma'_{\alpha\beta}(y(x))\pdif{y^\alpha}{x^\mu}(x)\pdif{y^\beta}{x^\nu}(x)=\Omega^2(x)\gamma_{\mn}(x), \]
we obtain
\begin{equation}
    \Omega^2(x)=\left(\sqrt{\frac{\gamma'(y(x))}{\gamma(x)}}\det\left(\pdif yx(x)\right)\right)^{2/d},\label{Omega2}
\end{equation}
where $\gamma:=\det\gamma_{\mn}$ and $\gamma':=\det\gamma'_{\alpha\beta}$. What we want to check can be rewritten
\[ \xi_M^\mu(\varphi^{-1}(y))\pdif{\ln\Omega^2}{x^\mu}(\varphi^{-1}(y))=\frac2d\Big[\nabla^{(\gamma')}\cdot\xi_N(y)-\nabla^{(\gamma)}\cdot\xi_M(\varphi^{-1}(y))\Big]. \]
Using (\ref{Omega2}), LHS reduces to
\begin{align*}
    (\text{LHS})&=\frac2d\xi_M^\mu(\varphi^{-1}(y))\pdif{}{x^\mu}\Big[\ln\sqrt{\gamma'(y)}-\ln\sqrt{\gamma(\varphi^{-1}(y))}+\ln\det\left(\pdif yx(\varphi^{-1}(y))\right)\Big]\\
    &=\frac2d\xi_M^\mu(\varphi^{-1}(y))\Big[\pdif{y^\alpha}{x^\mu}(\varphi^{-1}(y))\Gamma_{\alpha\beta}^{'\beta}-\Gamma_{\mn}^\nu+\pdif{x^\nu}{y^\alpha}(y)\pdif{^2y^\alpha}{x^\mu\partial x^\nu}(\varphi^{-1}(y))\Big]\\
    &=\frac2d\Big[\left(\pdif{}{x^\mu}\xi_M^\mu(\varphi^{-1}(y))+\xi_M^\mu(\varphi^{-1}(y))\pdif{x^\nu}{y^\alpha}(y)\pdif{^2y^\alpha}{x^\mu\partial x^\nu}(\varphi^{-1}(y))\right)+\Gamma_{\alpha\beta}^{'\beta}\xi_N^\alpha(y)\\
    &~~~~~~~-\pdif{}{x^\mu}\xi_M^\mu(\varphi^{-1}(y))-\Gamma_{\mn}^\nu\xi_M^\mu(\varphi^{-1}(y))\Big]\\
    &=\frac2d\Big[\pdif{}{y^\alpha}\xi_N^\alpha(y)+\Gamma_{\alpha\beta}^{'\beta}\xi_N^\alpha(y)-\pdif{}{x^\mu}\xi_M^\mu(\varphi^{-1}(y))-\Gamma_{\mn}^\nu\xi_M^\mu(\varphi^{-1}(y))\Big]=(\text{RHS}),
\end{align*}
where $\Gamma_{\mn}^\rho$ and $\Gamma_{\alpha\beta}^{'\gamma}$ are LC connections constructed from $\gamma$ and $\gamma'$, respectively. In the second line we used the famous formula
\[ \pdif{}{x^\mu}\ln\sqrt\gamma=\Gamma_{\mn}^\nu, \]
and similar formula for $\Gamma'$. In the last line, we used
\begin{align*}
    \pdif{}{y^\alpha}\xi_N^\alpha(y)&=\pdif{}{y^\alpha}\left(\xi_M^\mu(\varphi^{-1}(y))\pdif{y^\alpha}{x^\mu}(\varphi^{-1}(y))\right)\\
    &=\pdif{x^\nu}{y^\alpha}(y)\pdif{\xi_M^\mu}{x^\nu}(\varphi^{-1}(y))\pdif{y^\alpha}{x^\mu}(\varphi^{-1}(y))+\xi_M^\mu(\varphi^{-1}(y))\pdif{x^\nu}{y^\alpha}(y)\pdif{^2y^\alpha}{x^\nu\partial x^\mu}(\varphi^{-1}(y))\\
    &=\pdif{}{x^\mu}\xi_M^\mu(\varphi^{-1}(y))+\xi_M^\mu(\varphi^{-1}(y))\pdif{x^\nu}{y^\alpha}(y)\pdif{^2y^\alpha}{x^\nu\partial x^\mu}(\varphi^{-1}(y)).
\end{align*}

\section{$SU(1,1)$, $SL(2,\mbb R)$, and $SO(2,1)$}\label{suslso}
The group\footnote{For more details of the group, see, for example, \cite{CDP}.} $SU(1,1)$ or $SU(1,1;J)$ is defined as a collection of two by two complex matrices which preserve $J=\begin{pmatrix}1&0\\0&-1\end{pmatrix}$ and have determinant one:
\begin{equation}
    SU(1,1;J):=\{M\in M(2,2;\mbb C)|M^\dagger JM=J\&\det M=1\}.\label{SU(1,1;J)}
\end{equation}
A general element of the group has a form
\[ M=\begin{pmatrix}a&b^*\\b&a^*\end{pmatrix} \]
with $|a|^2-|b|^2=1$. Decomposing $a,b$ into real and imaginary parts
\[ a=t+iz,\quad b=x+iy, \]
one can write
\[ M=t1_2+iz\sigma_3+x\sigma_1+y\sigma_2,\quad t^2+x^2-y^2-z^2=1. \]
From this expression, one can easily learn that $SU(1,1)$ is not simply-connected because $\mbb S^1:=\{t^2+x^2=1\}$ is embedded. If one realizes $|a|\ge1$, it is convenient to use another parametrization of the group elements:
\begin{equation}
\begin{split}
	M(\eta,\phi,\psi)=\begin{pmatrix}\cosh\eta e^{i\phi}&\sinh\eta e^{-i\psi}\\\sinh\eta e^{i\phi}&\cosh\eta e^{-i\phi}\end{pmatrix},\\
	\eta\in[0,+\infty),\quad\phi\in[0,2\pi),\quad\psi\in[0,2\pi).
\end{split}\label{SU(1,1)element}
\end{equation}
Using the generators\footnote{These satisfy the standard commutation relations
\[ [k_1,k_2]=-ik_3,\quad[k_2,k_3]=ik_1,\quad[k_3,k_1]=ik_2. \]
If we absorb the imaginary unit $i$ and redefine $k_j':=k_j/i$, these satisfy the commutation relations we saw many times:
\[ [k_1',k_2']=-k_3',\quad[k_2',k_3']=k_1',\quad[k_3',k_1']=k_2'. \]}
\[ k_1:=i\frac{\sigma_1}2,\quad k_2:=i\frac{\sigma_2}2,\quad k_3:=\frac{\sigma_3}2, \]
and $k_\pm:=k_1\pm ik_2$, which obey commutation relations
\begin{equation}
	[k_+,k_-]=-2k_3,\quad[k_3,k_\pm]=\pm k_\pm,\label{su(1,1)}
\end{equation}
one can express (\ref{SU(1,1)element}) as
\begin{equation}
	M(\eta,\phi,\psi)=e^{i(\phi-\psi)k_3}e^{-2i\eta k_1}e^{i(\phi+\psi)k_3}.\label{Mk}
\end{equation}

Now, $J$ is unitary equivalent to $\tilde J:=\begin{pmatrix}0&i\\-i&0\end{pmatrix}$. In fact,
\[ U=\frac1{\sqrt2}\begin{pmatrix}e^{i\alpha}&-ie^{i\beta}\\-ie^{i\alpha}&e^{i\beta}\end{pmatrix}\quad(\alpha,\beta\in\mbb R) \]
does the job:
\[ \tilde J=UJU^{-1}. \]
Using this new Hermitian form, one obtains another group
\begin{equation}
	SU(1,1;\tilde J):=\{M\in M(2,2;\mbb C)|M^\dagger\tilde JM=\tilde J\&\det M=1\},\label{SU(1,1;tildeJ)}
\end{equation}
which is unitary equivalant to $SU(1,1)$. By an explicit computation, one can show that an arbitrary two by two matrix $M$ with $\det M=1$ satisfies
\[ M^t\tilde JM=\tilde J. \]
Therefore, we have
\[ \forall M\in SU(1,1;\tilde J),\quad M^*=M, \]
and we can also express the group as
\[ SU(1,1;\tilde J)=\{M\in M(2,2;\mbb C)|M^*=M\&\det M=1\}. \]
This is nothing but the definition of $SL(2,\mbb R)$:
\begin{equation}
    SU(1,1;\tilde J)=SL(2,\mbb R),\label{SUSL}
\end{equation}
proving unitary equivalence of $SU(1,1)$ and $SL(2,\mbb R)$. To summarize our consideration, we have
\begin{equation}
	USU(1,1)U^{-1}=SU(1,1;\tilde J)=SL(2,\mbb R).\label{SUSUSL}
\end{equation}

Let us consider a homomorphism from $SU(1,1)$. The Lie algebra $\mfrak{su}(1,1)$ is a three-dimensional vector space $V_{\text{Adj}(SU(1,1))}^3$ spanned by generators $K_1,K_2,$ and $K_3$ which is isomorphic to $\mbb R^3$ through the correspondence
\[ K_1\leftrightarrow\begin{pmatrix}1\\0\\0\end{pmatrix},\quad K_2\leftrightarrow\begin{pmatrix}0\\1\\0\end{pmatrix},\quad K_3\leftrightarrow\begin{pmatrix}0\\0\\1\end{pmatrix}. \]
Since the vector space $V_{\text{Adj}(SU(1,1))}^3$ is nothing but the representation space of the adjoint representation, defining $\text{Ad}(K_j)K_k:=[K_j,K_k]$, representation matrices can be computed from the commutation relations (\ref{su(1,1)}):
\[ \text{Ad}(K_1)=\begin{pmatrix}0&0&0\\0&0&-i\\0&-i&0\end{pmatrix},\quad\text{Ad}(K_2)=\begin{pmatrix}0&0&i\\0&0&0\\i&0&0\end{pmatrix},\quad\text{Ad}(K_3)=\begin{pmatrix}0&-i&0\\i&0&0\\0&0&0\end{pmatrix}. \]
Therefore, we have obtained the three-dimensional representation
\begin{equation}
	\text{Adj}:SU(1,1)\to\End(V_{\text{Adj}(SU(1,1))}^3),\label{AdjSU(1,1)}
\end{equation}
which is a homomorphism by definition, and the general representation matrix can be obtained from (\ref{Mk}) as
\begin{equation}
	R(\eta,\phi,\psi):=e^{i(\phi-\psi)\text{Ad}(K_3)}e^{-2i\eta\text{Ad}(K_1)}e^{i(\phi+\psi)\text{Ad}(K_3)}.\label{Rdef}
\end{equation}
A straightforward computation shows
\begin{equation}
	R(\eta,\phi,\psi)=\begin{pmatrix}c_-c_+-\cosh2\eta s_-s_+&c_-s_++\cosh2\eta s_-c_+&-\sinh2\eta s_-\\-s_-c_+-\cosh2\eta c_-s_+&-s_-s_++\cosh2\eta c_-c_+&-\sinh2\eta c_-\\\sinh2\eta s_+&-\sinh2\eta c_+&\cosh2\eta\end{pmatrix},\label{R}
\end{equation}
where $c_\pm:=\cos(\phi\pm\psi),s_\pm:=\sin(\phi\pm\psi)$. Using this expression, one can convince oneself that $R\in SO_+(2,1)$. In fact, it is easy to show the matrix has determinant one with the help of $\det M=e^{\tr\ln M}$. One can also show the matrix preserves $\mathbf v\cdot\mathbf w:=-v_1w_1-v_2w_2+v_3w_3$. For example, $\begin{pmatrix}0\\0\\1\end{pmatrix}$ is mapped to $\begin{pmatrix}-\sinh2\eta s_-\\-\sinh2\eta c_-\\\cosh2\eta\end{pmatrix}$, and this preserves $v_3w_3$. In addition, since $R_{33}=\cosh2\eta\ge1$, one concludes that $R$ is an element of $SO_+(2,1)$. Thus, (\ref{AdjSU(1,1)}) is a homomorphism from $SU(1,1)$ to $SO_+(2,1)$.

Let us study the kernel. To make $R_{33}=1$, we need $\eta=0$. Then, (\ref{R}) reduces to
\[ R(\eta=0,\phi,\psi)=\begin{pmatrix}\cos2\phi&\sin2\phi&0\\-\sin2\phi&\cos2\phi&0\\0&0&1\end{pmatrix}. \]
Thus, $(\eta,\phi,\psi)=(0,0,\psi)$ and $(\eta,\phi,\psi)=(0,\pi,\psi)$ give $+1_3\in SO_+(2,1)$. However, using (\ref{SU(1,1)element}), one notices that these parameters give $+1_2\in SU(1,1)$ and $-1_2\in SU(1,1)$, respectively. Our consideration gives a short exact sequence
\begin{equation}
	1\to\mbb Z_2\to SU(1,1)\to SO_+(2,1)\to1,\label{SUSO}
\end{equation}
or $SU(1,1)/\mbb Z_2\simeq SO_+(2,1)$. We use the element $-1_2\in SU(1,1)$ to judge whether a given group associated to the Lie algebra $\mfrak{su}(1,1)\cong\mfrak{so}(2,1)$ is $SU(1,1)$ or $SO(2,1)$; if $-id$ is an element of the group, the group must be $SU(1,1)$, and if it is not an element, the group must be $SO(2,1)$.

\section{Mass dimension, RG, and local counterterms on curved spaces}\label{massLRG}
Let us review why Euclidean or Minkowski spaces have (only one) mass dimensions. First of all, on $\mbb R^d$ or $\mbb R^{d-1,1}$, we have the dilation symmetry as a part of the spacetime symmetry. The generator $x\cdot\partial/\partial x$ of the symmetry enables us to count the number of coordinates. In other words, the presence of dilations makes the word `number of coordinates' meaningful. Then, a priori, there can exist $d$ length scales $L_j\ (j=1,\dots,d)$, however, since the spaces have rotation symmetries, we have to be blind to directions we are facing. The rotation symmetries unify the length scales $L_1=\cdots=L_d=L$, and we only matter the number of coordinates and not their directions. Coordinates are conjugate to momentum variables. So in order to make the number of coordinates conserved, momentum must have the opposite number $L^{-1}$. Writing $M=L^{-1}$, we can count the number of coordinates dimensionful parameters have in units of $M$. This is the mass dimension on $\mbb R^d$ or $\mbb R^{d-1,1}$.

On the other hand, as we saw in many examples, few metric spaces have dilations as a part of the spacetime symmetries. On metric spaces without dilations, we cannot initiate the first step above. So it is meaningless to talk about `mass dimensions' or `numbers of coordinates' on such spaces.

Accordingly, the usual RG would not work on curved spaces. Recalling the Wilsonian RG, one has to separate modes with short wavelength (a.k.a. slow or heavy modes) from those with long wavelength (a.k.a. fast or light modes). On flat spaces, this separation can be done globally because the length defined by the flat metric does not depend on position. However, on curved metric spaces, length does depend on points in general, and one cannot separate modes with short and long wavelength globally. Rather, one has to separate them locally. Thus, one is forced to study responses of QFTs $-$ more precisely, the Schwinger functional (a.k.a. the vacuum functional), which is given by log of the partition function $-$ to local scale transformations. This goal can be achieved by a method called local renormalization group (LRG) \cite{LRG} in which local scale transformations are realized by the Weyl transformations\footnote{In the gravitational physics, this transformation is called conformal transformation, and conformal transformations $-$ angle preserving diffeomorphisms $-$ we have been discussing are called conformal isometries.}
\[ \gamma_\mn(x)\mapsto e^{2\sigma(x)}\gamma_\mn(x). \]
Therefore, on generic curved spaces, one is obliged to consider LRG, not the usual RG.

What matters for our analysis in the context of LRG is the rule to write down local counterterms on curved spaces. On flat spaces, we use scaling dimensions which are eigenvalues of dilations, and make the local counterterms have scaling dimension zero. For example, in two dimensions, we can write a local counterterm
\[ S\ni\int d^2xg_{IJ}(\lambda)\delta^\mn\partial_\mu\lambda^I(x)\partial_\nu\lambda^J(x) \]
where a coupling constant $\lambda^I$ is promoted to a dimensionless background field $\lambda^I(x)$. This local counterterm contributes to contact terms of operators $O_I$ with scaling dimension two coupled to $\lambda^I$. On the other hand, on curved spaces, we do not have dilations in general as we saw. So scaling dimensions cannot be used to write down local counterterms. A natural generalization is to use Weyl weights. Under a Weyl transformation $\gamma_\mn(x)\mapsto e^{2\sigma(x)}\gamma_\mn(x)$, a field $\Phi(x)$ with Weyl weight $w$ transforms as
\[ \Phi(x)\mapsto e^{w\sigma(x)}\Phi(x). \]
In LRG, coupling constants are promoted to background fields,\footnote{As shown in \cite{shortening}, this cannot always be done. So here we simply assume this is possible.} and they have Weyl weights zero. Then, it is natural to require local counterterms have Weyl weights zero in LRG analysis corresponding to local counterterms have scaling (or mass) dimensions zero in the usual RG analysis. For example, in two dimensions, we can have local counterterms
\[ S\ni\int d^2x\sqrt\gamma\Big\{g_{IJ}(\lambda)\gamma^\mn(x)\partial_\mu\lambda^I(x)\partial_\nu\lambda^J(x)+c(\lambda)R(x)\Big\}, \]
where $R(x)$ is the Ricci scalar. Note that this criterion is consistent with the usual classification of conformal anomalies. For example, in four dimensions, we could have a term $\nabla^2R$ in the conformal anomaly, but we can tune this away with a local counterterm
\[ S\ni\int d^4x\sqrt\gamma b(\lambda)R^2(x) \]
while the other terms such as Euler density or Weyl tensor squared terms remain genuine contributions to the conformal anomaly. Furthermore, this criterion is also consistent with local counterterms associated with higher-form symmetries \cite{qsym}. $q$-form symmetries couple to $(q+1)$-form gauge fields. Like the usual one-form gauge fields, it is natural to assign them Weyl weights zero. Therefore, we can construct local counterterms consist of them by contracting with the LC tensor. For example, one-form symmetries couple to two-form gauge fields $B$, and in four dimensions we can write a local counterterm
\[ \int d^4x\sqrt\gamma\epsilon^{\mn\rs}B_\mn B_\rs\propto\int B\wedge B, \]
which has played important roles these days. We use this criterion to write down possible local counterterms in the main text.

But how we measure `dimensions' of operators? In LRG, we can define anomalous dimensions, so we read scaling behaviours of operators through the anomalous dimensions.

From these considerations, we can learn which parts of correlation functions are physical. We will consider local counterterms which would shift one- and higher-point functions separately. Below, we surpress `flavour' indices $I$.
\begin{enumerate}
\item Regardless of a presence of background gauge fields, on even-dimensional spaces with nontrivial Riemann tensors, all relevant local counterterms are given by
\[ S\ni\int d^dx\sqrt\gamma f(\lambda)R^{d/2}(x), \]
where $R^{d/2}$ is a suitable contraction of Riemann tensors. These local counterterms can shift one-point functions of local operators with Weyl weight $(-d)$, hence making $\la O_{w=-d}(x)\ra$ unphysical. Here $f(\lambda)$ is a functional of background fields $\lambda$ only which belongs to singlets of symmetries one would like to preserve. (Note that the functional does not depend on background gauge fields. Of course one can write local counterterms with background gauge fields, and these terms are considered below separately because they do not shift the one-point functions.) If there is no such functional, the one-point functions can become physical. If the Riemann tensors are trivial, the one-point functions $\la O_{w=-d}(x)\ra$ are physical. One-point functions of current operators $\la J(x)\ra$ are physical unless $d=2$.\footnote{In two dimensions, we can have a local counterterm
\[ S\ni\tr\int d^2x\sqrt\gamma\epsilon^\mn f(\lambda)F_\mn(x) \]
which threatens the physical property of the current one-point functions.}

On odd-dimensional spaces, one-point functions are physical.
\item Concerned with higher-point functions, one can write many local counterterms which make contact terms of the correlators unphysical. On even-dimensional spaces with nontrivial Riemann tensors and nontrivial background gauge fields, we have local counterterms
\[ S\ni\tr\int d^dx\sqrt\gamma f_{k,m}(\lambda)\Big(\gamma^\mn(x)D_\mu\lambda(x)D_\nu\lambda(x)\Big)^k\Big(\gamma^\rs(x)\gamma^{\delta\gamma}(x)F_{\rho\delta}(x)F_{\sigma\gamma}(x)\Big)^mR^{d-2k-4m}(x) \]
where $0\le k\le d/2,\ 0\le m\le\lfloor d/4\rfloor$ with $d-2k-4m\ge0$. $f_{k,m}$ is a functional of background fields. These local counterterms turn the contact terms of operators $O$, currents $J$, and energy-momentum tensors $T$ unphysical. When some background fields are trivial, we have fewer local counterterms; for example, if the Riemann tensors are trivial, we have local counterterms
\[ S\ni\tr\int d^dx\sqrt\gamma f_{k,m}(\lambda)\Big(\gamma^\mn(x)D_\mu\lambda(x)D_\nu\lambda(x)\Big)^k\Big(\gamma^\rs(x)\gamma^{\delta\gamma}(x)F_{\rho\delta}(x)F_{\sigma\gamma}(x)\Big)^m \]
with $2k+4m=d$, if background (scalar) fields are trivial, we have
\[ S\ni f_m\tr\int d^dx\sqrt\gamma \Big(\gamma^\rs(x)\gamma^{\delta\gamma}(x)F_{\rho\delta}(x)F_{\sigma\gamma}(x)\Big)^mR^{d-4m}(x) \]
with $d-4m\ge0$, where $f_m$ are constants, and so on. We can also use LC tensors to construct local counterterms
\[ S\ni\tr\int d^dx\sqrt\gamma\epsilon^{\mu_1\cdots\mu_d}f_k(\lambda)D_{\mu_1}\lambda(x)\cdots D_{\mu_k}\lambda(x)F_{\mu_{k+1}\mu_{k+2}}(x)\cdots F_{\mu_{d-1}\mu_d}(x) \]
with $k=0,2,\dots,d$. These local counterterms make contact terms of operators coupled to background fields in the local counterterms unphysical.

On odd-dimensional spaces, we have to use LC tensors, and there are local counterterms such as
\[ S\ni\tr\int d^dx\sqrt\gamma\epsilon^{\mu_1\cdots\mu_d}f_k(\lambda)D_{\mu_1}\lambda(x)\cdots D_{\mu_k}\lambda(x)F_{\mu_{k+1}\mu_{k+2}}(x)\cdots F_{\mu_{d-1}\mu_d}(x) \]
with $k=1,3,\dots,d$. Thus, these local counterterms vanish when background (scalar) fields are annihilated by covariant derivatives. One can also write Chern-Simons terms constructed of background gauge or gravitational fields. These local counterterms render some parts of contact terms unphysical.
\end{enumerate}

\section{More examples}\label{moreex}
In this appendix, we present more examples, focusing on unorientable manifolds possibly with boundaries. Since correlation functions are either known or not much constrained, we only compute the identity components of the conformal groups.

\subsection{Klein bottle}
A typical example of an unorientable manifold is the Klein bottle. One way to see the manifold is as a fibre bundle over $\mbb S^1$:
\begin{equation}
    \text{KB}=(\text{KB},\pi,\mbb S^1,\mbb S^1,\mbb Z_2).\label{KBfibre}
\end{equation}
Although the total space is unorientable, each patch is orientable, and one can define a metric on it. In other words, one can define a metric locally. Since the Klein bottle locally looks like a two-torus, the metric is given by
\begin{equation}
\begin{split}
    ds^2=l^2d\tau^2&+r^2d\theta^2,\\
    \tau\in[0,2\pi),\quad&\theta\in[0,2\pi),
\end{split}\label{KB}
\end{equation}
where $\tau$ is a coordinate system which labels the base $\mbb S^1$ direction, and $\theta$ parametrizes the fibre $\mbb S^1$.

To derive boundary conditions CKVs obey, pick a point $p\in\text{KB}$. Then, consider a move along a nontrivial cycle once. (i) If the cycle is along the fibre direction, we simply come back to the original point $p'$. Thus, the single-valuedness of given local coordinate systems $x$ require
\begin{equation}
    x(p)\stackrel!=x(p').\label{KBtheta}
\end{equation}
Along the way, the coordinate $\theta$ swipes $[0,2\pi)$ once. Therefore, we get
\begin{align*}
    x(p')&\equiv(\tau(p'),\theta(p'))\\
    &=(\tau(p),\theta(p)+2\pi)\stackrel!=(\tau(p),\theta(p)),
\end{align*} 
hence a boundary condition
\begin{equation}
    (\tau,\theta)\sim(\tau,\theta+2\pi).\label{KBbctheta}
\end{equation}
(ii) If the cycle is along the base direction, we achieve a point $p''$ such that $\pi(p)=\pi(p'')$ while $\tau$ swipes $[0,2\pi)$ once. Although $p$ and $p''$ have the same base coordinate, we have to perform a $\mbb Z_2$ transformation on the fibre $\mbb Z_2:\theta\mapsto-\theta$. Since the points $p$ and $p''$ are identical, we get
\begin{equation}
    x(p'')\stackrel!=x(p),\label{KBtau}
\end{equation}
or
\begin{align*}
    x(p'')&\equiv(\tau(p''),\theta(p''))\\
    &=(\tau(p)+2\pi,-\theta(p))\stackrel!=(\tau(p),\theta(p)),
\end{align*}
hence another boundary condition
\begin{equation}
    (\tau,\theta)\sim(\tau+2\pi,-\theta).\label{KBbctau}
\end{equation}

Since a conformal group is a subgroup of the diffeomorphism group, we would like to consider a coordinate transformation\footnote{Although both $x$ and $x'$ are bijective, the difference $\xi$ can be neither injection nor surjection. For example, if $x'=x$, $\xi=0$, and it is neither injective nor surjective.}
\[ x\mapsto x'(x):=x+\xi(x). \]
Since both $x$ and $x'$ serve as local coordinate systems, they both satisfy the boundary conditions. Then, we have
\begin{align*}
    x'(p')&\equiv(\tau'(p'),\theta'(p'))=(\tau'(p),\theta'(p)+2\pi)\equiv(\tau(p)+\xi^\tau(x(p)),\theta(p)+\xi^\theta(x(p))+2\pi)\\
    &\equiv x(p')+\xi(x(p'))\equiv(\tau(p)+\xi^\tau(x(p')),\theta(p)+2\pi+\xi^\theta(x(p'))).
\end{align*}
Comparing two expressions on RHSs with $x(p')=(\tau(p),\theta(p)+2\pi)$, we get
\begin{equation}
    \xi_\mu(\tau,\theta+2\pi)=\xi_\mu(\tau,\theta).\label{KBCKVtheta}
\end{equation}
A similar argument yields
\begin{align*}
    x'(p'')&\equiv(\tau'(p''),\theta'(p''))=(\tau'(p)+2\pi,-\theta'(p))\equiv(\tau(p)+\xi^\tau(x(p))+2\pi,-\theta(p)-\xi^\theta(x(p)))\\
    &\equiv(\tau(p'')+\xi^\tau(x(p'')),\theta(p'')+\xi^\theta(x(p'')))=(\tau(p)+2\pi+\xi^\tau(x(p'')),-\theta(p)+\xi^\theta(x(p''))),
\end{align*}
or comparing the two expressions on RHSs we obtain another boundary condition
\begin{equation}
    (\xi_\tau(\tau+2\pi,-\theta),\xi_\theta(\tau+2\pi,-\theta))=(\xi_\tau(\tau,\theta),-\xi_\theta(\tau,\theta)).\label{KBCKVtau}
\end{equation}

Now we are ready to solve CKE. Since the Klein bottle is locally the same as a two-torus, we obtain the same CKE as the case:
\begin{equation}
\begin{split}
    0&=\partial_\tau\xi_\tau(x)-\left(\frac lr\right)^2\partial_\theta\xi_\theta(x),\\
    0&=\partial_\tau\xi_\theta(x)+\partial_\theta\xi_\tau(x).
\end{split}\label{CKEKB}
\end{equation}
Because of the boundary condition (\ref{KBCKVtheta}), $\xi_\mu$ must have a form
\begin{equation}
    \xi_\mu(\tau,\theta)=\sum_{m\in\mbb Z}e^{im\theta}\xi_\mu^{(m)}(\tau).\label{KBansatz}
\end{equation}
The reality condition $\xi_\mu=\xi_\mu^*$ is given by
\begin{equation}
    \forall m\in\mbb Z,\forall\tau\in[0,2\pi),\quad\xi_\mu^{(m)}(\tau)=\xi_\mu^{(-m)*}(\tau).\label{xi_muKB}
\end{equation}
Substituting the ansatz (\ref{KBansatz}) in (\ref{CKEKB}), one obtains
\begin{equation}
    \forall m\in\mbb Z,\forall\tau\in[0,2\pi),\quad\partial_\tau\begin{pmatrix}\xi_\tau^{(m)}(\tau)\\\xi_\theta^{(m)}(\tau)\end{pmatrix}=im\begin{pmatrix}0&\left(\frac lr\right)^2\\-1&0\end{pmatrix}\begin{pmatrix}\xi_\tau^{(m)}(\tau)\\\xi_\theta^{(m)}(\tau)\end{pmatrix}.\label{CKEKB'}
\end{equation}
The matrix on RHS can be diagonalized as before, and the PDE can be solved with ease:
\begin{equation}
    \forall m\in\mbb Z,\forall\tau\in[0,2\pi),\quad\begin{pmatrix}\xi_\tau^{(m)}(\tau)\\\xi_\theta^{(m)}(\tau)\end{pmatrix}=\begin{pmatrix}\frac{il}r\left(e^{lm\tau/r}c_\tau^{(m)}-e^{-lm\tau/r}c_\theta^{(m)}\right)\\e^{lm\tau/r}c_\tau^{(m)}+e^{-lm\tau/r}c_\theta^{(m)}\end{pmatrix},\label{xi_muKBsol}
\end{equation}
where $c_\mu^{(m)}$ are complex constants.

Here, on the ansatz, the other boundary condition (\ref{KBCKVtau}) imposes
\begin{equation}
    \forall m\in\mbb Z,\forall\tau\in[0,2\pi),\quad\xi_\tau^{(-m)}(\tau+2\pi)=\xi_\tau^{(m)}(\tau)\&\xi_\theta^{(-m)}(\tau+2\pi)=-\xi_\theta^{(m)}(\tau).\label{KBCKVtau'}
\end{equation}
On the solution (\ref{xi_muKBsol}), this means
\begin{equation}
    0=c_\tau^{(m)}(e^{4\pi lm/r}-1),\quad c_\theta^{(m)}=-e^{-2\pi lm/r}c_\tau^{(-m)}.\label{c_muKB}
\end{equation}
The first condition says only $c_\tau^{(0)}$ can be nonzero. Then, the second condition implies only $c_\theta^{(0)}$ can be nonzero, which is given by
\begin{equation}
    c_\theta^{(0)}=-c_\tau^{(0)}.\label{c_muKB'}
\end{equation}
Furthermore, the reality condition (\ref{xi_muKB}) on the solution (\ref{xi_muKBsol}) reduces to
\begin{equation}
    \forall m\in\mbb Z,\quad c_\theta^{(m)}=c_\tau^{(-m)*}.\label{xi_muKB'}
\end{equation}
Combinining the two conditions (\ref{c_muKB'}) and (\ref{xi_muKB'}), it turns out that $c_\tau^{(0)}$ is pure imaginary:
\[ c_\tau^{(0)*}=c_\theta^{(0)}=-c_\tau^{(0)}. \]
Collecting all the results, we arrive
\begin{equation}
    \begin{pmatrix}\xi_\tau(\tau,\theta)\\\xi_\theta(\tau,\theta)\end{pmatrix}=\begin{pmatrix}\frac{2il}rc_\tau^{(0)}\\0\end{pmatrix}.\label{xiKB}
\end{equation}
Note that $\xi_\mu$ is real as required because $c_\tau^{(0)}$ is pure imaginary. Thus, conformal transformations on $\text{KB}$ are generated by constant shifts in the $\tau$-direction. More explicitly, they are generated by
\begin{equation}
    \xi(\tau,\theta)=\xi^\tau(\tau,\theta)\partial_\tau=\frac{2i}{lr}c_\tau^{(0)}\partial_\tau.\label{KBgen}
\end{equation}
Since $\tau+2\pi$ and $\tau$ are the same point on the base space, they are identified, and the group generated by $P_\tau:=\partial_\tau$ is $U(1)$. Because the conformal factor is trivial, (the identity component of) the conformal group reduces to (the identity component of) the isometry group. To summarize, we have shown
\begin{equation}
    \text{Conf}_0(\text{KB})\simeq U(1)\simeq\text{Isom}_0(\text{KB}),\label{confKB}
\end{equation}
which reproduces the well-known result mentioned in \cite{Pol}. A reflection
\[ I_\theta:\theta\mapsto-\theta \]
preserves the metric (\ref{KB}), and forms $\mbb Z_2$. Thus, the (full) conformal group must contain
\[ U(1)\times\mbb Z_2\subset\text{Conf}(\text{KB}). \]

\subsection{M\"obius strip}
As in the case of the Klein bottle, one way to see this manifold with boundaries is as a fibre bundle over $\mbb S^1$:
\begin{equation}
    \text{MS}=(\text{MS},\pi,\mbb S^1,[-1,1],\mbb Z_2).\label{MSfibre}
\end{equation}
Since this space is locally the same as the cylinder, we can locally define a metric
\begin{equation}
\begin{split}
    ds^2=r^2&d\tau^2+dy^2,\\
    \tau\in[0,2\pi),&\quad y\in[-1,1],
\end{split}\label{MS}
\end{equation}
where $\tau$ is the base $\mbb S^1$ direction, and $y$ parametrizes the fibre $[-1,1]$.

To derive boundary conditions satisfied by CKVs, we follow the same logic as in the previous example; we consider a coordinate transformation $x\mapsto x'(x):=x+\xi(x)$, follow a transportation from a point $p\in\text{MS}$ along a nontrivial cycle to the identical point $p'$, and decompose $x'(p')$ in two ways. Then, we obtain
\begin{align*}
    x'(p')&\equiv(\tau'(p'),y'(p'))=(\tau'(p)+2\pi,-y'(p))\equiv(\tau(p)+\xi^\tau(x(p))+2\pi,-y(p)-\xi^y(x(p)))\\
    &\equiv x(p')+\xi(x(p'))=(\tau(p)+2\pi+\xi^\tau(x(p')),-y(p)+\xi^y(x(p'))),
\end{align*}
or comparing two expressions on RHSs using $x(p')=(\tau(p)+2\pi,-y(p))$, we get
\begin{equation}
    (\xi_\tau(\tau+2\pi,-y),\xi_y(\tau+2\pi,-y))=(\xi_\tau(\tau,y),-\xi_y(\tau,y)).\label{MSCKVtau}
\end{equation}
To preserve the boundaries $y=\pm1$, it is natural to impose another boundary condition
\begin{equation}
    \forall\tau\in[0,2\pi),\quad\xi_y(\tau,y=1)\stackrel!=0\stackrel!=\xi_y(\tau,y=-1).\label{MSCKVy}
\end{equation}

Now we solve the CKE. Since $\xi_\mu$ is periodic up to sign in $\tau$-direction, it is convenient to take
\begin{equation}
    \xi_\mu(\tau,y)=\sum_{m\in\mbb Z}e^{im\tau}\xi_\mu^{(m)}(y),\label{MSansatz}
\end{equation}
and let the `mode' $\xi_\mu^{(m)}(y)$ take care of signs. Then, similar computations as before give
\begin{equation}
    \begin{pmatrix}\xi_\tau(\tau,y)\\\xi_y(\tau,y)\end{pmatrix}=\sum_{m\in\mbb Z}e^{im\tau}\begin{pmatrix}-ir\left(e^{my/r}c_\tau^{(m)}-e^{-my/r}c_\tau^{(-m)*}\right)\\e^{my/r}c_\tau^{(m)}+e^{-my/r}c_\tau^{(-m)*}\end{pmatrix},\label{xi_muMSsol}
\end{equation}
in which the reality condition is already implemented. On the solution, the boundary condition (\ref{MSCKVtau}) imposes
\begin{equation}
    \forall m\in\mbb Z,\quad c_\tau^{(-m)*}=-c_\tau^{(m)}.\label{c_muMS}
\end{equation}
The other boundary condition (\ref{MSCKVy}) requires
\begin{equation}
    \forall m\in\mbb Z,\quad-e^{2m/r}c_\tau^{(m)}=c_\tau^{(-m)*}=-e^{-2m/r}c_\tau^{(m)}.\label{c_muMS2}
\end{equation}
This means only $c_\tau^{(0)}$ can be nonzero, which is pure imaginary due to the same condition. Collecting all the results, we arrive
\begin{equation}
    \begin{pmatrix}\xi_\tau(\tau,y)\\\xi_y(\tau,y)\end{pmatrix}=\begin{pmatrix}-2irc_\tau^{(0)}\\0\end{pmatrix}.\label{xiMS}
\end{equation}
This is again real as required because $c_\tau^{(0)}$ is pure imaginary. Thus, conformal transformations on $\text{MS}$ are constant shifts in the $\tau$-direction which are generated by
\begin{equation}
    \xi(\tau,y)=\xi^\tau(\tau,y)\partial_\tau=-\frac{2ic_\tau^{(0)}}r\partial_\tau,\label{MSgen}
\end{equation}
and they form a compact group $U(1)$ due to the periodic identification $\tau+2\pi\sim\tau$. Since the conformal factor is trivial, the group reduces to (the identity component of) the isometry group. To conclude, we have shown
\begin{equation}
    \text{Conf}_0(\text{MS})\simeq U(1)\simeq\text{Isom}_0(\text{MS}).\label{confMS}
\end{equation}
This again reproduces the result mentioned in \cite{Pol}. As in the previous example, a reflection
\[ I_y:y\mapsto-y \]
forms a subgroup $\mbb Z_2$ of the (full) conformal group:
\[ U(1)\times\mbb Z_2\subset\text{Conf}(\text{MS}). \]

\subsection{$\mbb{RP}^d$}
This space is given by
\begin{equation}
    \mbb{RP}^d:=(\mbb R^d\cup\{\infty\})/\sim,\label{RPd}
\end{equation}
where two points $p$ and $p'$ are identified $p\sim p'$ if
\begin{equation}
    x^\mu(p')=-\frac{x^\mu(p)}{\delta_\rs x^\rho(p)x^\sigma(p)}.\label{simRPd}
\end{equation}
Since the metric is flat, we can borrow the well-known form of CKVs on $\mbb R^d\cup\{\infty\}$
\begin{equation}
    \xi^\mu(x)=a^\mu+m^\mn x_\nu+\lambda x^\mu+(2b^\nu x_\nu x^\mu-\delta_\rs x^\rho x^\sigma b^\mu),\label{xiRd}
\end{equation}
where $a,m,\lambda,$ and $b$ parametrize translation, rotation, dilation, and special conformal transformation, respectively. On $\mbb{RP}^d$, we have to impose a boundary condition originating from the identification (\ref{simRPd}). We use our simple trick based on the single-valuedness of local coordinate systems. Calculating $x'(p')$ in two ways, we obtain
\begin{align*}
    x'(p')&\equiv(x^{'1}(p'),\dots,x^{'d}(p'))=\left(-\frac{x^{'1}(p)}{|x'(p)|^2},\dots,-\frac{x^{'d}(p)}{|x'(p)|^2}\right)\\
    &\equiv\left(-\frac{x^1(p)+\xi^1(p)}{|x(p)+\xi(x(p))|^2},\dots,-\frac{x^d(p)+\xi^d(p)}{|x(p)+\xi(x(p))|^2}\right)\\
    &\equiv(x^1(p')+\xi^1(x(p')),\dots,x^d(p')+\xi^d(x(p')))\\
    &=\left(-\frac{x^1(p)}{|x(p)|^2}+\xi^1(x(p')),\dots,-\frac{x^d(p)}{|x(p)|^2}+\xi^d(x(p'))\right),
\end{align*}
or using $x^\mu(p')=-x^\mu(p)/|x(p)|^2$, we obtain a boundary condition
\begin{equation}
    \xi^\mu\left(-\frac x{|x|^2}\right)=\frac{x^\mu}{|x|^2}-\frac{x^\mu+\xi^\mu(x)}{|x+\xi(x)|^2}.\label{RPdbc}
\end{equation}
LHS can be computed easily:
\[ (\text{LHS})=a^\mu-\frac1{|x|^2}m^\mn x_\nu-\frac\lambda{|x|^2}x^\mu+\left(\frac{2b^\nu x_\nu}{|x|^4}x^\mu-\frac1{|x|^2}b^\mu\right). \]
On the other hand, up to the first order in each parameter, RHS reduces to
\[ (\text{RHS})=\left(\frac{2a^\nu x_\nu}{|x|^4}x^\mu-\frac1{|x|^2}a^\mu\right)-\frac1{|x|^2}m^\mn x_\nu+\frac\lambda{|x|^2}x^\mu+b^\mu. \]
One notices at once that $a$ by itself cannot survive the boundary condition (\ref{RPdbc}), but since it has the form of special conformal transformation on RHS, it can survive if we set $a^\mu=b^\mu$. This is also the case for $b$; it cannot survive by itself, but since it is nothing but a translation in RHS, it can survive if we set $b=a$. Thus, translation and special conformal transformation survive the boundary condition by mixing them. This result was expected because special conformal transformations are realized as translations associated with inversions $I:x^\mu\mapsto-x^\mu/|x|^2$. One can easily read that rotations $m$ survive the boundary condition while dilations $\lambda$ do not. Thus, CKV on $\mbb{RP}^d$ is given by
\begin{equation}
	\xi(x)=\xi^\mu(x)\partial_\mu=a^\mu(\partial_\mu+2x_\mu x^\nu\partial_\nu-x^\nu x_\nu\partial_\mu)+\frac12m^\mn(x_\nu\partial_\mu-x_\mu\partial_\nu).\label{RPdgen}
\end{equation}

Let us work out the group generated by the vector field. Defining
\[ Q_\mu:=\frac12\left(\partial_\mu+2x_\mu x^\nu\partial_\nu-x^\nu x_\nu\partial_\mu\right),\quad M_\mn:=x_\nu\partial_\mu-x_\mu\partial_\nu, \]
one can easily check they satisfy commutation relations
\begin{align*}
	[Q_\mu,Q_\nu]=M_\mn,\quad[Q_\mu,M_\rs]=-\delta_{\mu\rho}Q_\sigma+\delta_{\mu\sigma}Q_\rho,\\
	[M_\mn,M_\rs]=\delta_{\mu\rho}M_{\nu\sigma}-\delta_{\mu\sigma}M_{\nu\rho}-\delta_{\nu\rho}M_{\mu\sigma}+\delta_{\nu\sigma}M_{\mu\rho}.
\end{align*}
This is nothing but the Lie algebra $\mfrak{so}(d,1)$ because if one defines\footnote{As long as we are concerned with the algebra, both signs of $J_{\mu0}$ work.}
\[ J_\mn:=M_\mn,\quad J_{\mu0}:=\pm iQ_\mu, \]
the generator $J_{MN}=-J_{NM}$ satisfy the defining commutation relations of $\mfrak{so}(d,1)$
\[ [J_{MN},J_{RS}]=\eta_{MR}J_{NS}-\eta_{MS}J_{NR}-\eta_{NR}J_{MS}+\eta_{NS}J_{MR} \]
with $\eta_{MN}=\text{diag}(-1,+1,\dots,+1)$ and $M=0,1,\dots,d$. So we conclude
\begin{equation}
    \text{Conf}_0(\mbb{RP}^d)\simeq SO(d,1),\label{confRPd}
\end{equation}
in accord with \cite{realpro}.

Since the forms of correlation functions on $\mbb{RP}^d$ were discussed in the papers, we do not repeat here. (Basically, one uses rotation symmetry to restrict $x_j$ dependence to $x_j\cdot x_k$ dependence, and then imposes inversion covariance of conformal tensors.)

\end{document}